\UseRawInputEncoding 
\documentclass[preprint,12pt,fleqn]{elsarticle}

\usepackage{lineno,hyperref}
\modulolinenumbers[5]

\journal{Journal of \LaTeX\ Templates}

\usepackage{amsmath}
\usepackage{amsthm}
\usepackage{amssymb}
\usepackage{eucal}
\usepackage{mathrsfs}
\usepackage{subfigure}
\usepackage{mathtools} 
\usepackage{nccmath} 
\usepackage{fix-cm}
\usepackage{bm}

\def \div{\mbox{\rm div}}

\def \Ga {\Gamma}

\def \ep {\varepsilon}

\def \Om {\Omega}

\begin{document}

\begin{frontmatter}

\title{Self-support topology optimization considering distortion for metal additive manufacturing}
\author[mymainaddress]{Takao Miki\corref{mycorrespondingauthor}}
\cortext[mycorrespondingauthor]{Corresponding author}
\ead{mikit@orist.jp}
\address[mymainaddress]{Osaka Research Institute of Industrial Science and Technology, 7-1, Ayumino-2, Izumi-city, Osaka, 594-1157, Japan}
\begin{abstract}
This paper proposes a self-support topology optimization method that considers distortion to improve the manufacturability of additive manufacturing.
First, a self-support constraint is proposed that combines an overhang angle constraint with an adjustable degree of the dripping effect and a thermal constraint for heat dissipation in the building process.
Next, we introduce a mechanical model based on the inherent strain method in the building process and propose a constraint that can suppress distortion.
An optimization problem is formulated to satisfy all constraints, and an optimization algorithm based on level-set-based topology optimization is constructed. 
Finally, two- and three-dimensional optimization examples are presented to validate the effectiveness of the proposed topology optimization method.
\end{abstract}
\end{frontmatter}
\section{Introduction}
The combination of topology optimization \cite{bendsoe1988generating,suzuki1991homogenization} and additive manufacturing (AM) technology enables innovative product development, which is not possible with conventional manufacturing.
This technology is becoming more prevalent for creating high-performance and lightweight integral parts, particularly in the aviation and automotive industries \cite{emmelmann2011laser,zhu2016topology,jankovics2019customization}.
Among the various AM methods, laser powder bed fusion (LPBF) is the most widely used method for metal-based AM.
The LPBF process creates a three-dimensional (3D) object by irradiating a metal powder bed with a laser based on sliced two-dimensional (2D) data from a CAD file and repeating the melting and solidification process layer-by-layer.
Although the LPBF process can create complex geometric parts, it is associated with manufacturing problems, such as overhang limitations, residual stress, distortion, and overheating.
One way of avoiding these issues is attaching a support structure to the part.
The support structure not only supports the overhang shape but also suppresses distortion and overheating.
In general, the support structure must be removed after manufacturing, for example, by machining.
Therefore, because the consumption of support material leads to an increase not only in the cost of the material but also in the cost of its removal, it is desirable to obtain a support-free shape during the design phase. 

In this context, topology optimization has received considerable attention as an effective design tool for AM.
Topology optimization is a method of creating an optimal shape under the governing physics to maximize structural performance.
In recent years, topology optimization methods that consider the manufacturability challenges in AM have been actively investigated.
The most noticeable issue in AM that limits manufacturability is the overhang limitation.
In other words, there is a limit to the overhang angle, the angle between the downward-facing surface and the base plate.
If the overhang angle exceeds a threshold angle, the overhang shape may collapse under its own weight. In this study, this overhang is termed the overhanging region.
Several self-support topology optimization methods have been proposed to eliminate overhanging regions. These approaches use two main approaches: a density filter and an explicit angle constraint.
Density filtering is a common technique used in density-based topology optimization frameworks to ensure a self-supporting structure \cite{gaynor2016topology,langelaar2016topology,langelaar2017additive}.
The optimal structures obtained using this approach have sharp inner corners, and several methods have been proposed to prevent their occurrence \cite{van2018continuous,garaigordobil2018new,thore2019penalty}.
These methods generate rounded shapes and do not strictly satisfy the angle constraint.
The explicit angle constraint is a technique that can be applied to both density- and level-set-based topology optimization frameworks.
This technique can be easily incorporated into the above frameworks using a gradient of the element density or level-set function.
However, the constraint on the normal direction of the geometry is known to create downward convex shapes that cannot be manufactured.
In the density-based framework, Qian \cite{qian2017undercut} proposed a suppression method that added perimeter and grayness constraints but reported that the parameters of each constraint are not easy to set.
Using the level-set-based framework, Wang et al. \cite{wang2018level} facilitated the detection of the downward convex shapes by formulating the overhang angle constraint as a domain integral but did not discuss how to suppress it.
Allaire et al. \cite{allaire2017structural} successfully suppressed the downward convex shapes by imposing mechanical constraints to prevent the part that is to be manufactured from collapsing under its weight during the building process.
However, this method does not sufficiently eliminate overhanging regions and may adversely affect structural compliance because it considers the self-weight of the entire geometry.
Therefore, it is necessary to establish a method for eliminating the overhanging region and downward convex shape when the explicit angle constraint technique is applied.

Recently, residual stress, distortion, and overheating have been the focus of attention.
Residual stress and distortion are caused by the inelastic strain generated in the process of melting and solidification of the metal material, which leads to manufacturing failure and deterioration of strength and dimensional accuracy.
Furthermore, shapes that block heat flow during laser irradiation cause porosity and degrade surface quality, a phenomenon that is known as overheating.
AM process methods for predicting part-scale residual stress and distortion can be summarized into two methods: the inherent strain method \cite{keller2014new,setien2019empirical,chen2019inherent,liang2019modified,prabhune2020fast} and thermo-mechanical analysis method \cite{papadakis2014numerical,hodge2014implementation,li2016multiscale,mukherjee2017improved,denlinger2017thermomechanical,chiumenti2017numerical,li2017efficient}.
The inherent strain method is an elastic or elasto-plastic analysis method that uses a strain field that is obtained experimentally or analytically.
This method can efficiently predict part-scale residual stress and distortion because it is computationally inexpensive, but it cannot predict overheating, as thermal information is not generated when it is applied.
The thermo-mechanical analysis method predicts the residual stress and distortion using thermal information obtained by thermal analysis.
Although this method can predict overheating, it is computationally expensive because it requires coupled and nonlinear analyses to be performed.
To suppress residual stress, distortion, and overheating, several methods have been proposed that incorporate the aforementioned AM analytical models into a topology optimization framework \cite{wildman2017topology,allaire2018taking,MIKI2021103558,allaire2018optimizing,zhou2019topology,wang2020optimizing,misiun2021topology,miki2022topology,xu2022residual,ranjan2022controlling}.
However, few studies have combined the these methods with the overhang angle constraint.

This paper provides a new self-support topology optimization that considers distortion.
The remainder of this paper is organized as follows.
Section \ref{sec:2} presents a brief overview of the level-set-based topology optimization method.
In Section \ref{sec:3}, we formulate an overhang angle constraint using a Helmholtz-type partial differential equation (PDE) and propose a method to suppress the downward convex shape using a thermal model.
In Section \ref{sec:4}, we introduce the AM process model based on the inherent strain method and present a formulation of the distortion constraint.
We then formulate an optimization problem considering multiple constraints in Section \ref{sec:5}.
In Section \ref{sec:6}, we construct an optimization algorithm using the finite element method (FEM).
Section \ref{sec:7} presents 2D and 3D design examples to demonstrate the utility of the proposed optimization method.
Finally, section \ref{sec:8} presents the conclusions of this study.
\section{Level-set-based topology optimization}\label{sec:2}
Topology optimization is a material distribution problem that determines the presence or absence of a material within a fixed design domain $D \subset \mathbb{R}^{N}$.
The fixed design domain $D$ is composed of the material domain $\mathit{\Omega}$ and void domain $D \backslash \mathit{\Omega}$, which are distinguished by a characteristic function.
This study employs level-set-based topology optimization \cite{yamada2010topology} using a reaction-diffusion equation.
In this method, the structural boundary $\partial \mathit{\Omega}$ is expressed by the isosurface of the level-set function.
Furthermore, the material domain $\mathit{\Omega}$ and void domain $D \backslash \mathit{\Omega}$ are distinguished by the sign of the level set function, as follows:
\begin{equation}
\left\{ \begin{array} { l l }
{ \phi ( \bm{x} ) > 0 } & { \text { for } \bm{x} \in \mathit{\Omega} } \\
{ \phi ( \bm{x} ) = 0 } & { \text { for } \bm{x} \in \partial \mathit{\Omega} } \\
{ \phi ( \bm{x} ) < 0 } & { \text { for } \bm{x} \in D\backslash \mathit{\Omega} }.
\end{array} \right.
\label{eq:lsf1}
\end{equation}
The characteristic function is expressed using the level-set function $\phi$ as follows:
\begin{equation}
\chi_{\phi}=\left\{\begin{array}{ll}
1 & \text { for } \phi(x) \geq 0 \\
0 & \text { for } \phi(x)<0.
\end{array}\right. 
\label{eq:lsf2}
\end{equation}
The optimal material distribution $\chi_{\phi}$, that is, the distribution of the level set function $\phi$, is determined by solving the time evolution equation as follows:
\begin{equation}
	\frac { \partial \phi(t) } { \partial t } = - K(J'-\tau\nabla^{2}\phi),
	\label{eq:lsf3}
\end{equation}
where $t$ is the fictitious time, $K$ is a positive parameter, $J'$ is the topological derivative \cite{yamada2010topology,amstutz2006new,allaire2005structural} of the target optimization problem, and $\tau>0$ is a regularization parameter.
The diffusion term not only ensures the smoothness of the level set function $\phi$ but also controls the geometric complexity of the optimal configuration by adjusting the regularization parameter $\tau$.
We set $K=0.7$ and $\tau=5.0\times10^{-4}$.
\section{Self-support constraint}\label{sec:3}
This section introduces an overhang angle constraint in the level-set-based framework and a method for suppressing the downward convex shape.
\subsection{Smoothed characteristic function based on Helmholtz-type PDE}\label{sec:3.1}
Allaire et al. \cite{allaire2017structural} and Wang et al. \cite{wang2018level} used the gradient of the level set function to directly formulate the overhang angle constraint.
In this study, a Helmholtz-type partial differential equation (PDE) is introduced to limit the overhang angle using the gradient of the projected characteristic function $\chi_{\phi}$.
The physical variable $\psi\in H^{1}(D)$ and its governing equation are defined as follows: 
\begin{align}
\begin{cases}
-a L^{2}\nabla^{2}\psi+\psi=\chi_{\phi} \hspace{0.5cm}&\text{in}\hspace{0.3cm}D
\\
\bm{n}\cdot\nabla\psi=0 \hspace{3.85cm}&\text{on}\hspace{0.3cm}\partial\mathit{\Omega},
\end{cases}
\label{eq:ovhg1}
\end{align}
where $a \in \mathbb{R}^{+}$ is the isotropic diffusion coefficient, $L$ is the representative length, and $\bm{n}$ is the outward normal vector.
The diffusion coefficient $a$ affects the transition width $\psi$.
In other words, the evaluation of the overhanging region can be controlled by adjusting the diffusion coefficient $a$.
Furthermore, it can also control the downward convex shape.
\subsection{Overhang angle constraint function and its derivative}\label{sec:3.2}
\begin{figure}[htbp]
	\begin{center}
		\includegraphics[width=8.0cm]{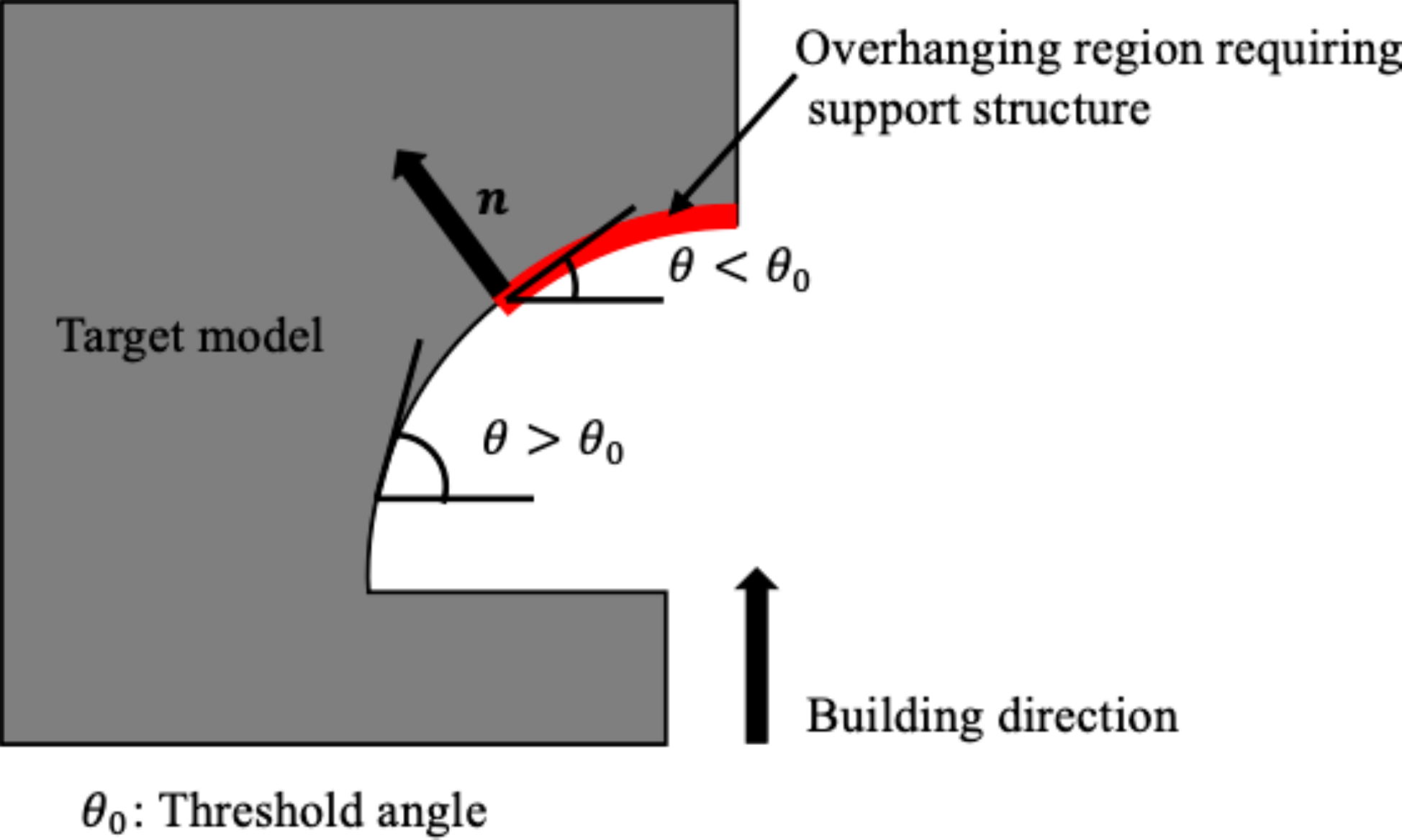}
		\caption{Schematic of the overhanging region in additive manufacturing.}
		\label{fig:ovhg_1}
	\end{center}
\end{figure}
Fig. \ref{fig:ovhg_1} shows an example of the overhanging region in additive manufacturing.
A support structure is required to build an overhanging region above a certain angle to the horizontal plane.
In the explicit overhang angle constraint, a common method for detecting overhanging regions is evaluating the angle between the normal vector $\bm{n}$ from the structural boundary $\partial \mathit{\Omega}$ and the building direction.
In this study, the overhanging region is detected by directly evaluating the inner product of the normal vector $\bm{n}_{\psi}:=\nabla\psi$ from the structural boundary $\partial \mathit{\Omega}$ and threshold angle vectors $\bm{d}_{1}$ and $\bm{d}_{2}$.
For the 2D problem, the threshold angle vectors $\bm{d}_{1}$ and $\bm{d}_{2}$ are given by
\begin{equation}
\bm{d}_{1}=\left(\begin{array}{c}
	-\cos \theta_{0} \\
	-\sin \theta_{0}
\end{array}\right), \quad \bm{d}_{2}=\left(\begin{array}{c}
	\cos \theta_{0} \\
	-\sin \theta_{0}
\end{array}\right).
\label{eq:ovhg2}
\end{equation}
An overhanging region is detected when both inner products take positive values.
Therefore, the condition for constraining the overhang angle is given by
\begin{align}
{\int_{D} R(\nabla\psi\cdot\bm{d}_{1})R(\nabla\psi\cdot\bm{d}_{2})\mathrm{~d}\mathit\Omega=0},
\label{eq:ovhg3}
\end{align}
where $R(s):=(s+|s|) / 2$ denotes the ramp function.
Furthermore, the overhang angle constraint normalizes the above equation and is defined as
\begin{align}
{G_{o}=\int_{D} R(\sqrt{a}L\nabla\psi\cdot\bm{d}_{1})R(\sqrt{a}L\nabla\psi\cdot\bm{d}_{2})\mathrm{~d}\mathit\Omega},
\label{eq:ovhg4}
\end{align}
\begin{figure}[htbp]
	\begin{center}
		\centering
		\subfigure[$\psi$]{\includegraphics[width=6.0cm]{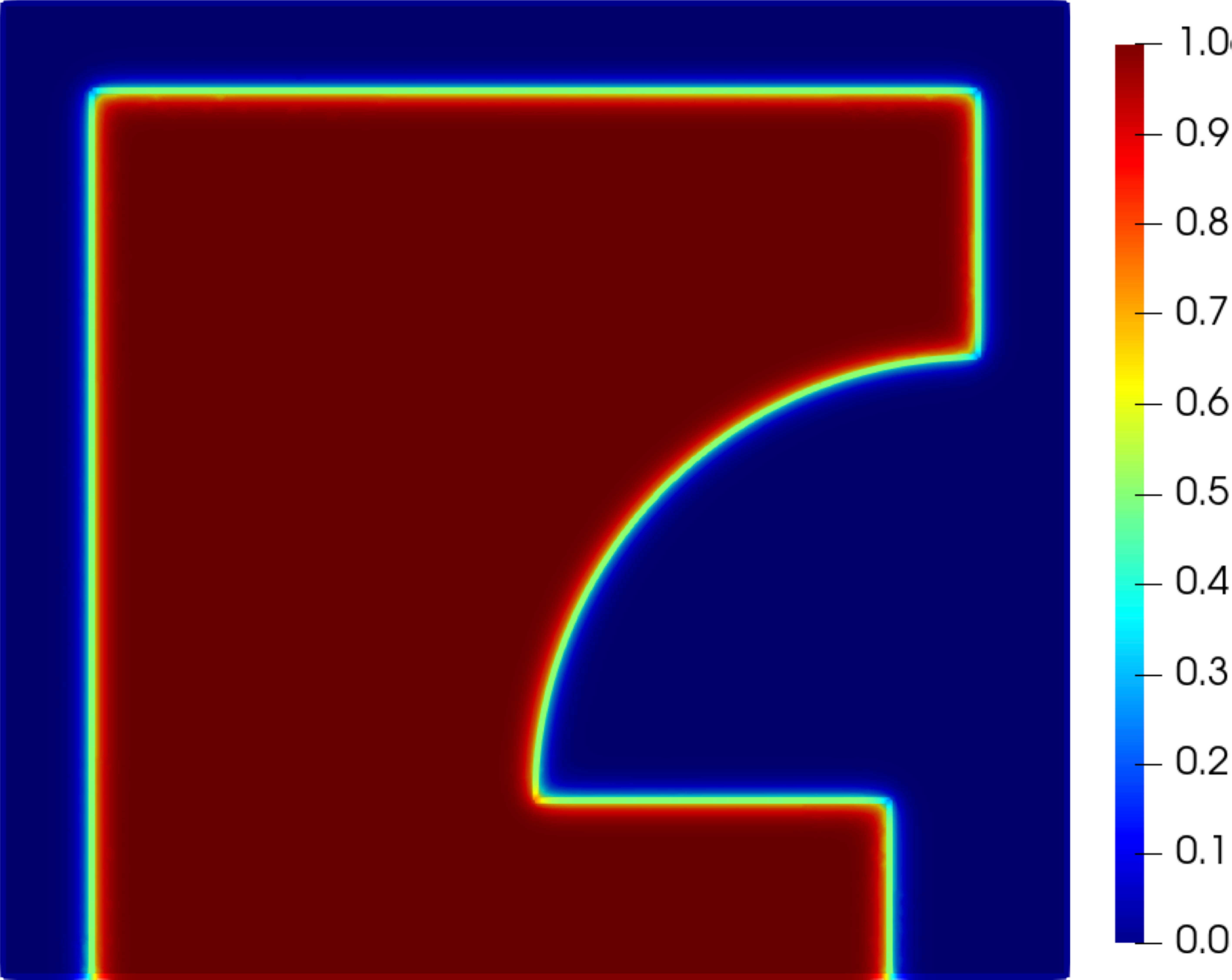}}
		\hspace{0.5cm}
		\subfigure[$\sqrt{a}L\nabla\psi\cdot\bm{d}_{1}$]{\includegraphics[width=6.0cm]{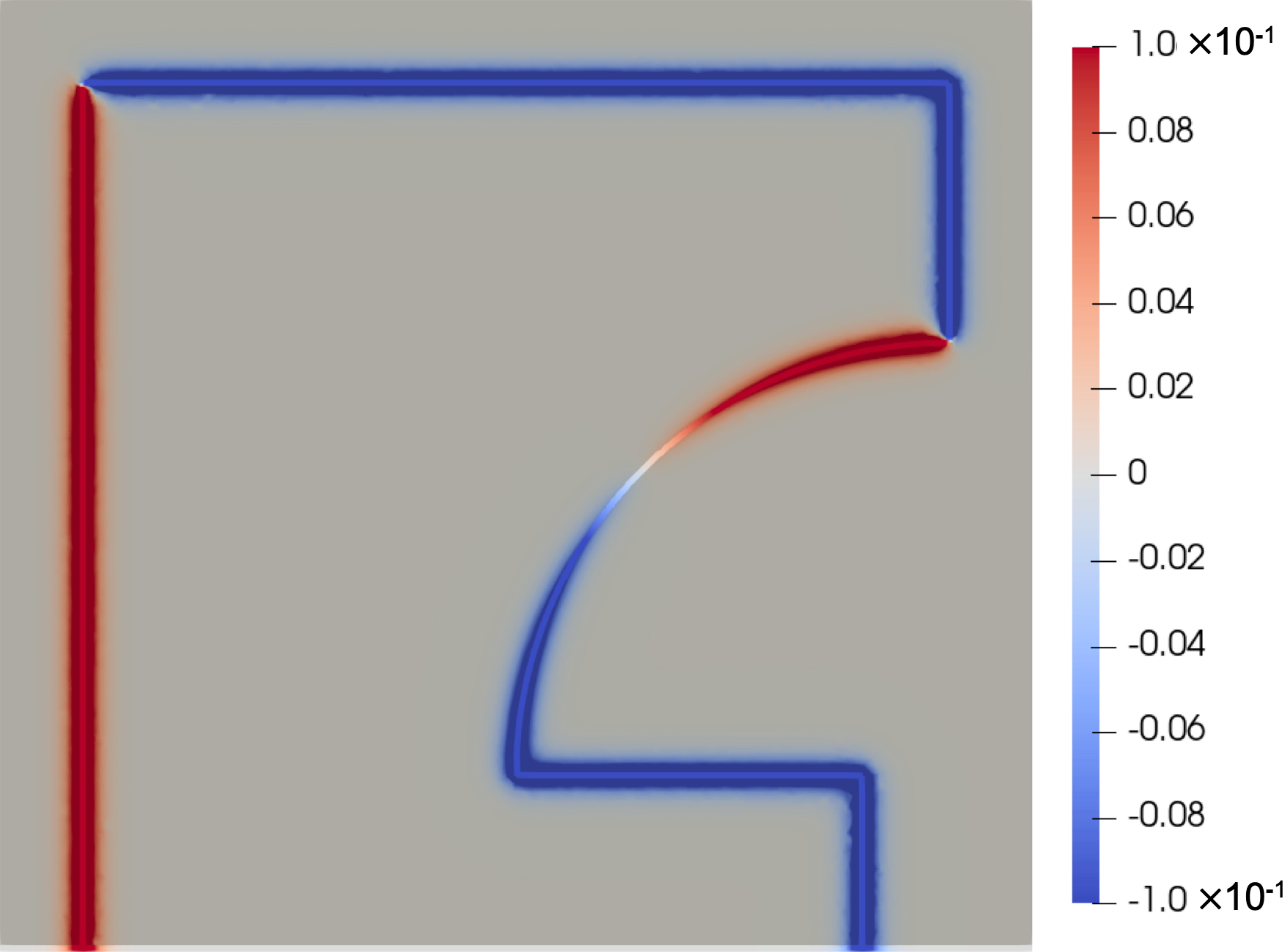}}
						\hspace{0.5cm}
		\subfigure[$\sqrt{a}L\nabla\psi\cdot\bm{d}_{2}$]{\includegraphics[width=6.0cm]{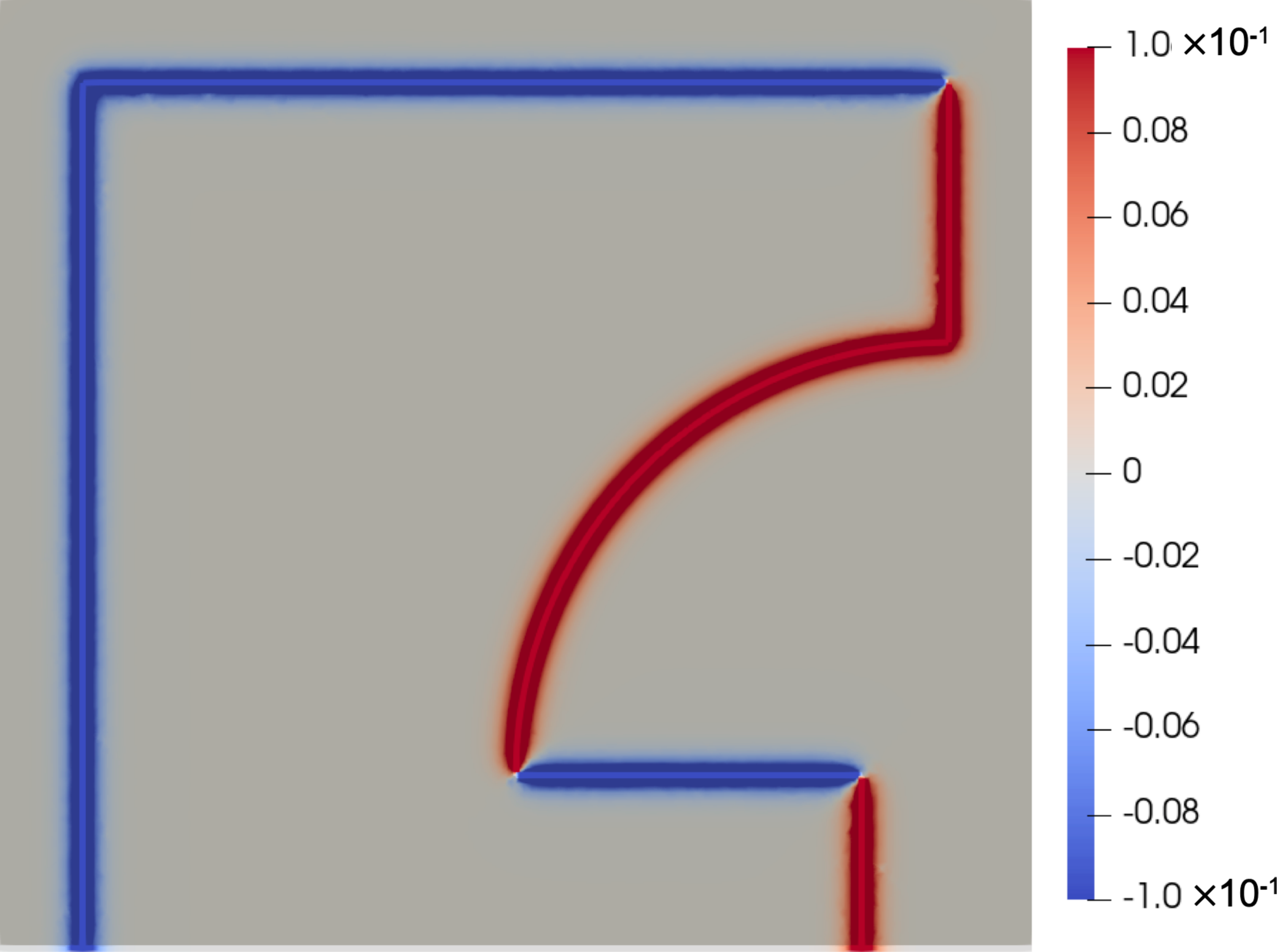}}
						\hspace{0.5cm}
		\subfigure[$R(\sqrt{a}L\nabla\psi\cdot\bm{d}_{1})R(\sqrt{a}L\nabla\psi\cdot\bm{d}_{2})$, \protect\linebreak$a=1\times10^{-4}$]{\includegraphics[width=6.0cm]{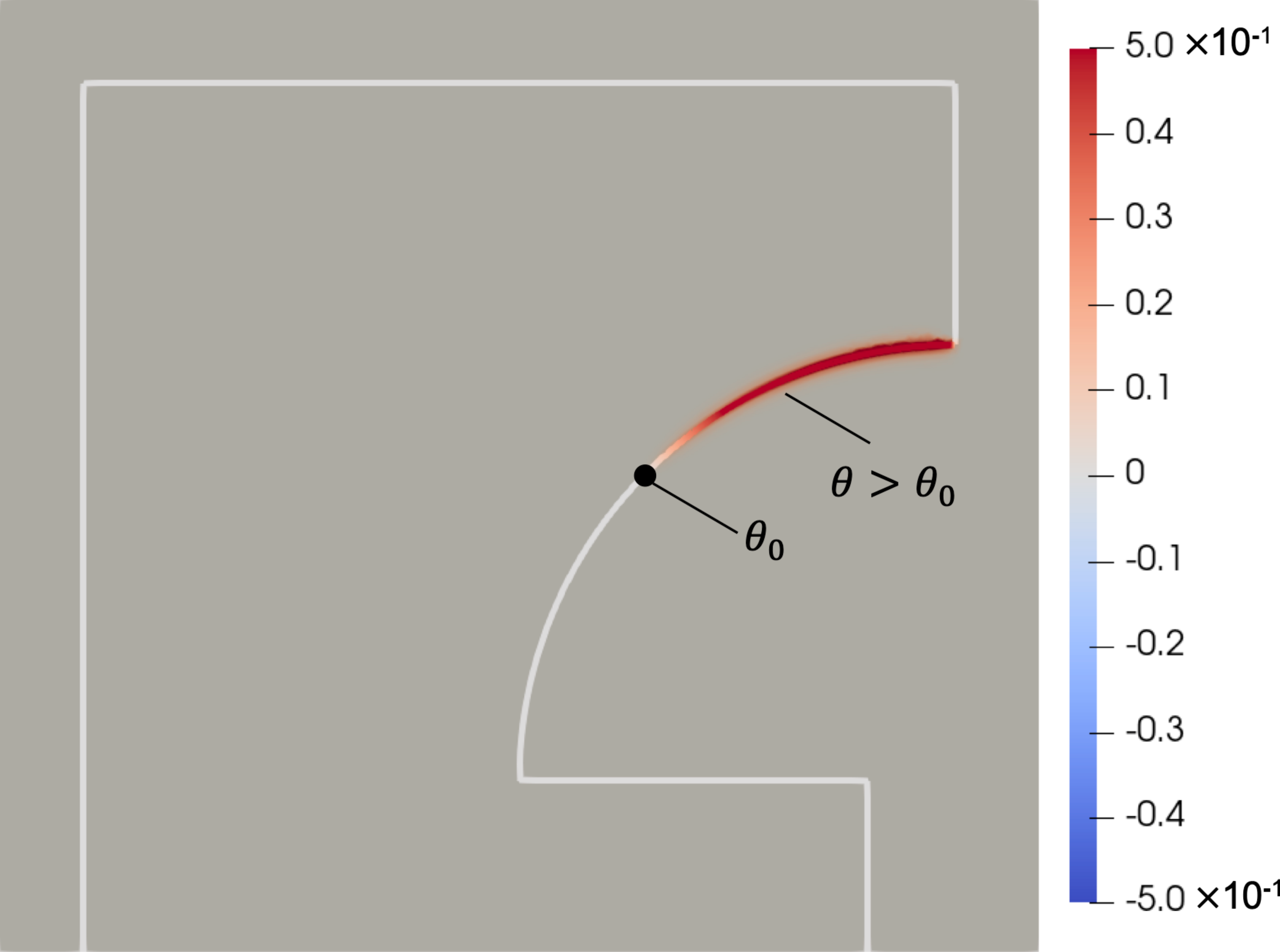}}
						\hspace{0.5cm}
		\subfigure[$(R(\sqrt{a}L\nabla\psi\cdot\bm{d}-\sqrt{a}L|\nabla\psi|\cos\theta_{0}))^{2}$, \protect\linebreak$a=1\times10^{-4}$]{\includegraphics[width=6.0cm]{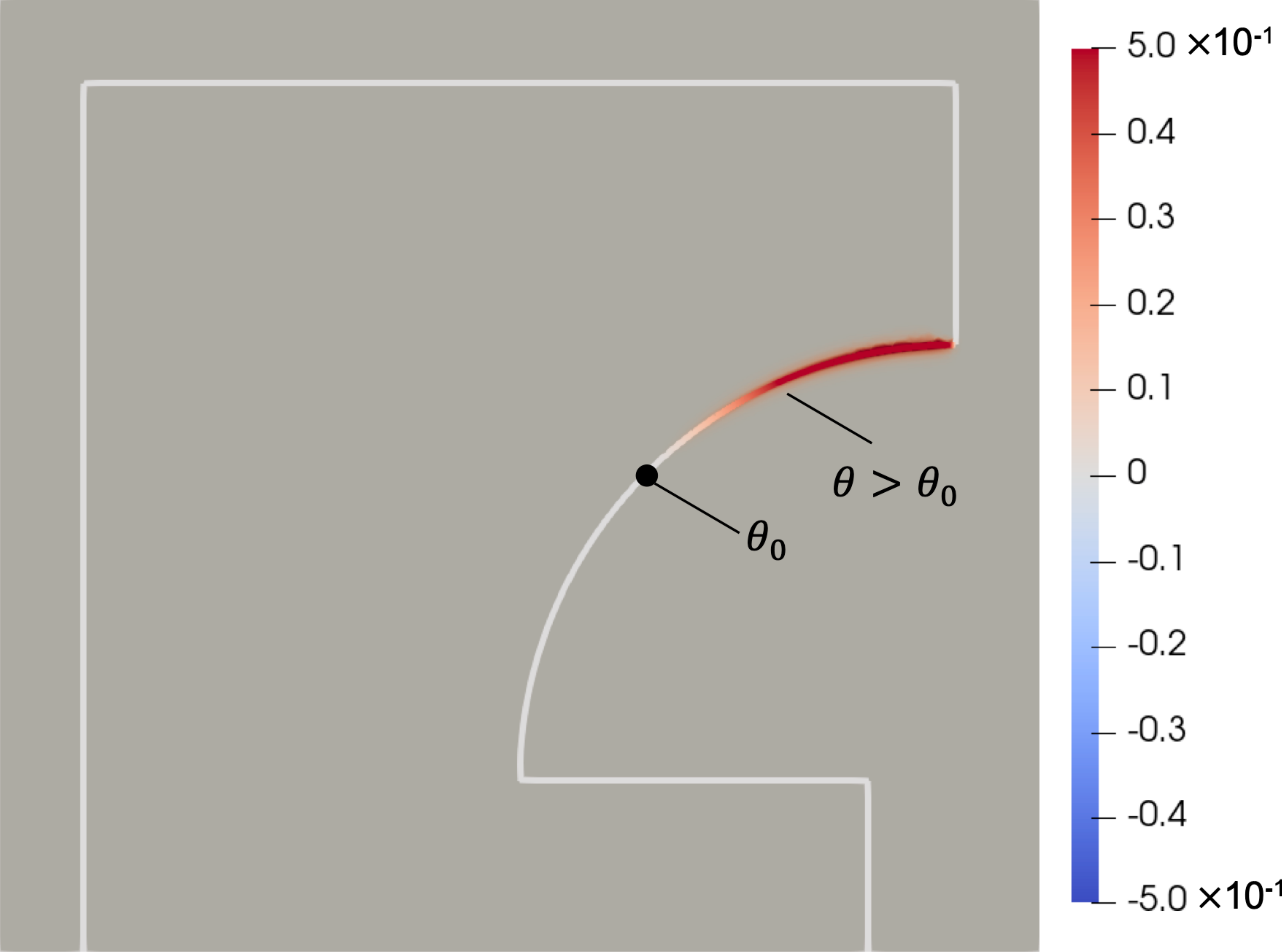}}
						\hspace{0.5cm}
		\subfigure[$R(\sqrt{a}L\nabla\psi\cdot\bm{d}_{1})R(\sqrt{a}L\nabla\psi\cdot\bm{d}_{2})$, \protect\linebreak$a=1\times10^{-3}$]{\includegraphics[width=6.0cm]{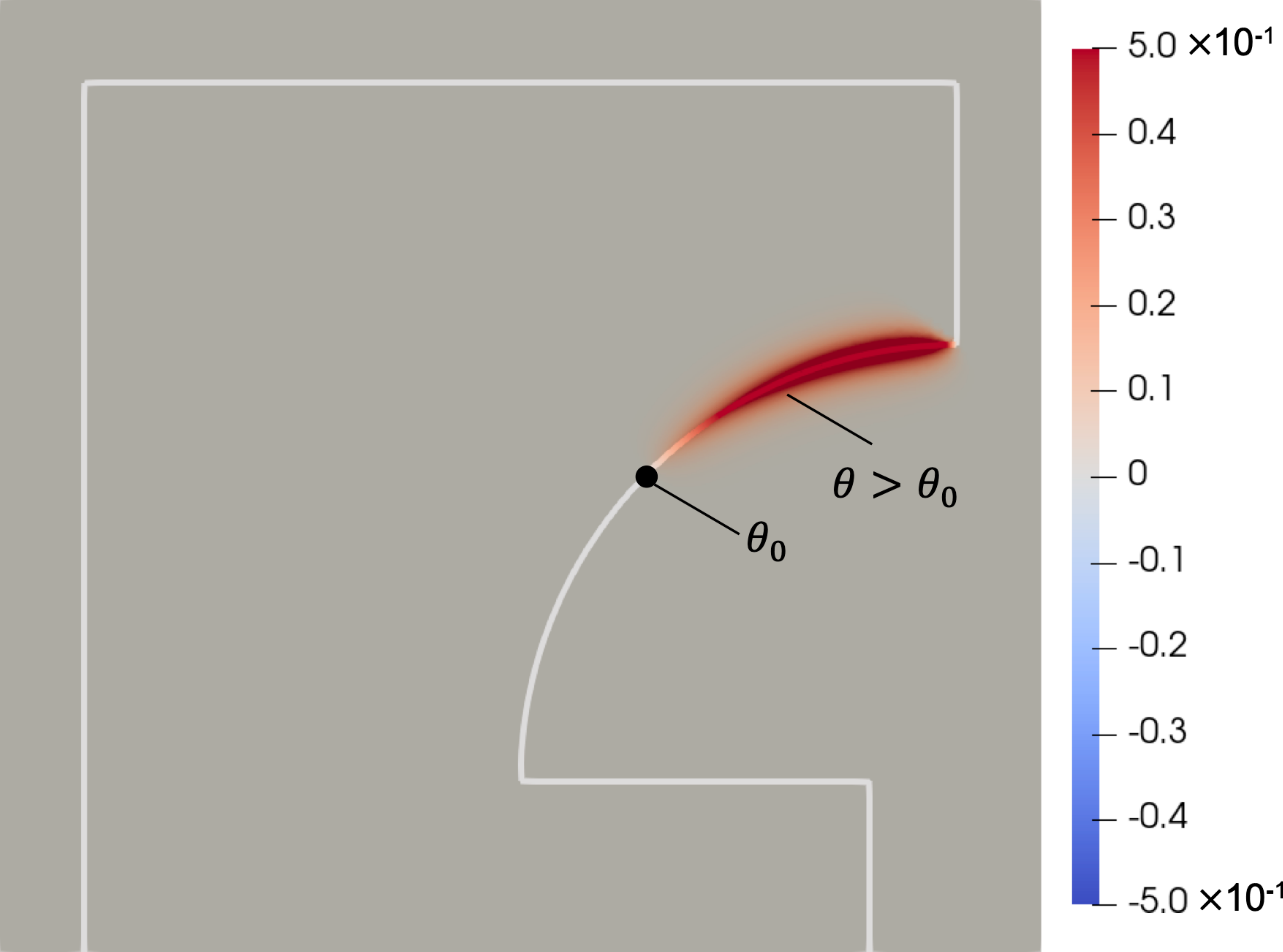}}
		\caption{Numerical example of the overhang angle constraint.}
		\label{fig:ovhg_2}
	\end{center}
\end{figure}
To validate the formulated PDE and overhang angle constraint, a numerical example using the FEM is provided and is shown in Fig. \ref{fig:ovhg_2}.
The analysis domain $D$ consists of two domains with the target model shown in Fig. \ref{fig:ovhg_1} as the material domain $\mathit{\Omega}$ and the other region as the void domain $D\backslash \mathit{\Omega}$.
The domain $D$ has dimensions of $1.0\times1.2$ and is discretized into a mesh of second-order triangular elements.
The diffusion coefficient $a$ is set to 1$\times10^{-4}$, representative length $L$ is set to 1.0, and threshold angle $\theta_{0}$ is set to 45$^\circ$.
Fig. \ref{fig:ovhg_2}(a) shows the distribution of $\psi$ that was obtained by solving Eq. \ref{eq:ovhg1}. $\psi$ smoothly transitions from the material domain to the void domain.
Figs. \ref{fig:ovhg_2} (b), (c), and (d) indicate that the region where both inner products assume positive values is the overhanging region.
Fig. \ref{fig:ovhg_2} (e) shows the overhanging region obtained from the comparison of the cosine value with the inner product of the normal vector $\bm{n}_{\psi}$ and the building direction $\bm{d}$.
Compared with the aforementioned conventional method, the proposed constraint function is able to detect overhanging regions near the threshold angle.
This implies that during the optimization process, the overhanging regions near the threshold angle are also optimized to satisfy the constraint.
Fig. \ref{fig:ovhg_2} (f) shows the result with the diffusion coefficient $a$ set to 1$\times10^{-3}$.
This result indicates that the diffusion coefficient $a$ affects the evaluation area of the overhanging region.
The overhang angle constraint in the 3D problem is given by adding the threshold angle vectors of an orthogonal plane, as follows:
\begin{equation}
\begin{split}
G_{o}=\int_{D} R(\sqrt{a}L\nabla\psi\cdot\bm{d}_{1})&R(\sqrt{a}L\nabla\psi\cdot\bm{d}_{2})\\
	+&R(\sqrt{a}L\nabla\psi\cdot\bm{d}_{3})R(\sqrt{a}L\nabla\psi\cdot\bm{d}_{4})\mathrm{~d}\mathit\Omega,
\label{eq:ovhg5}
\end{split}
\end{equation}
where
\begin{align}
\bm{d}_{1}=\left(\begin{array}{c}
-\cos \theta_{0} \\
-\sin \theta_{0} \\
0
\end{array}\right), 
\quad 
\bm{d}_{2}=\left(\begin{array}{c}
\cos \theta_{0} \\
-\sin \theta_{0} \\
0
\end{array}\right),\nonumber
\\
\quad
\bm{d}_{3}=\left(\begin{array}{c}
\cos \theta_{0} \\
0 \\
-\sin \theta_{0}
\end{array}\right),
\quad
\bm{d}_{4}=\left(\begin{array}{c}
\cos \theta_{0} \\
0 \\
-\sin \theta_{0}
\end{array}\right).
\label{eq:ovhg6}
\end{align}
\begin{figure}[htbp]
	\begin{center}
		\includegraphics[width=6.0cm]{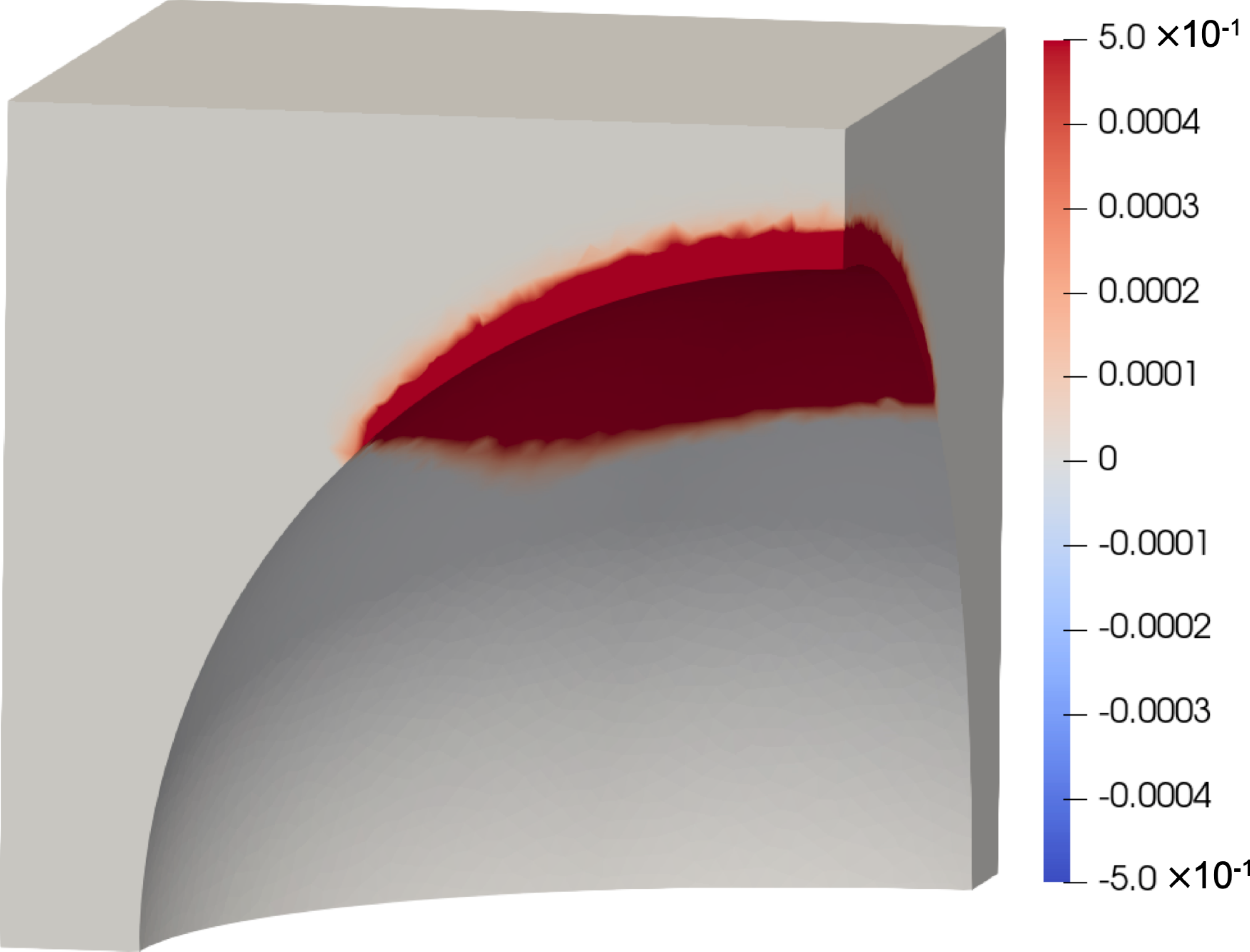}
		\caption{Numerical example of 3D model.}
		\label{fig:ovhg_3}
	\end{center}
\end{figure}
Fig. \ref{fig:ovhg_3} presents the results of the evaluation of the overhanging region with the 1/4 hemispheric model using Eq. \ref{eq:ovhg5}.
The overhanging region can be detected in the 3D case as well as in the 2D case.
Next, the topological derivative of the overhang angle constraint is derived using the adjoint variable method.
The adjoint variable $\tilde{\psi}\in H^{1}(D)$ and its adjoint equation are defined as follows:
\begin{align}
\begin{cases}
-a L^{2}\nabla^{2}\tilde{\psi}+\tilde{\psi}= \displaystyle\frac{\partial G_{o}}{\partial \psi}&\text{in}\hspace{0.3cm}D
\\
\bm{n}\cdot\nabla\tilde{\psi}=0 \hspace{3.85cm}&\text{on}\hspace{0.3cm}\partial\mathit{\Omega},
\end{cases}
\label{eq:ovhg7}
\end{align}
where
\begin{equation}
\displaystyle\frac{\partial G_{o}}{\partial \psi}=\left\{\begin{array}{ll}
\begin{split}
	\nabla\cdot \left[
	H(\sqrt{a}L\nabla\psi\cdot\bm{d}_{1})R(\sqrt{a}L\nabla\psi\cdot\bm{d}_{2})\bm{d}_{1}\right.\\
	\left.+R(\sqrt{a}L\nabla\psi\cdot\bm{d}_{1})H(\sqrt{a}L\nabla\psi\cdot\bm{d}_{2})\bm{d}_{2}
	\right]\tilde{\psi}&\hspace{1cm} \text { if } N=2\\
\\[1ex]
\nabla\cdot \left[
H(\sqrt{a}L\nabla\psi\cdot\bm{d}_{1})R(\sqrt{a}L\nabla\psi\cdot\bm{d}_{2})\bm{d}_{1}\right.\\
	\left.+R(\sqrt{a}L\nabla\psi\cdot\bm{d}_{1})H(\sqrt{a}L\nabla\psi\cdot\bm{d}_{2})\bm{d}_{2}
\right]\tilde{\psi}\\
\quad+\nabla\cdot \left[H(\sqrt{a}L\nabla\psi\cdot\bm{d}_{3})R(\sqrt{a}L\nabla\psi\cdot\bm{d}_{4})\bm{d}_{3}\right.\\
	\left.+R(\sqrt{a}L\nabla\psi\cdot\bm{d}_{3})H(\sqrt{a}L\nabla\psi\cdot\bm{d}_{4})\bm{d}_{4}
\right]\tilde{\psi}
&\hspace{1cm} \text { if } N=3 .
\end{split}
\end{array}\right.
\label{eq:ovhg8}
\end{equation}
Here, $H(s):=\mathrm{d}R(s)/\mathrm{d}s$ is the Heaviside function and $N$ is the number of spatial dimensions.
As only the source term in Eq. \ref{eq:ovhg1} is affected by the characteristic function $\chi_{\phi}$, the topological derivative \cite{carpio2008solving} of the overhang angle constraint $G_{o}$ is derived as follows:
\begin{equation}
G_{o}'=-\tilde{\psi}.
\label{eq:ovhg9}
\end{equation}
\subsection{Thermal model for the downward convex shapes}\label{sec:3.3}
\begin{figure}[htbp]
	\begin{center}
		\includegraphics[width=13.0cm]{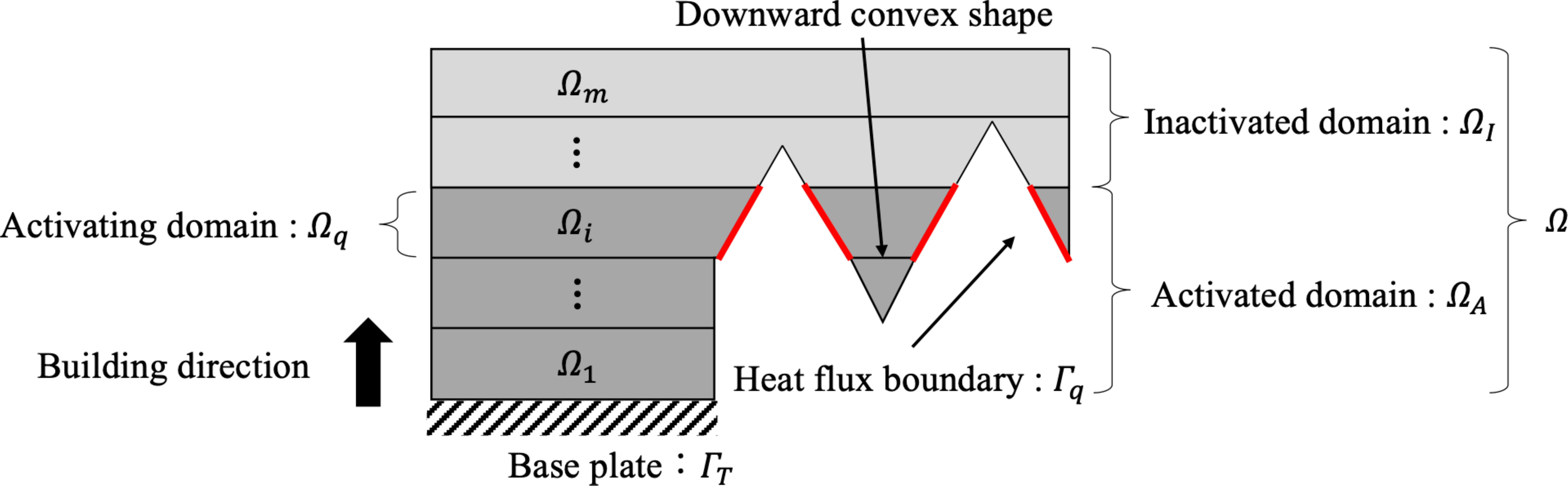}
		\caption{Domains and boundaries of the thermal model in the building process.}
		\label{fig:analysisdomain}
	\end{center}
\end{figure}

As mentioned in the introduction, the above overhang angle constraint creates downward convex shapes.
In actual manufacturing, these shapes impede heat flow, causing overheating, which generates undesirable defects such as porosity and degraded surface quality. 
Therefore, we propose a method to suppress the downward convex shapes by considering heat dissipation in the AM building process.
Several analytical models have been proposed to detect overheating during thermal processes. 
For the steady-state heat conduction problem, Ranjan et al.\cite{ranjan2017controlling} proposed a model in which heat flux is applied layer-by-layer, whereas Wang et al.\cite{wang2020optimizing} proposed a model in which heat flux is applied only to the overhang boundaries, and not layer-by-layer.
In this study, we consider a thermal model that applies heat flux to the overhang boundary at each layer, as shown in Fig. \ref{fig:analysisdomain}.
First, to represent the building process, the domain $\mathit{\Omega}$ is divided into $m$ layers with a fixed thickness in the building direction.
The domain $\mathit{\Om}$ is defined by each domain $\mathit{\Om}_{i}$ for $1\leq i\leq m$.
Moreover, the domain $\mathit{\Om}$ consists of three subdomains: the activated domain $\mathit{\Omega}_{A}$, inactive domain $\mathit{\Omega}_{I}$, and activating domain $\mathit{\Omega}_{q}$, which is determined by the activation status of $\mathit{\Omega}_{i}$.
The activated domain $\mathit{\Om_{A}}$ is occupied by an isotropic material with thermal conductivity $k$, and a fixed temperature boundary condition is applied to the base plate $\mathit{\Ga_{T}}$.
Then, the temperature field $T_{i}\in H^{1}(\mathit{\Omega}_{A})$ with heat flux $q$ applied to the overhang boundary $\mathit{\Ga_{q}}$ of the added layer $\Omega_{q}$ is represented by the following governing equation:
\begin{equation}
\left\{
\begin{alignedat}{3}
\hspace{2mm}&\operatorname{div}(k \nabla T_{i})=0&\hspace{5mm}&\text { in }&& \mathit{\Omega_{A}} \\
\hspace{2mm}&(k \nabla T_{i}) \cdot \bm{n}=q&\hspace{5mm}&\text { on }&& \mathit{\Gamma_{q}} \\
\hspace{2mm}&T_{i}=T_{amb}&\hspace{5mm}&\text {on }&& \mathit{\Gamma_{T}}, \\
\end{alignedat}
\right.
\label{eq:drp1}
\end{equation}
for all indices $i = 1,2,\ldots,m$, where $T_{amb}$ is the temperature of the base plate, which acts as a heat sink.
\subsection{Thermal constraint function and its derivative}\label{sec:3.4}
The thermal constraint function that improves the heat dissipation of each domain $\mathit{\Omega}_{i}$ is defined as follows:
\begin{equation}
G_{t}=\sum_{i=1}^{m}\int_{\mathit{\Omega}_{i}}\left(T_{i} - T_{amb}\right)^{2}\mathrm{~d}\mathit\Omega.
\label{eq:drp2}
\end{equation}
Next, the topological derivative of the thermal constraint is derived using the adjoint variable method.
The adjoint variable $\tilde{T}\in H^{1}(\mathit{\Omega}_{A})$ and its adjoint equation are defined as follows:
\begin{equation}
\left\{
\begin{alignedat}{3}
\hspace{2mm}&\operatorname{div}(k \nabla \tilde{T}_{i})=2\left(T_{i} - T_{amb}\right)&\hspace{5mm}&\text { in }&& \mathit{\Omega_{A}}, \\
\hspace{2mm}&(k \nabla \tilde{T}_{i}) \cdot \bm{n}=0&\hspace{5mm}&\text { on }&& \mathit{\Omega_{A}}\setminus\mathit{\Gamma_{T}}, \\
\hspace{2mm}&\tilde{T}_{i}=T_{amb}&\hspace{5mm}&\text {on }&& \mathit{\Gamma_{T}}, \\
\end{alignedat}
\right.
\label{eq:drp3}
\end{equation}
Subsequently, the topological derivative of the thermal constraint $G_{t}$ is derived as follows \cite{novotny2003topological,giusti2010topological}:
 \begin{equation}
 G_{t}'=\sum_{i=1}^{m}-k\nabla T_{i}\cdot\nabla\tilde{T}_{i}.
 \label{eq:drp4}
 \end{equation}
\section{Distortion constraint}\label{sec:4}
\subsection{Mechanical model based on the inherent strain method}\label{sec:4.1}
\begin{figure}[htbp]
	\begin{center}
		\includegraphics[width=13.0cm]{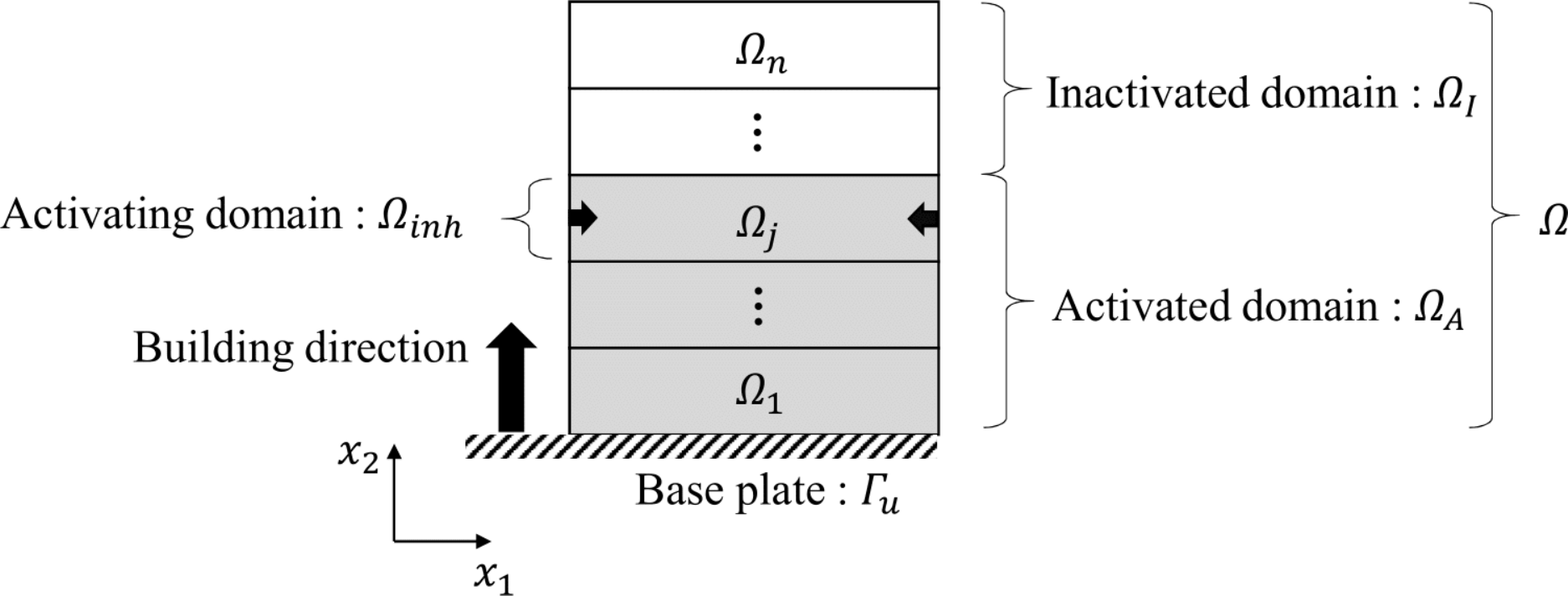}
		\caption{Domains and boundaries of the mechanical model in the building process.}
		\label{fig:analysisdomain2}
	\end{center}
\end{figure}

The mechanical model for predicting the part-scale residual stress and distortion uses the inherent strain method, which applies the strain for each layer, as shown in Fig. \ref{fig:analysisdomain2}.
The domain is divided into $n$ layers with a fixed thickness, as in the thermal model, and consists of three subdomains.
The domain $\mathit{\Om}$ is defined by each domain $\mathit{\Om}_{j}$ for $1\leq j\leq n$.
Note that each subdomain region is determined by the activation status of $\mathit{\Omega}_{j}$.
The activated domain $\mathit{\Om_{A}}$ is occupied by an isotropic elastic material, and a fixed displacement boundary condition is applied to the base plate $\mathit{\Ga_{u}}$.
The displacement $\bm{u}_{j} \in H^{1}(\mathit{\Om}_{A})^{N} $ with inherent strain $\bm{\ep}^{inh}$ applied to the domain $\mathit{\Om_{inh}}$ is represented by the following governing equation:
\begin{align}
\begin{cases}
\hspace{2mm}-\div(\bm{\sigma}_{j})=0\hspace{5mm}&\text{in}\hspace{2mm}\mathit{\Om_{A}}\\
\hspace{2mm}\bm{\sigma}_{j}=\mathbb{C}\bm{\ep}(\bm{u}_{i})-\mathbb{C}\bm{\ep}^{inh},\hspace{5mm}&\\
\hspace{2mm}\bm{u}_{j}=\bm{0}\hspace{5mm}&\text{on}\hspace{2mm}\mathit{\Ga_{u}}\\
\hspace{2mm}-\bm{\sigma}_{j}\cdot\bm{n}=\bm{0}\hspace{5mm}&\text{on}\hspace{2mm}\partial \mathit{\Om_{A}} \setminus \mathit{\Ga_{u}},
\label{eq:dst1}
\end{cases}
\end{align}
for all indices $j = 1,2,\ldots,n$, where $\bm{\sigma}$ denotes the Cauchy stress tensor, $\mathbb{C}$ denotes the elasticity tensor, and $\bm{\ep}(\bm{u})$ denotes the total strain tensor defined by $\bm{\ep}(\bm{u})\coloneqq\frac{1}{2}(\nabla\bm{u}+(\nabla\bm{u})^{\top})$. 
Note that the above equation represents an elastic analysis that does not consider plastic deformation.
The inherent strain $\bm{\ep}^{inh}$ in the domain $\mathit{\Om_{inh}}$ is defined as follows:
\begin{equation}
\bm{\ep}^{inh} ( \bm{x} ) = \left\{ \begin{array} { l l } { \bm{\ep}^{inh} } & { \text { for } \bm{x} \in \mathit{\Omega_{inh}} }, \\ { \bm{0} } & { \text { otherwise } }, \end{array} \right.
\label{eq:dst2}
\end{equation}
where $\bm{x}$ denotes a point located in $\mathit{\Omega_{A}}$.
In this study, the inherent strain component is set as $\ep^{inh}_{x}=\ep^{inh}_{y}=-0.0025$ and $\ep^{inh}_{z}=0$ in the 3D case.
In addition, the domain is divided in the building direction at 1 mm per layer for computational cost. 
For a detailed discussion of the methods for identifying the inherent strain $\bm{\ep}^{inh}$, the building process algorithm using the FEM, and the number of layers $n$ suitable for topology optimization, refer to Miki et al. \cite{MIKI2021103558}.
The residual stress and distortion after the end of the building process are expressed as follows:
\begin{align}
\bm{\sigma} &= \sum_{j=1}^{n}\bm{\sigma}_{j} \hspace{5mm}\text { for } \bm{x} \in \mathit{\Omega_{A}}, \label{eq:dst3}\\
\bm{u} &= \sum_{j=1}^{n}\bm{u}_{j} \hspace{5mm}\text { for } \bm{x} \in \mathit{\Omega_{A}}. \label{eq:dst4}
\end{align}
\subsection{Distortion constraint function and its derivative}\label{sec:4.2}
The distortion constraint function is defined using the P-norm function, as follows:
\begin{equation}
G_{u}=\sum_{j=1}^{n}\left(\int_{\mathit{\Omega}_{A}}\left|\bm{u}_{j}\right|^b \mathrm{~d} \mathit{\Omega}\right)^{1/b},\\
\label{eq:dst5}
\end{equation}
where $b\geq 2$ is the penalization parameter set to $5$ in this study.
Next, the topological derivative of the distortion constraint is derived using the adjoint variable method.
The adjoint variable $\tilde{\bm{u}}_{j}\in H^{1}(\mathit{\Omega}_{A})^{N}$ and its adjoint equation are defined as follows:
\begin{align}
\begin{cases}
\hspace{2mm}-\div(\mathbb{C}\bm{\ep}(\tilde{\bm{u}_{j}}))=-\left(\int_{\mathit{\Om_{A}}} \left|\bm{u}_{j}\right|^{b}\mathrm{~d}\mathit{\Omega}\right)^{1/b-1}\left|\bm{u}_{j}\right|^{b-2}\bm{u}_{j}\hspace{5mm}&\text{in}\hspace{2mm}\mathit{\Om_{A}},\\
\hspace{2mm}\tilde{\bm{u}_{j}}=\bm{0}\hspace{5mm}&\text{on}\hspace{2mm}\mathit{\Ga_{u}},\\
\hspace{2mm}-(\mathbb{C}\bm{\ep}(\tilde{\bm{u}_{j}}))\cdot\bm{n}=\bm{0}\hspace{5mm}&\text{on}\hspace{2mm}\partial \mathit{\Om_{A}} \setminus \mathit{\Ga_{u}},
\label{eq:dst6}
\end{cases}
\end{align}
for all indices $j = 1,2,\ldots,n$ and the topological derivative of the distortion constraint $G_{u}$ is derived as follows \cite{MIKI2021103558,giusti2017topology}:
\begin{equation}
G_{u}'=\sum_{j=1}^{n}\left(-\bm{\ep}(\bm{u}_{j}):\mathbb{A}:\bm{\ep}(\tilde{\bm{u}_{j}})+\bm{\ep}^{inh}:\mathbb{A}:\bm{\ep}(\tilde{\bm{u}_{j}})\right),
\label{eq:dst7}
\end{equation}
where the constant fourth-order tensor $\mathbb{A}$ is given by
\begin{equation}
\mathbb{A}_{i j k l}=\frac{3(1-\nu)}{2(1+\nu)(7-5 \nu)}\left\{\frac{-\left(1-14 \nu+15 \nu^{2}\right) E}{(1-2 \nu)^{2}} \delta_{i j} \delta_{k l}+5 E\left(\delta_{i k} \delta_{j l}+\delta_{i l} \delta_{j k}\right)\right\},
\end{equation}
where $E$, $\nu$, and $\delta$ are the Young's modulus, Poisson's ratio, and Kronecker delta, respectively. 
\section{Formulation of the optimization problem}\label{sec:5}
\subsection{Minimum mean compliance problem}\label{sec:5.1}
First, we incorporate the aforementioned constraints into the minimum mean compliance problem.
In this problem, the material domain $\mathit{\Om}$ is fixed at the boundary $\mathit{\Ga_{v}}$, and traction $\bm{t}$ is applied at the boundary $\mathit{\Ga_{t}}$.
The displacement field is denoted by $\bm{v} \in H^{1}(\mathit{\Om})^{N} $ in the static equilibrium state, and the objective function is defined as
\begin{equation}
J_{v}=\int_{\mathit{\Ga_{t}}}\bm{t}\cdot\bm{v}\mathrm{d}\mathit{\Ga}.
\label{eq:op1}
\end{equation}
The optimization problem is formulated as an unconstrained problem by including the constraint function as a penalty term in the objective function, as follows:
\begin{alignat}{3}
\inf_{\phi}\hspace{17mm} &J= (1-\alpha)\dfrac{ J_{v}\int_{D}\mathrm{~d}\mathit\Omega}{\int_{D}| J_{v}|\mathrm{~d}\mathit\Omega}+\alpha\dfrac{ G_{u}\int_{D}\mathrm{~d}\mathit\Omega}{\int_{D}| G_{u}|\mathrm{~d}\mathit\Omega}\nonumber\\
&\hspace{8mm}+\beta G_{o}+\gamma \dfrac{ G_{t}\int_{D}\mathrm{~d}\mathit\Omega}{\int_{D}| G_{t}|\mathrm{~d}\mathit\Omega}\label{eq:op2}\\
\text{subject to}:\hspace{2mm}&E_{v}=\int_{\Gamma_{t}} \bm{t} \cdot \tilde{\bm{v}} \mathrm{~d} \Gamma-\int_{\mathit{\Om}} \ep(\bm{v}): \mathbb{C}: \ep(\tilde{\bm{v}}) \mathrm{~d} \mathit\Omega=0\label{eq:op3}\\
&\hspace{5mm}\text { for } \forall \tilde{\bm{v}} \in \mathcal{V}, \bm{v} \in \mathcal{V}\nonumber\\
&E_{u}=\int_{\mathit{\Om}_{inh}} \bm{\ep}^{inh}: \mathbb{C}: \ep(\tilde{\bm{u}_{j}}) \mathrm{~d} \mathit\Omega-\int_{\mathit{\Om}_{A}} \ep(\bm{u}_{j}): \mathbb{C}: \ep(\tilde{\bm{u}}) \mathrm{~d} \mathit\Omega=0\label{eq:op4}\\
&\hspace{5mm}\text { for } \forall \tilde{\bm{u}_{j}} \in \mathcal{U}, \bm{u}_{j} \in \mathcal{U}\nonumber\\
&E_{\psi}=-\int_{D}aL^{2}\nabla\psi\cdot\nabla\tilde{\psi} \mathrm{~d} \mathit\Omega-\int_{D}\psi \tilde{\psi}\mathrm{~d} \mathit\Omega-\int_{D}\chi_{\phi}\tilde{\psi} \mathrm{~d}\mathit\Omega=0\label{eq:op5}\\
&\hspace{5mm}\text { for } \forall \tilde{\psi} \in \mathcal{S}, \psi \in \mathcal{S}\nonumber\\
&E_{t}=\int_{\Gamma_{q}} q \tilde{T_{i}} \mathrm{~d} \Gamma-\int_{\mathit{\Om}_{A}} \nabla T_{i}\cdot\nabla\tilde{T_{i}} \mathrm{~d} \mathit\Omega=0\label{eq:op6}\\
&\hspace{5mm}\text { for } \forall \tilde{T_{i}} \in \mathcal{T}, T_{i} \in \mathcal{T}\nonumber\\
&G=\int_{D}\tilde\chi \hspace{1mm}\mathrm{~d}\mathit\Omega-V_\text{max}\leq 0,\nonumber
\end{alignat}
for all indices $i = 1,2,\ldots,m$, $j = 1,2,\ldots,n$, where $0\leq\alpha\leq1$, $\beta$, and $\gamma$ are the weighting parameters. 
In the above formulation, $G$ represents the volume constraint, and $V_{max}$ is the upper limit of the material volume in $D$.
Furthermore, functional spaces $\mathcal{V}$, $\mathcal{U}$, $\mathcal{S}$, and $\mathcal{T}$ are defined as follows:
\begin{align}
&\mathcal{V}:=\left\{\tilde{\bm{v}} \in H^{1}(\mathit{\Om})^{N}, \hspace{1mm}  \tilde{\bm{v}}=\bm{0} \hspace{1mm} \text { on } \hspace{1mm} \mathit\Ga_{v}\right\}\\
&\mathcal{U}:=\left\{\tilde{\bm{u}_{j}} \in H^{1}(\mathit{\Om_{A}})^{N}, \hspace{1mm}  \tilde{\bm{u}_{j}}=\bm{0} \hspace{1mm} \text { on } \hspace{1mm} \mathit\Ga_{u}\right\}\\
&\mathcal{S}:=\left\{\tilde{\psi} \in H^{1}(\mathit{D})\right\}\\
&\mathcal{T}:=\left\{\tilde{T_{i}} \in H^{1}(\mathit{\Om_{A}}), \hspace{1mm}  \tilde{T_{i}}=T_{amb} \hspace{1mm} \text { on } \hspace{1mm} \mathit\Ga_{T}\right\}
\end{align}

The minimum mean compliance problem is known as a self-adjoint problem.
Therefore, the adjoint variable is equivalent to the displacement field $\bm{v}$, and the topological derivative of the minimum mean compliance problem is derived as follows \cite{garreau2001topological,feijoo2005topological}:
\begin{equation}
J_{v}'=-\bm{\ep}(\bm{v}):\mathbb{A}:\bm{\ep}(\bm{v}). \label{eq:op7}\\
\end{equation}
\subsection{Thermal diffusion problem}\label{sec:5.2}
Next, we consider the steady-state heat conduction problem with internal heat generation. 
In this problem, the heat source $Q$ is applied to the design domain $D$, and the temperature $p=p_{amb}$ is fixed at the boundary $\mathit{\Ga_{p}}$.
The temperature field is denoted by $p \in H^{1}(\mathit{\Om}) $ in the static equilibrium state, and the objective function is defined as
\begin{equation}
J_{p}=\int_{D}Qp\mathrm{~d}\mathit{\Omega}.
\label{eq:op8}
\end{equation}
This objective function is called the thermal compliance \cite{gersborg2006topology}.
Subsequently, by replacing the objective function in Eq. \ref{eq:op2} and the governing equation in Eq. \ref{eq:op3}, the optimization problem is formulated as follows:
\begin{alignat}{3}
\inf_{\phi}\hspace{17mm} &J= (1-\alpha)\dfrac{ J_{p}\int_{D}\mathrm{~d}\mathit\Omega}{\int_{D}| J_{p}|\mathrm{~d}\mathit\Omega}+\alpha\dfrac{ G_{u}\int_{D}\mathrm{~d}\mathit\Omega}{\int_{D}| G_{u}|\mathrm{~d}\mathit\Omega}\nonumber\\
&\hspace{8mm}+\beta G_{o}+\gamma \dfrac{ G_{t}\int_{D}\mathrm{~d}\mathit\Omega}{\int_{D}| G_{t}|\mathrm{d}\mathit\Omega}\label{eq:op9}\\
\text{subject to}:\hspace{2mm}&E_{p}=\int_{D} Q \tilde{p} \mathrm{~d} \mathit\Omega-\int_{\mathit{\Omega}} k \nabla p \cdot \nabla \tilde{p} \mathrm{~d} \mathit\Omega=0\label{eq:op10}\\
&\hspace{5mm}\text { for } \forall \tilde{p} \in \mathcal{P}, p \in \mathcal{P}\nonumber\\
&E_{u}=0\nonumber\\
&\hspace{5mm}\text { for } \forall \tilde{\bm{u}_{j}} \in \mathcal{U}, \bm{u}_{j} \in \mathcal{U}\nonumber\\
&E_{\psi}=0\nonumber\\
&\hspace{5mm}\text { for } \forall \tilde{\psi} \in \mathcal{S}, \psi \in \mathcal{S}\nonumber\\
&E_{t}=0\nonumber\\
&\hspace{5mm}\text { for } \forall \tilde{T_{i}} \in \mathcal{T}, T_{i} \in \mathcal{T}\nonumber\\
&G\leq 0.\nonumber
\end{alignat}
The functional space $\mathcal{p}$ is defined as follows:
\begin{equation}
\mathcal{P}:=\left\{\tilde{p} \in H^{1}(\mathit{\Om_{A}}), \hspace{1mm} \tilde{p}=p_{amb} \hspace{1mm} \text { on } \hspace{1mm} \mathit\Ga_{p}\right\}
\label{eq:op10-1}
\end{equation}

This optimization problem is also a self-adjoint problem and is similar to the minimum mean compliance problem.
Therefore, the topological derivative of this optimization problem is derived as follows:
\begin{equation}
 J_{p}'=-k\nabla p\cdot\nabla p. \label{eq:op11}\\
\end{equation}
\section{Numerical implementation}\label{sec:6}
\subsection{Optimization algorithm}\label{sec:6.1}
The optimization algorithm is as follows:
\begin{description}
	\item[Step1.] The initial value of the level set function $\phi$ is set to the fixed design domain $D$.
	\item[Step2.] The governing equations for the target physics and each state variable defined in Eqs. \ref{eq:ovhg1}, \ref{eq:drp1}, and \ref{eq:dst1} are solved by the FEM.
	\item[Step3.] The objective function $J$ with respect to the target physics is calculated.
	\item[Step4.] If the objective function converges, the optimization procedure is terminated; otherwise, each adjoint variable defined in Eq. \ref{eq:ovhg7}, \ref{eq:drp3}, and \ref{eq:dst6} is solved by the FEM, and the sum of the topological derivative $J'$ is calculated.
	\item[Step5.] The level set function is updated using the time-evolution equation given by Eq. \ref{eq:lsf3}; then, the optimization procedure returns to the second step.
\end{description}
This optimization algorithm is implemented using the open-source PDE solver FreeFEM++\cite{MR3043640}.
\subsection{Numerical scheme for the governing equation}
In this study, the material and void domains are distinguished using the ersatz material approach \cite{allaire2004structural} from the perspective of computational cost.
Specifically, we assume that the void domain has a small material property and that the boundary between the material and void domains has a smoothly distributed material property.
Subsequently, the extended elastic tensor $\tilde{\mathbb{C}}$ and thermal conductivity $\tilde{k}$ for solving the governing equations in the fixed design domain $D$ are defined as follows:
\begin{align}
&\tilde{\mathbb{C}}(\phi;w) = \left\{(1-c)H_{\phi}(\phi;w)+c\right\}\mathbb{C}\label{eq:ni1}\\
&\tilde{k}(\phi;w) = \left\{(1-c)H_{\phi}(\phi;w)+c\right\}k,\label{eq:ni2}
\end{align}
where $H_{\phi}(\phi;w)$ is defined as
\begin{equation}
H_{\phi}(\phi;w) := \left\{\begin{array}{ll}
1 & \text { for } \phi>w, \\
\frac{1}{2}+\frac{\phi}{w}\left(\frac{15}{16}-\frac{\phi^{2}}{w^{2}}\left(\frac{5}{8}-\frac{3}{16} \frac{\phi^{2}}{w^{2}}\right)\right) & \text { for }-w \leq \phi \leq w, \\
0 & \text { for } \phi<-w.
\end{array}\right.\label{eq:ni3}
\end{equation}
Here, $w$ represents the width of the transition, and $c$ is the ratio of the material properties for the material and void domains.
Moreover, Eqs. \ref{eq:ovhg1} and \ref{eq:ovhg7} are solved by replacing the characteristic function $\chi_{\phi}$ with the following Heaviside function:
\begin{equation}
\chi_{\phi} = H_{\phi}(\phi;\xi).\label{eq:ni4}
\end{equation}
In our implementation, we set $c=1.0\times10^{-3}$, $w=0.5$ and $\xi=0.9$.

Next, an approximate solution method for the heat conduction equation defined in Eq. \ref{eq:op6} is introduced.
Here, we replace the boundary integral with a domain integral using the Dirac delta function $\delta(\boldsymbol{x})$, as follows:
\begin{equation}
\int_{\Gamma} \xi(\boldsymbol{x}) \mathrm{d} \Gamma \approx \int_{\mathit\Omega} \xi(\boldsymbol{x}) \delta(\boldsymbol{x}) \mathrm{~d} \mathit\Omega,
\label{eq:ovhg10}
\end{equation}
The delta function $\delta(\boldsymbol{x})$ is expressed using the Heaviside function $H_{\psi}(\psi;w)$, as follows:
\begin{equation}
\delta(\boldsymbol{x}) = \nabla H_{\psi}(\psi;w)\cdot \bm{n}_{\psi},
\label{eq:ovhg10_1}
\end{equation}
where $\bm{n}_{\psi}$ denotes the normal vector for $H_{\psi}(\psi;w)$.
The above equation can be rewritten as
\begin{equation}
\delta(\boldsymbol{x}) = \frac{\mathrm{d} H_{\psi}(\psi;w)}{\mathrm{d} \psi}\nabla \psi \cdot \frac{\nabla \psi}{|\nabla\psi|}=\frac{\mathrm{d} H_{\psi}(\psi;w)}{\mathrm{d} \psi}|\nabla\psi|.
\label{eq:ovhg10_2}
\end{equation}
Then, only the overhang boundary is extracted using the inner product of the normal vector $\bm{n}_{\psi}$ and the building direction $\bm{d}$, as follows:
\begin{equation}
\int_{\mathit{\Gamma}_{q}} \xi(\bm{x})\mathrm{d}\mathit{\Gamma} \approx \int_{\mathit{\Omega}_{q}} \xi(\bm{x})\frac{\mathrm{d} H_{\psi}(\psi;w)}{\mathrm{d} \psi}|\nabla\psi|H(\bm{n}_{\psi}\cdot\bm{d}) \mathrm{d}\mathit{\Omega}.
\label{eq:ovhg11}
\end{equation}
Substituting the above equation into Eq. \ref{eq:op6} yields
\begin{equation}
E_{t}=\int_{\mathit{\Omega}_{q}} q \tilde{T_{i}} \frac{\mathrm{d} H_{\psi}(\psi;w)}{\mathrm{~d} \psi}|\nabla\psi|H(\bm{n}_{\psi}\cdot\bm{d})\mathrm{~d}\mathit{\Omega}-\int_{\mathit{\Om}_{A}} \nabla T_{i}\cdot\nabla\tilde{T_{i}} \mathrm{~d} \mathit{\Omega}=0
\end{equation}
\subsection{Smoothed ramp function for the derivative of overhang constraint}
Because the derivative of the ramp function cannot be defined at $s=0$, a small parameter $\epsilon_{s}\in \mathbb{R}^{+}$ is introduced and the function is defined as an approximate ramp function  $\tilde R$, as follows:
\begin{equation}
 \tilde R(s;\epsilon_{s}):=\frac{1}{2}\left(s+\sqrt{s^{2}+\epsilon_{s}}\right).
\end{equation}
In our implementation, we set $\epsilon_{s}=1\times10^{-4}$.
\section{Numerical examples}\label{sec:7}
\subsection{Verification of the self-support constraint}\label{sec:7.1}
\subsubsection{Effect of overhang angle constraint parameters}\label{sec:7.1.1}
\begin{figure}[htbp]
	\begin{center}
		\includegraphics[width=11cm]{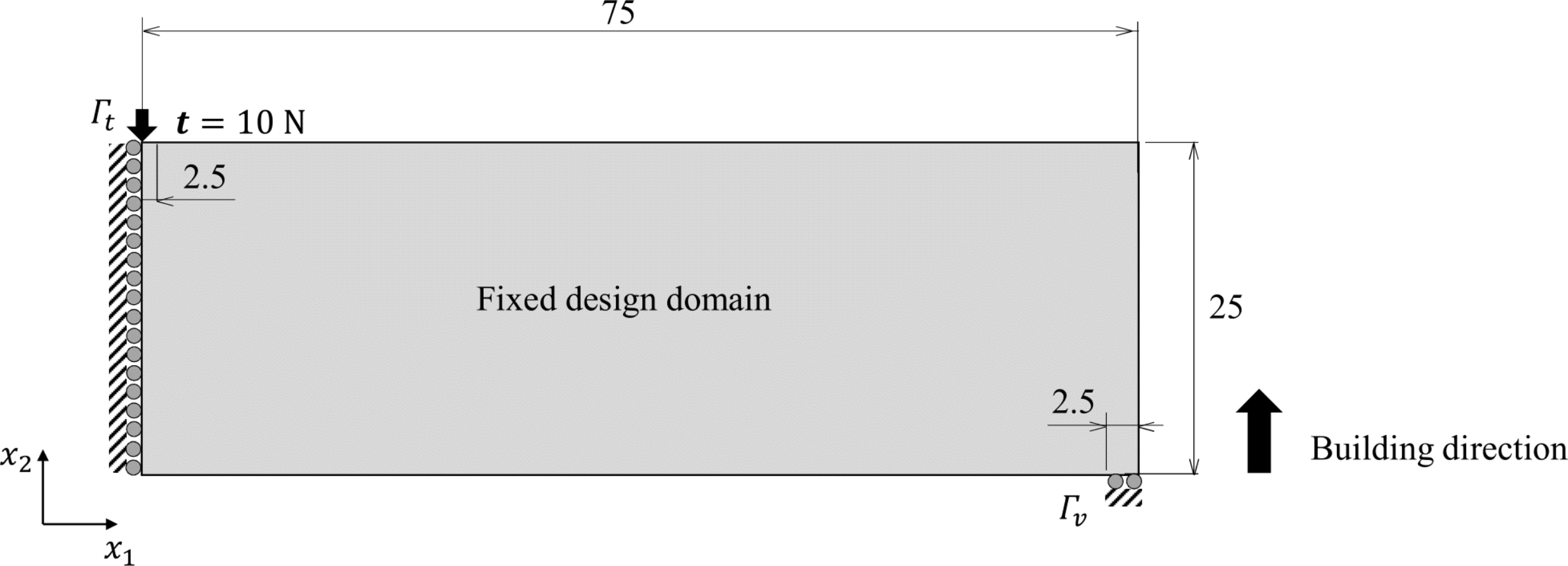}
		\caption{Problem setting for the MBB beam, with the dimensions in mm.}
		\label{fig:MBB}
	\end{center}
\end{figure}

This subsection presents optimization examples with only the imposed overhang angle constraint.
The optimization example considers the minimum mean compliance problem for a symmetric 2D MBB beam, as shown in Fig. \ref{fig:MBB}.
The material adopted in this study is AlSi10Mg, which has a Young's modulus of 75 GPa and Poisson's ratio of 0.34.
The upper limit of the allowable volume is set to 50\% of the fixed design domain.
The representative length in Eq. \ref{eq:ovhg1} is set to $L=$25 mm.
The threshold angle is set to $\theta_{0}=45^{\circ}$.
Here, we examine the effects of the parameters related to the overhang angle constraint $\beta$ and diffusion coefficient $a$ in Eq. \ref{eq:ovhg1}. 
\begin{figure}[htbp]
	\begin{center}
		\centering
		\subfigure[Without constraints; $J_{v~ref}=1.13$]{\includegraphics[width=6.5cm]{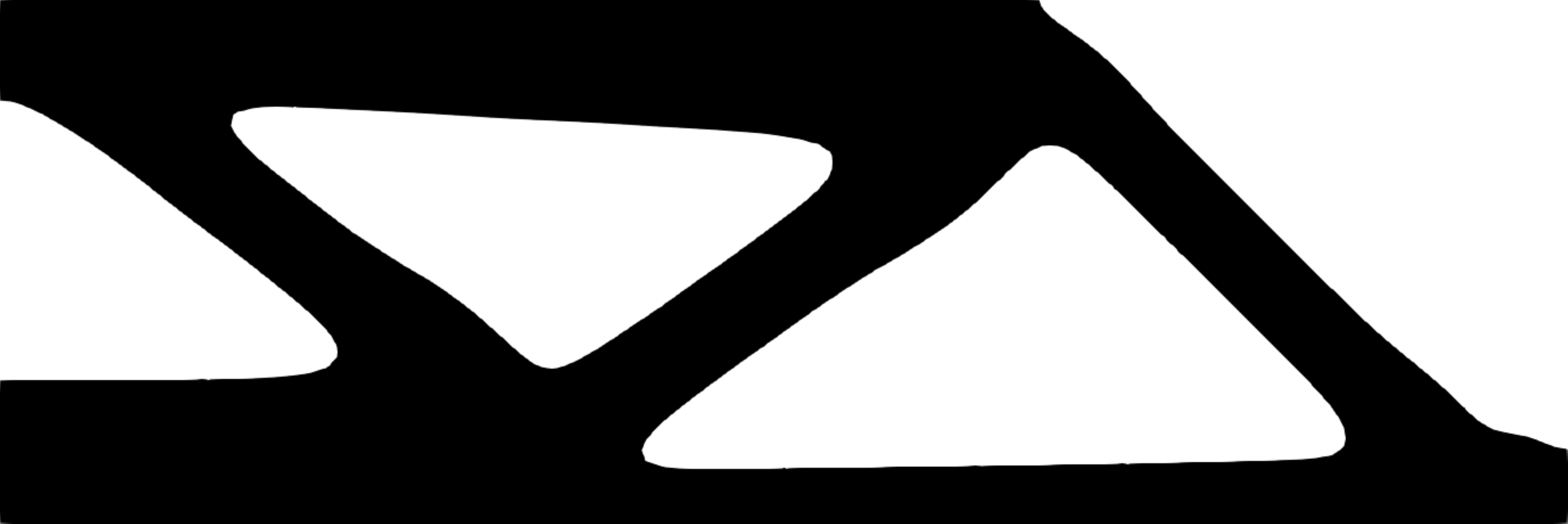}}
		\subfigure[$\beta=5$, $a=1\times10^{-4}$]{\includegraphics[width=6.5cm]{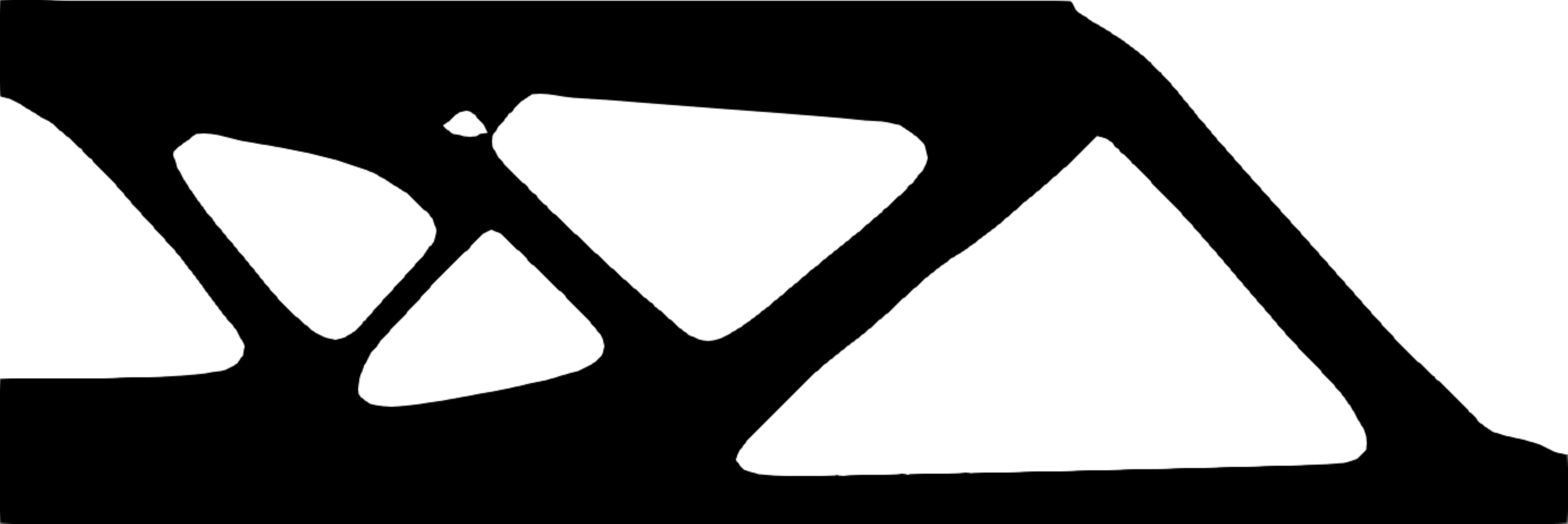}}
		\subfigure[$\beta=20$, $a=1\times10^{-4}$]{\includegraphics[width=6.5cm]{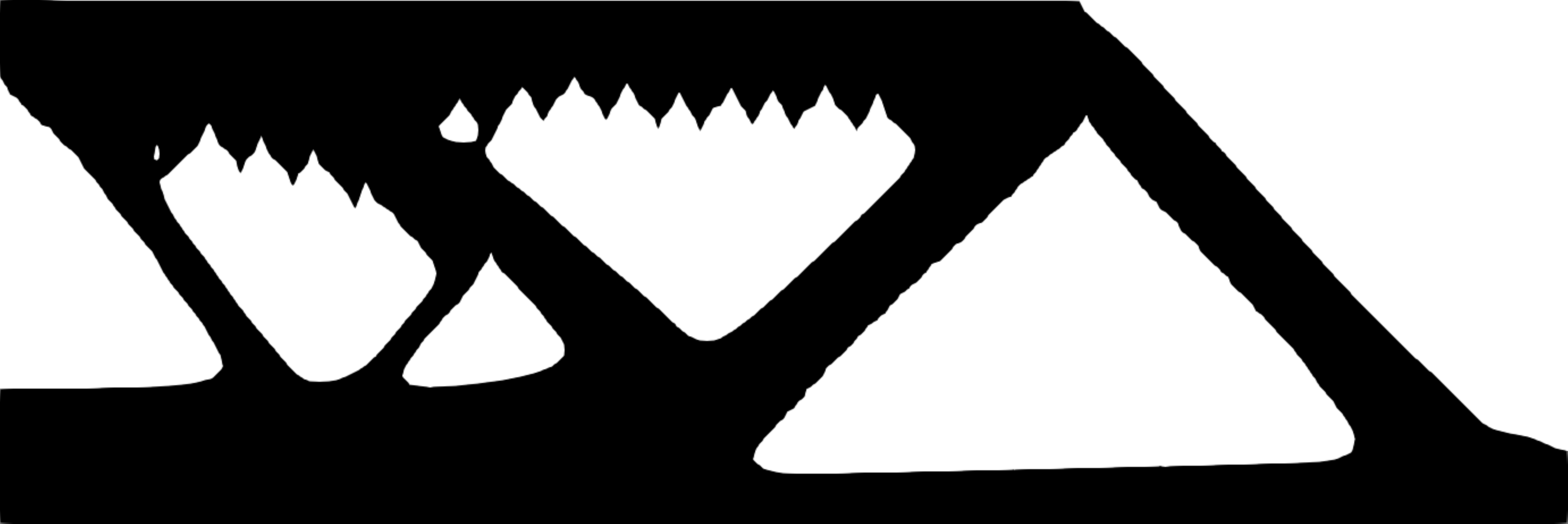}}
		\caption{Optimized MBB beams under different overhang angle constraint parameters $\beta$.}
		\label{fig:ne1}
	\end{center}
\end{figure}
\begin{figure}[htbp]
	\begin{center}
		\centering
		\subfigure[$\beta=20$, $a=5\times10^{-4}$]{\includegraphics[width=6.5cm]{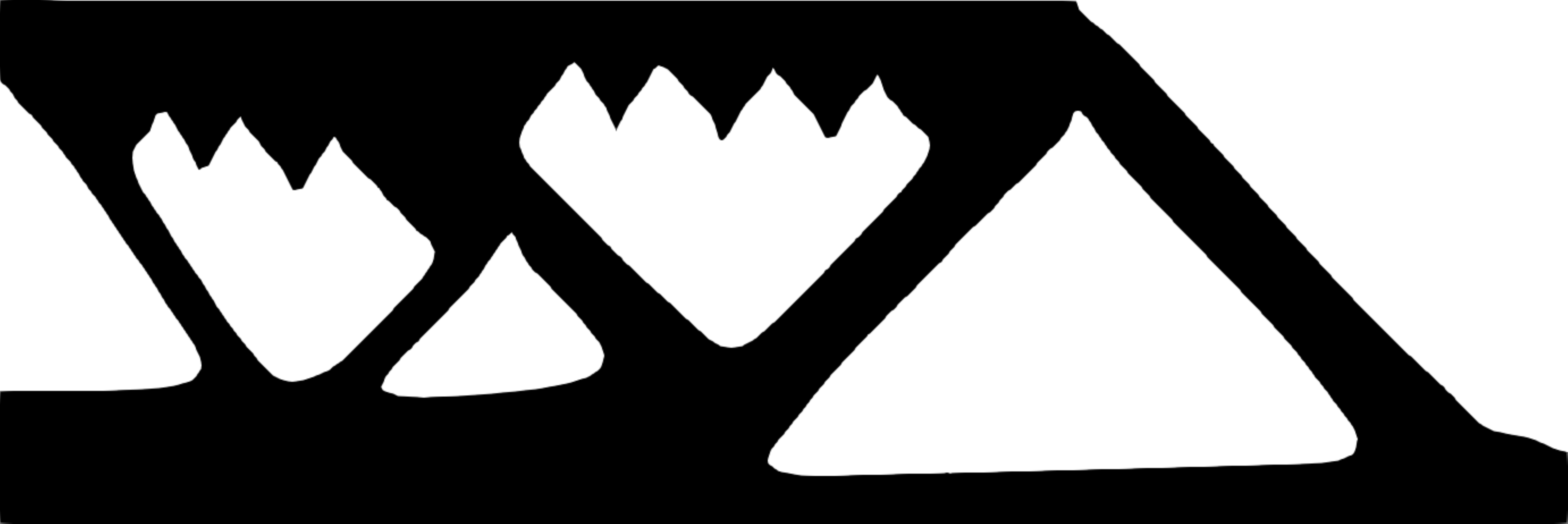}}
		\subfigure[$\beta=20$, $a=1\times10^{-3}$]{\includegraphics[width=6.5cm]{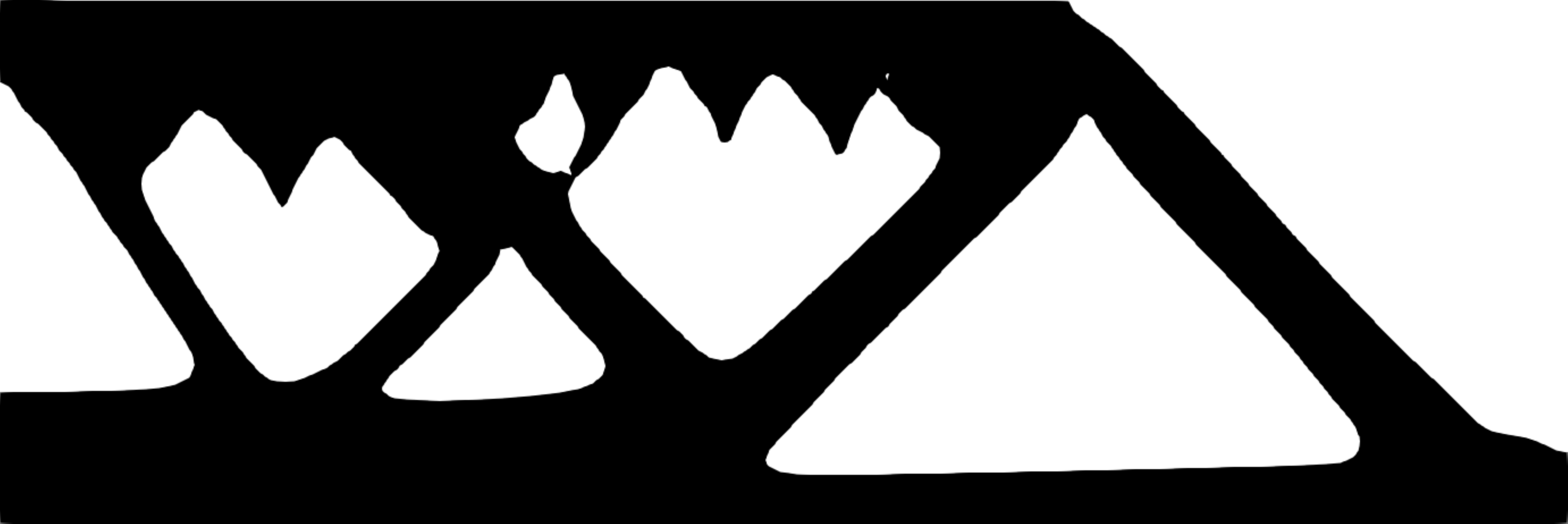}}
		\subfigure[$\beta=20$, $a=5\times10^{-3}$]{\includegraphics[width=6.5cm]{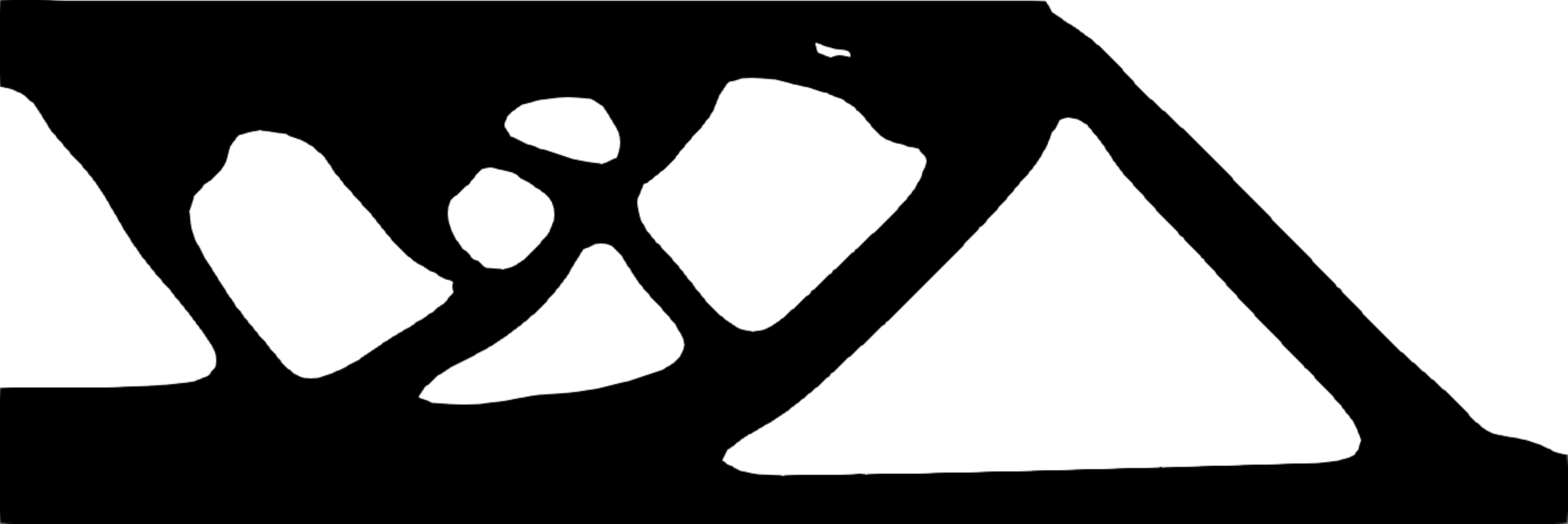}}
		\caption{Optimized MBB beams under different diffusion coefficients $a$.}
		\label{fig:ne2}
	\end{center}
\end{figure}
Figs. \ref{fig:ne1} and \ref{fig:ne2} show the set of optimization results obtained under different overhang angle constraints $\beta$ and diffusion coefficients $a$.
As shown in Fig. \ref{fig:ne1}, increasing $\beta$ eliminates the overhanging region, but creates many downward convex shapes.
By contrast, increasing the diffusion coefficient $a$ suppresses the downward convex shapes but creates an overhanging region, as shown in Fig. \ref{fig:ne2}.
This is because the diffusion coefficient affects the evaluation area of the overhanging region, as shown in Fig. \ref{fig:ovhg_2}.
In other words, if $a$ is set larger than $5\times10^{-3}$, the evaluation area is too large.
Therefore, the diffusion coefficient $a$ should be set in the range $1\times10^{-4}$ to $1\times10^{-3}$.
In the following optimization examples, $\beta=20$ and $a=5\times10^{-4}$ are set.
This result shows that constraining only the overhang angle is insufficient.
Therefore, to satisfy the self-support constraint, it is necessary to combine the angle constraint with other constraints, as in the proposed thermal constraint.
\subsubsection{Effect of the thermal constraint parameters}\label{sec:7.1.2}
This subsection presents optimization examples in which a thermal constraint is added to the overhang angle constraint to suppress the downward convex shapes.
Each parameter related to Eq. \ref{eq:drp1} is set as follows.
The thermal conductivity of the heat-conductive material is 119 W/mK.
The applied heat flux $q$ is set to 10 W, and the base plate temperature is set to $T_{amb}=0^{\circ}C$.
Here, we examine the effects of the parameters related to the thermal constraint $\gamma$ and the number of layers $m$ in the fixed design domain.
\begin{figure}[htbp]
	\begin{center}
		\centering
		\subfigure[$\gamma=0.1$; $J_{v}/J_{v~ref}=101\%$]{\includegraphics[width=6.5cm]{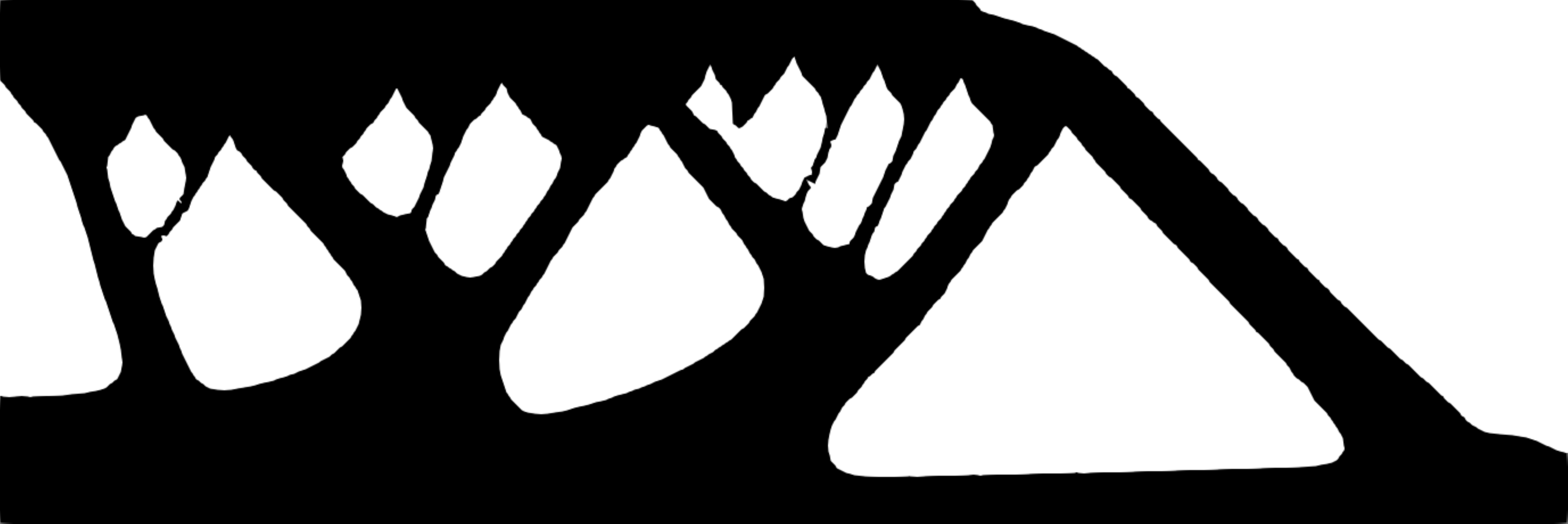}}
		\subfigure[$\gamma=0.2$; $J_{v}/J_{v~ref}=103\%$]{\includegraphics[width=6.5cm]{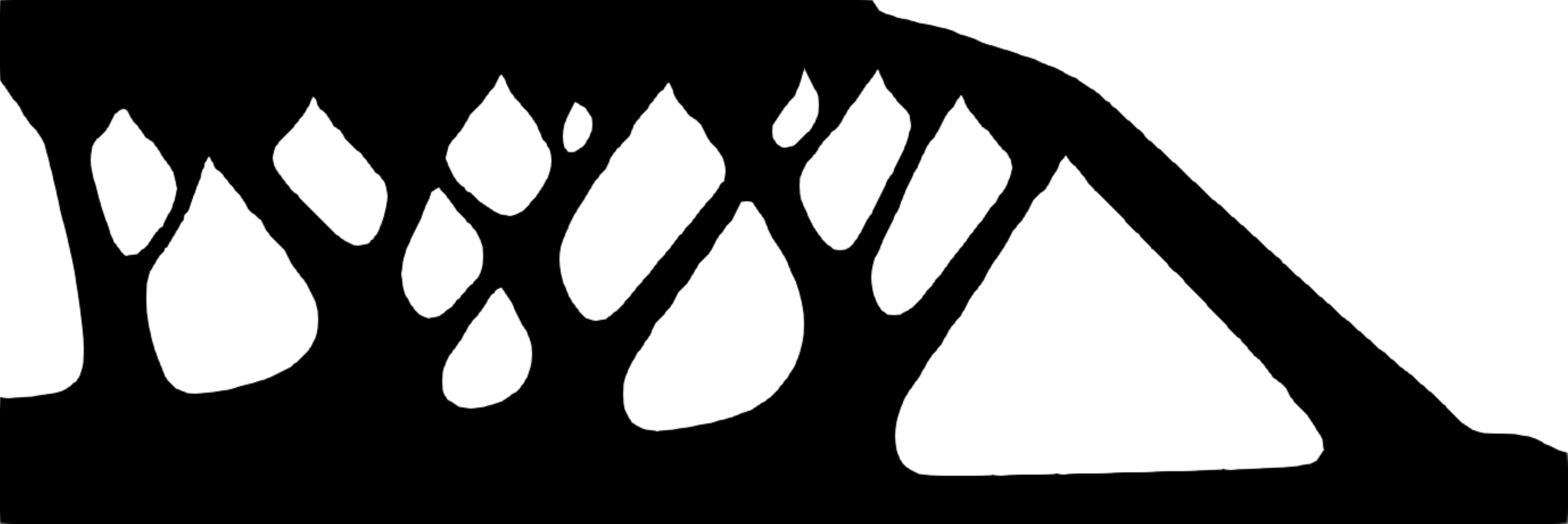}}
		\subfigure[$\gamma=0.4$; $J_{v}/J_{v~ref}=112\%$]{\includegraphics[width=6.5cm]{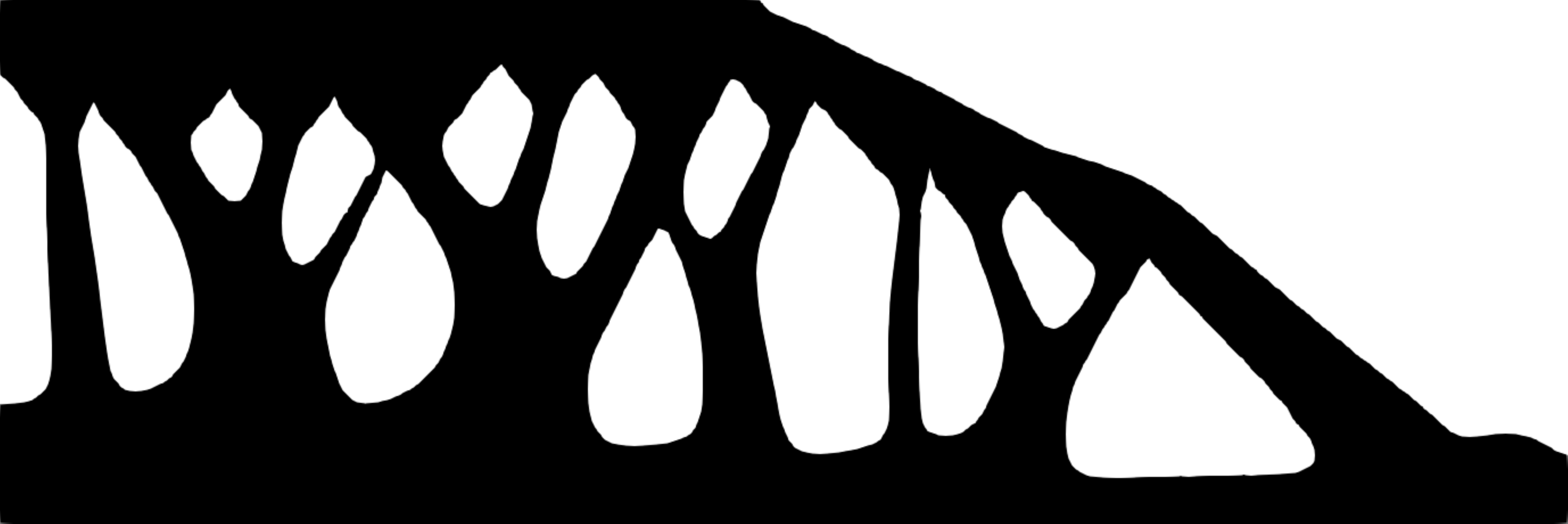}}
		\caption{Optimized MBB beams under different thermal constraint parameters $\gamma$.}
		\label{fig:ne3}
	\end{center}
\end{figure}
\begin{figure}[htbp]
	\begin{center}
		\centering
		\subfigure[$m=50$; $J_{v}/J_{v~ref}=103\%$]{\includegraphics[width=6.5cm]{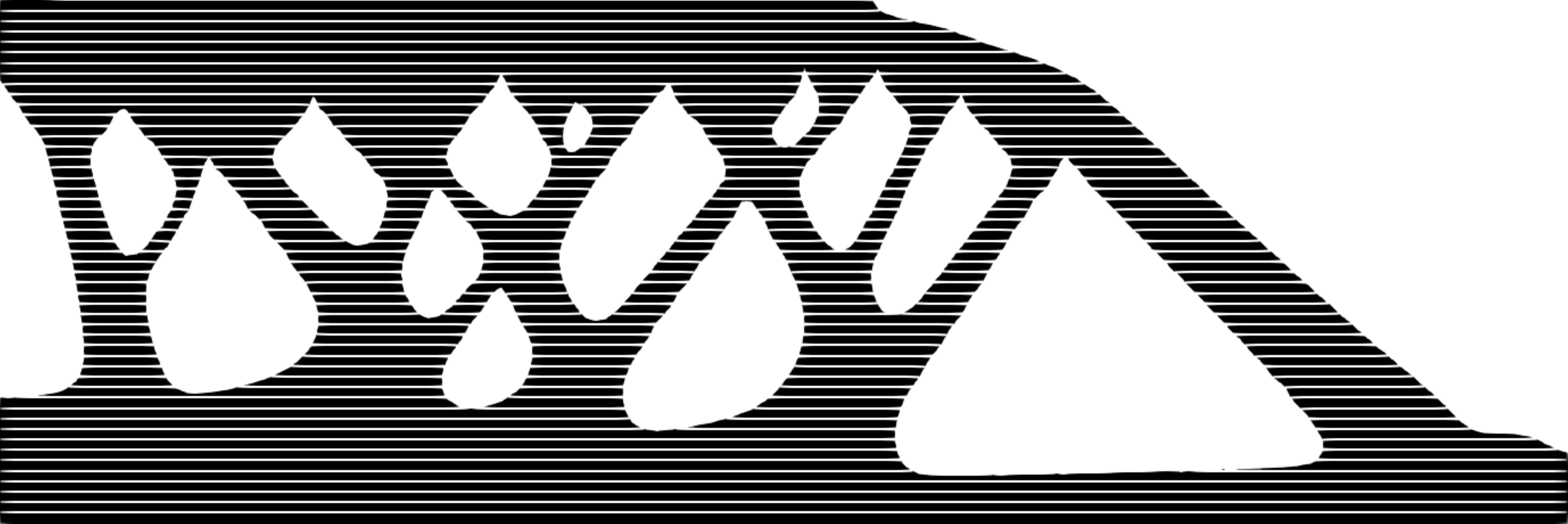}}
		\subfigure[$m=25$; $J_{v}/J_{v~ref}=104\%$]{\includegraphics[width=6.5cm]{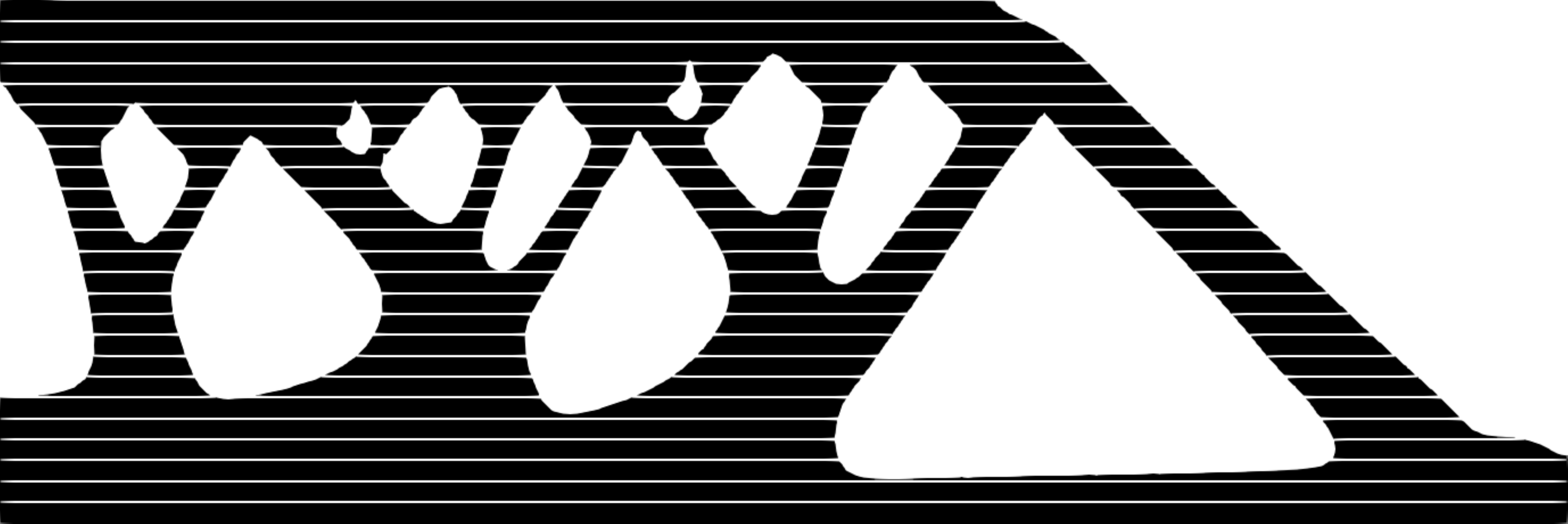}}
		\subfigure[$m=10$; $J_{v}/J_{v~ref}=103\%$]{\includegraphics[width=6.5cm]{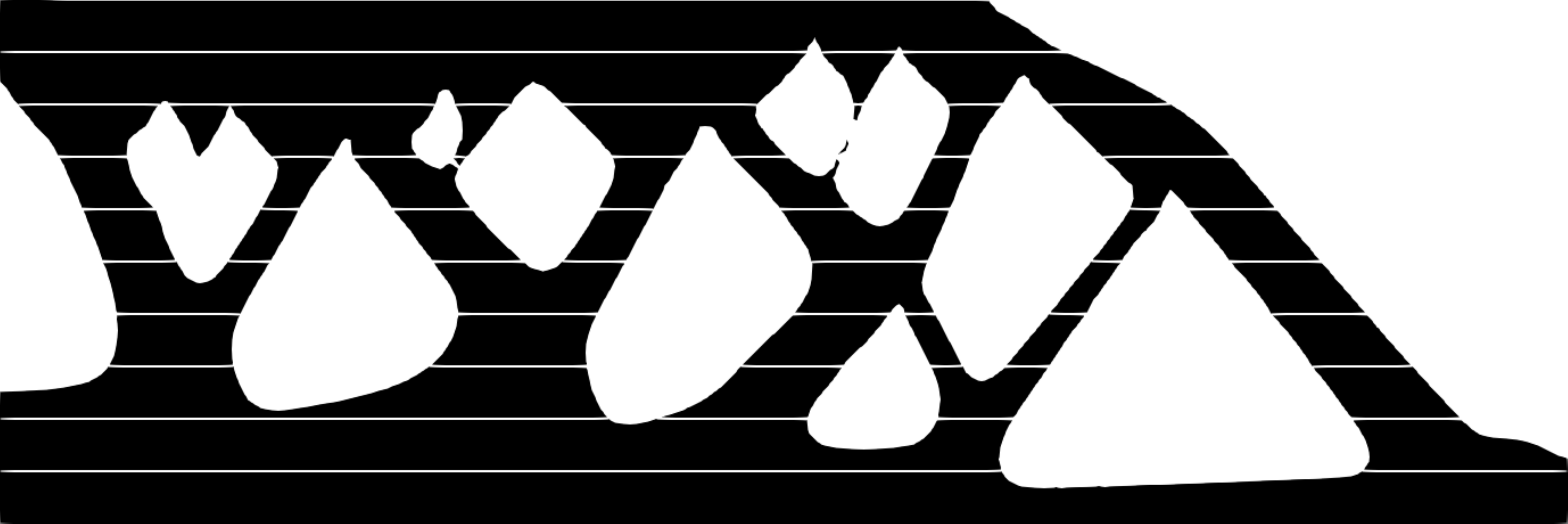}}
		\caption{Optimized MBB beams under the different layer number $m$.}
		\label{fig:ne4}
	\end{center}
\end{figure}
Fig. \ref{fig:ne3} shows the optimization results obtained under different thermal constraints $\gamma$, where the fixed design domain is divided into $m=$ 50 layers.
It can be observed that, by setting $\gamma$ larger than 0.2, the downward convex shapes is completely suppressed.
Furthermore, as $\gamma$ increases, each member becomes thicker, which is expected to improve heat dissipation.
However, setting $\gamma$ between 0.2 and 0.4 is appropriate, as increasing $\gamma$ beyond 0.4 deteriorates the compliance $J_{v}$ by more than 10\%.
Fig. \ref{fig:ne4} shows the optimization results obtained under different numbers of layers $m$, where $\gamma$ is set to 0.2.
The white lines in the figure indicate layer boundaries.
The downward convex shapes is no longer suppressed when the number of layers is less than $m=$ 10.
This result reveals that, if the number of layers is set to a size that can divide the downward convex shape, it can be suppressed by considering thermal constraints.
In the optimization following examples, $\gamma=$ 0.2.
\subsubsection{Effect of building direction}\label{sec:7.1.3}
Here, optimization examples for different building directions are presented to demonstrate the effectiveness of the proposed self-support constraint.
\begin{figure}[htbp]
	\begin{center}
		\includegraphics[width=9cm]{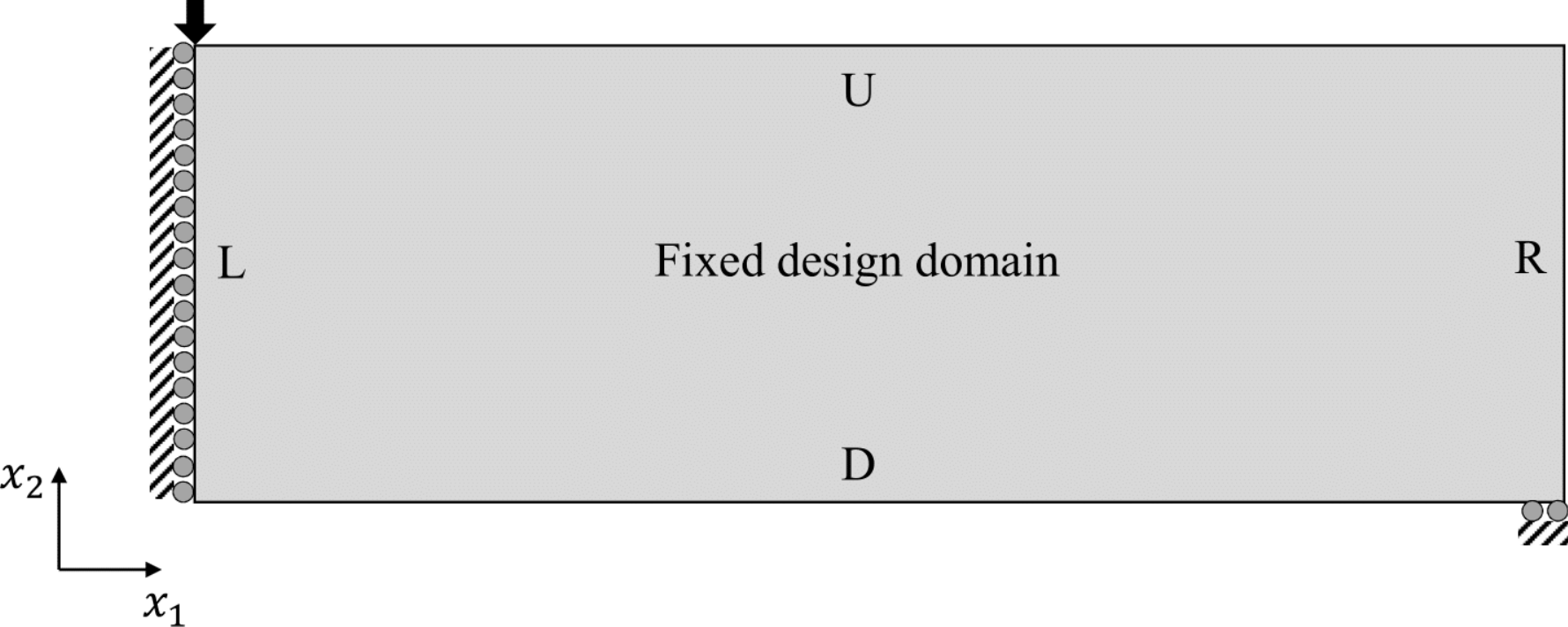}
		\caption{Problem setting for the building direction of the MBB beam.}
		\label{fig:MBB2}
	\end{center}
\end{figure}
As shown in Fig. \ref{fig:MBB2}, the building direction considers four cases in which each side U, D, L, and R of the fixed design domain is regarded as the base plate.
The number of layers in the fixed design domain is set to 25 for U and D and 50 layers for L and R.
\begin{figure}[htbp]
	\begin{center}
		\centering
		\subfigure[U case; $J_{v}/J_{v~ref}=107\%$]{\includegraphics[width=6.5cm]{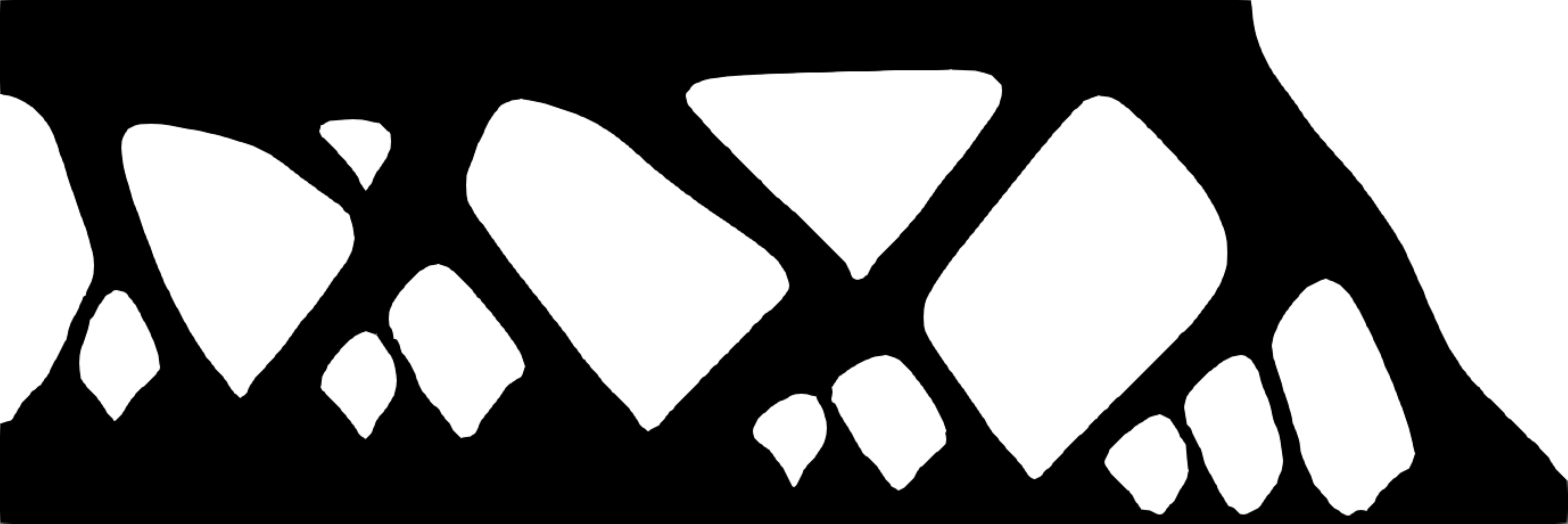}}
		\subfigure[D case; $J_{v}/J_{v~ref}=104\%$]{\includegraphics[width=6.5cm]{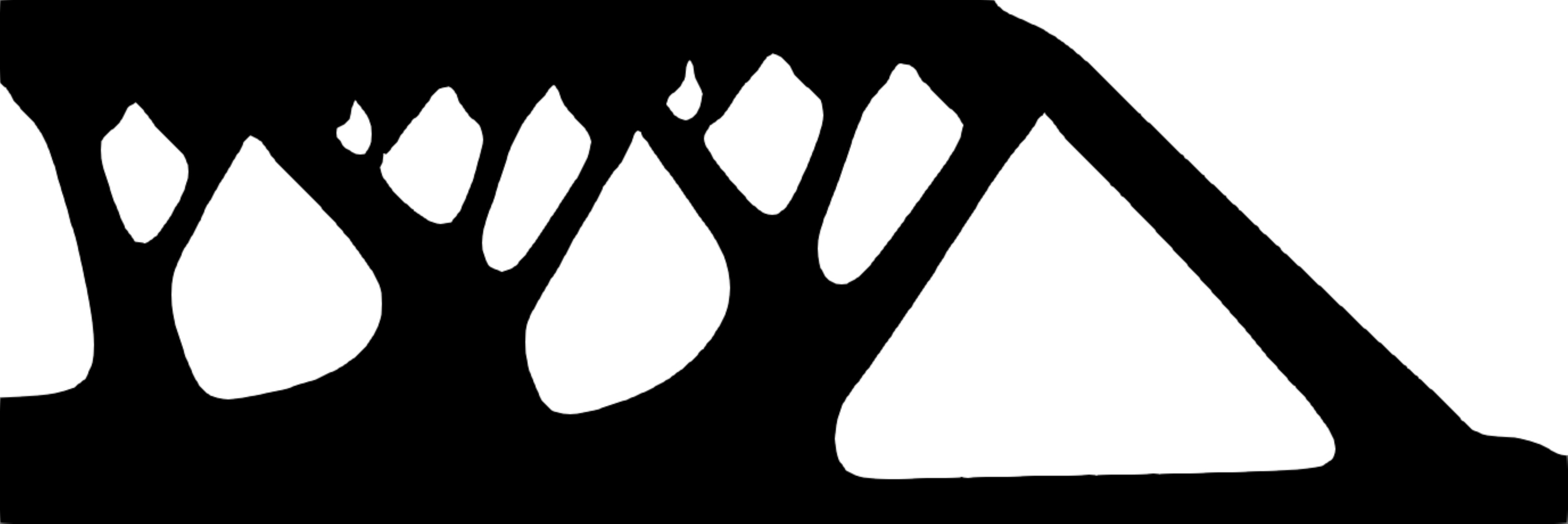}}
		\subfigure[L case; $J_{v}/J_{v~ref}=101\%$]{\includegraphics[width=6.5cm]{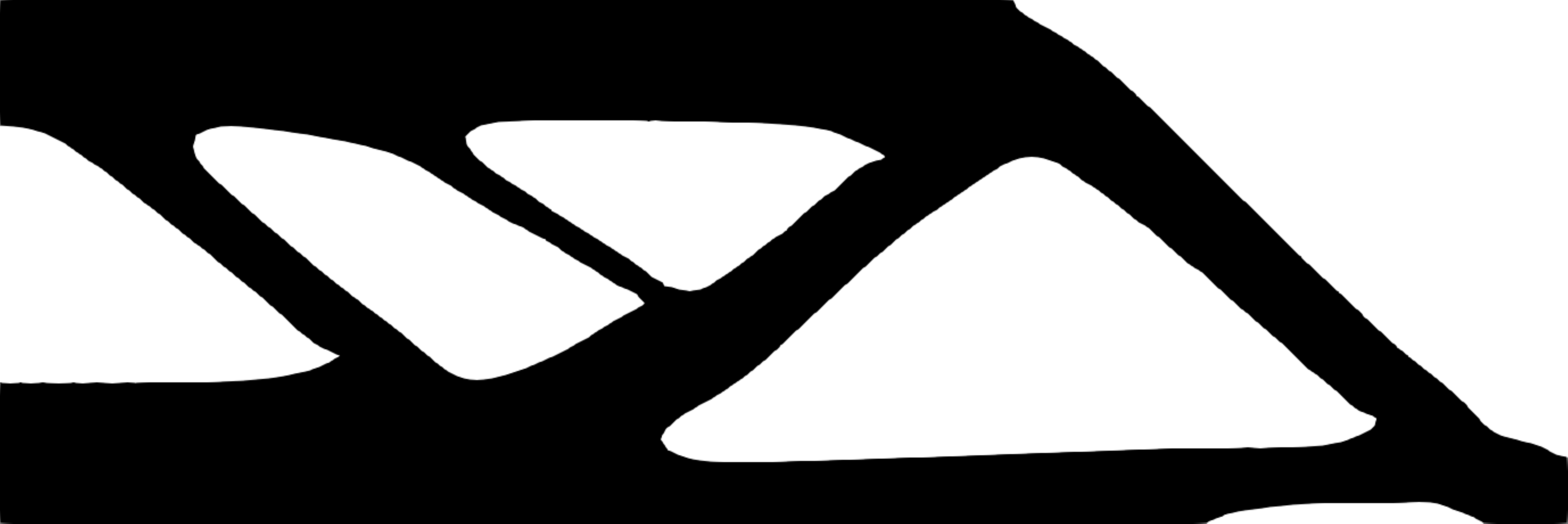}}
		\subfigure[R case; $J_{v}/J_{v~ref}=103\%$]{\includegraphics[width=6.5cm]{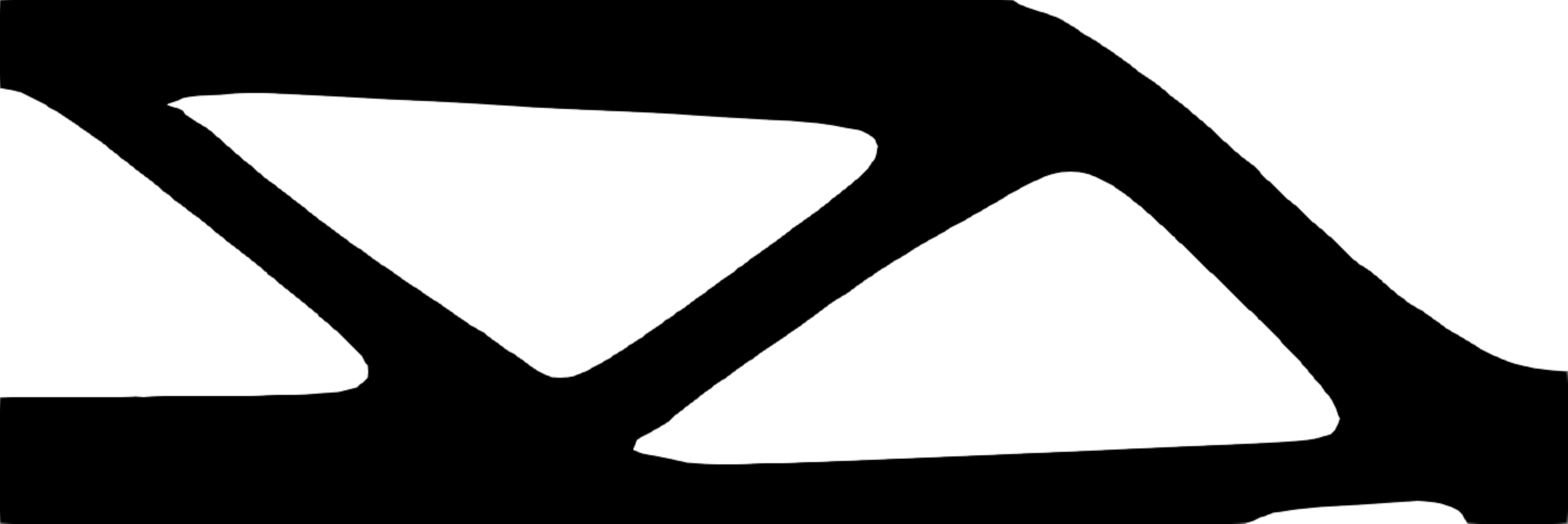}}
		\caption{Optimized MBB beams under different building directions.}
		\label{fig:ne5}
	\end{center}
\end{figure}
Fig. \ref{fig:ne5} shows the optimization results obtained under different building directions.
All optimal shapes suppress the downward convex shapes and satisfy the overhang angle constraint for any building direction.
In particular, the shape of L achieves a compliance equivalent to that of an unconstrained shape.
Thus, selecting an appropriate building direction will result in a printable design without compromising structural performance.
\subsubsection{Effect of the overhang angle}\label{sec:7.1.4}
\begin{figure}[htbp]
	\begin{center}
		\includegraphics[width=11cm]{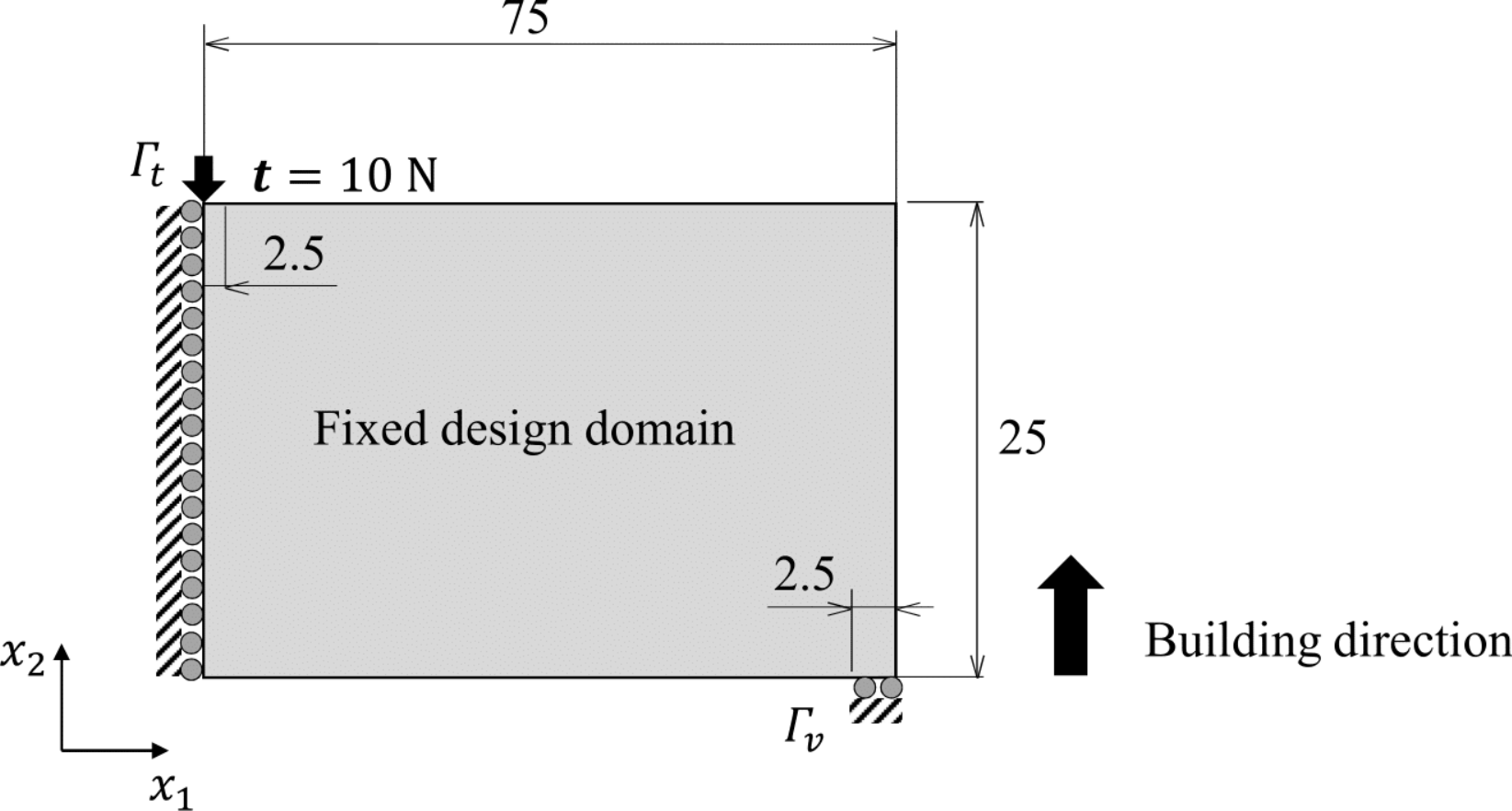}
		\caption{Problem setting for the cantilever beam, with dimensions in mm.}
		\label{fig:Canti}
	\end{center}
\end{figure}

This subsection presents optimization examples for different threshold angles to further demonstrate the effectiveness of the proposed method.
Therefore, the cases of threshold angles $\theta_{0}=30^{\circ}$, $45^{\circ}$, and $60^{\circ}$ are considered here. 
Examples include the 2D MBB and cantilever beam shown in Fig. \ref{fig:Canti}.
Each parameter is set to the same value as that of the MBB beam.
\begin{figure}[htbp]
	\begin{center}
		\centering
		\subfigure[$\theta_{0}=30^{\circ}$; $J_{v}/J_{v~ref}=101\%$]{\includegraphics[width=6.5cm]{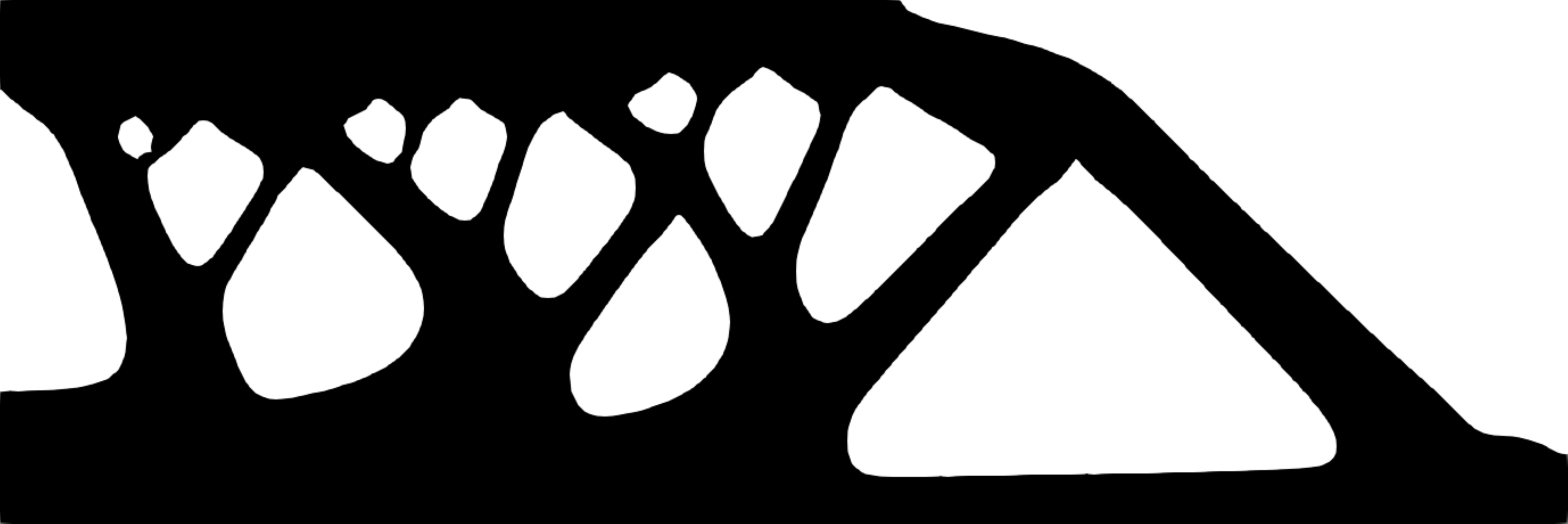}}
		\subfigure[$\theta_{0}=60^{\circ}$; $J_{v}/J_{v~ref}=106\%$]{\includegraphics[width=6.5cm]{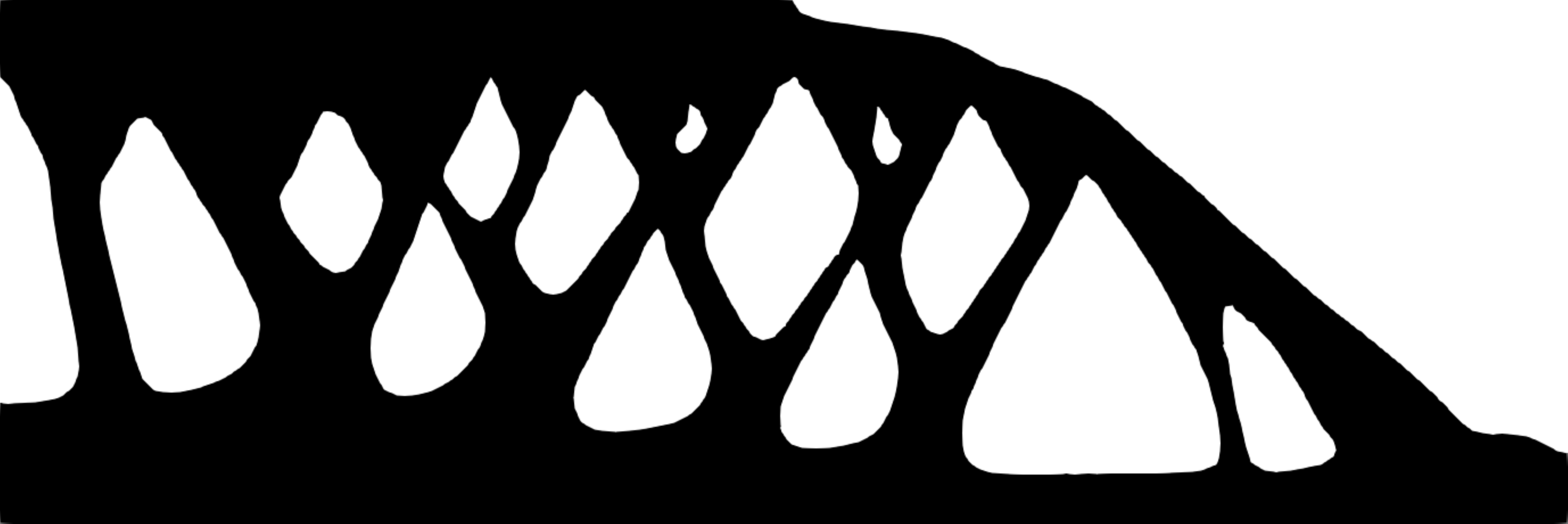}}
		\caption{Optimized MBB beams under different threshold angles.}
		\label{fig:ne6}
	\end{center}
\end{figure}
\begin{figure}[htbp]
	\begin{center}
		\centering
		\subfigure[Without constraints; $J_{v~ref}=0.64$]{\includegraphics[width=6cm]{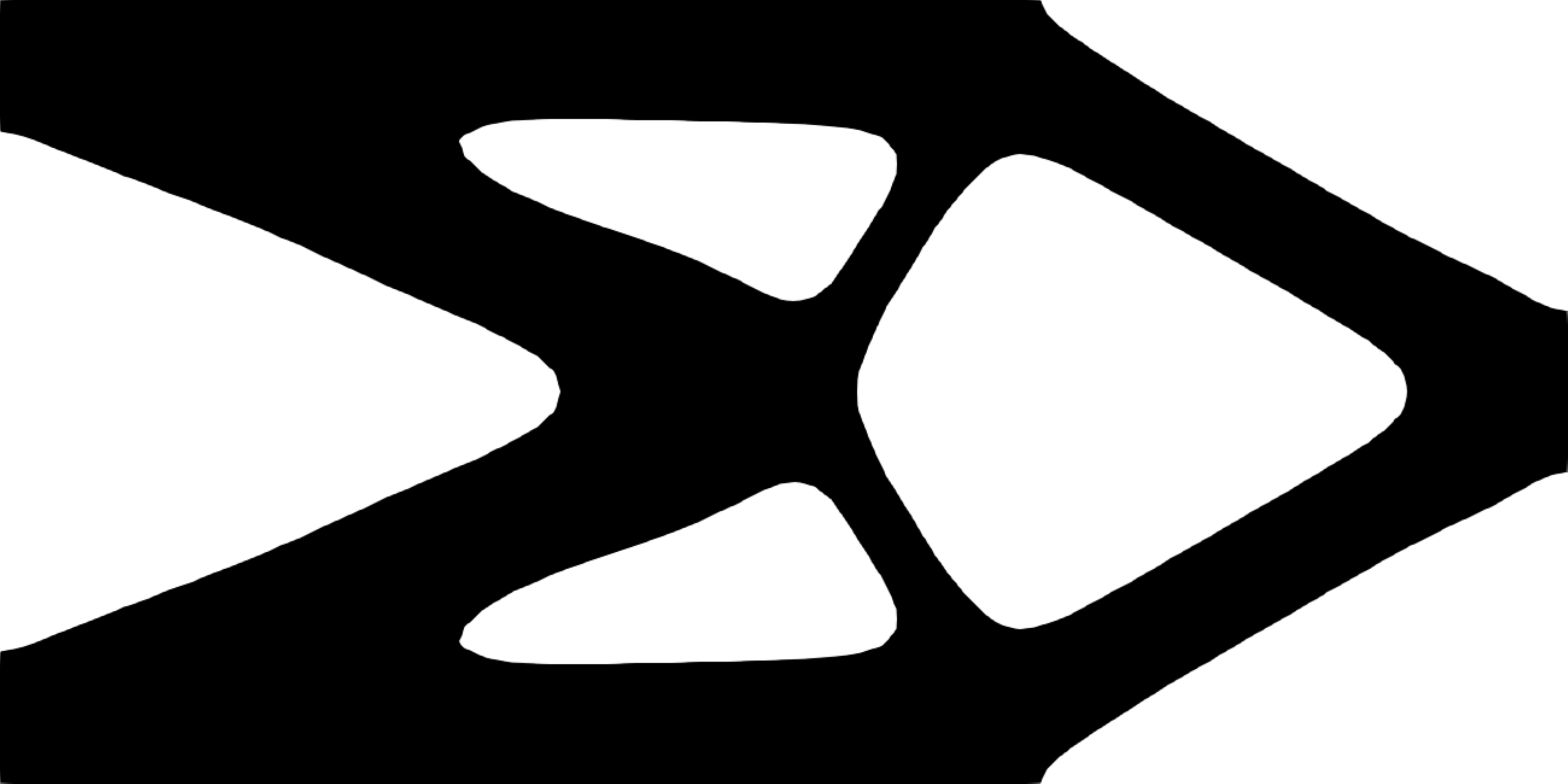}}
		\hspace{0.5cm}
		\subfigure[$\theta_{0}=30^{\circ}$; $J_{v}/J_{v~ref}=103\%$]{\includegraphics[width=6cm]{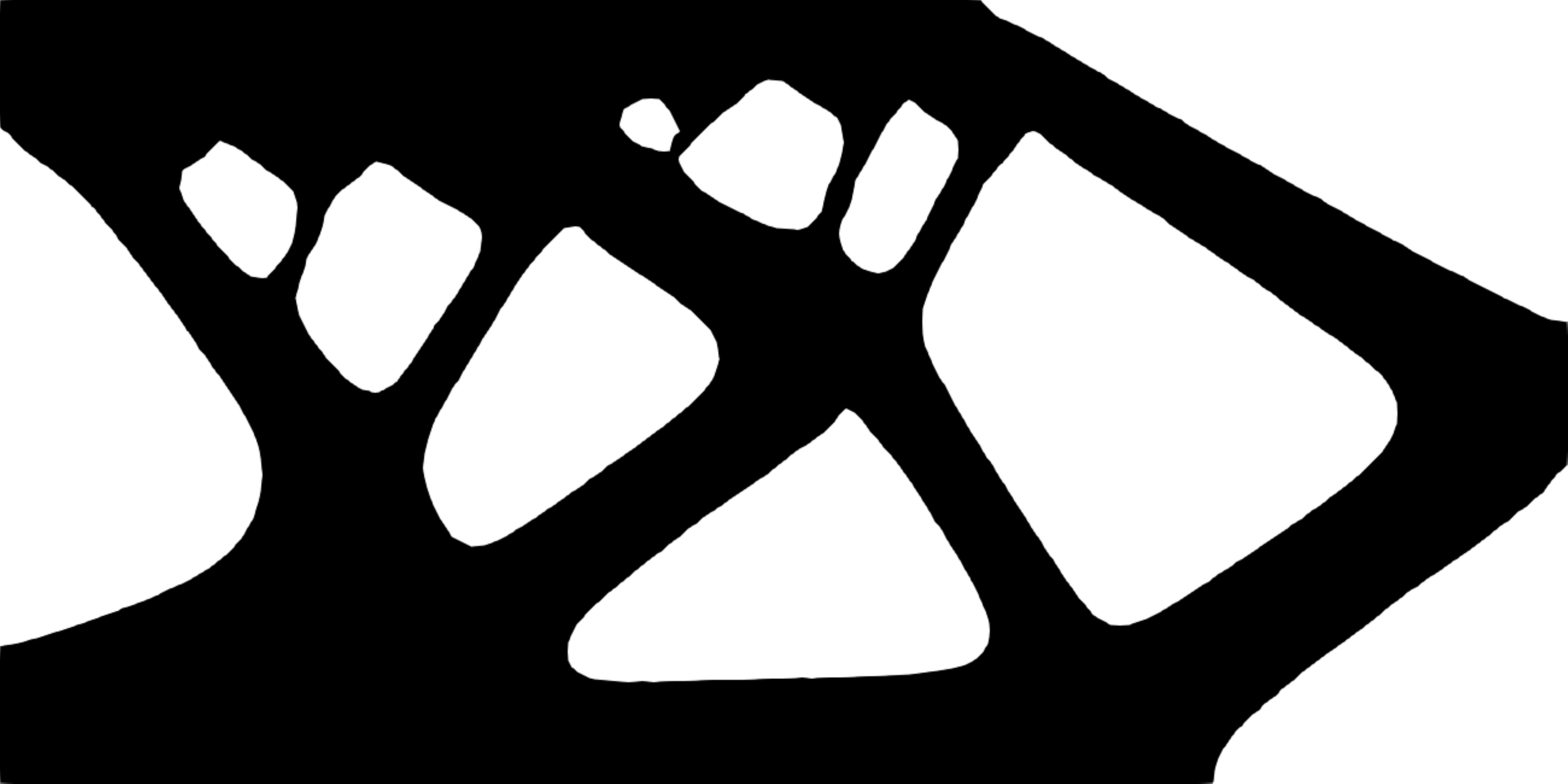}}
		\subfigure[$\theta_{0}=45^{\circ}$; $J_{v}/J_{v~ref}=108\%$]{\includegraphics[width=6cm]{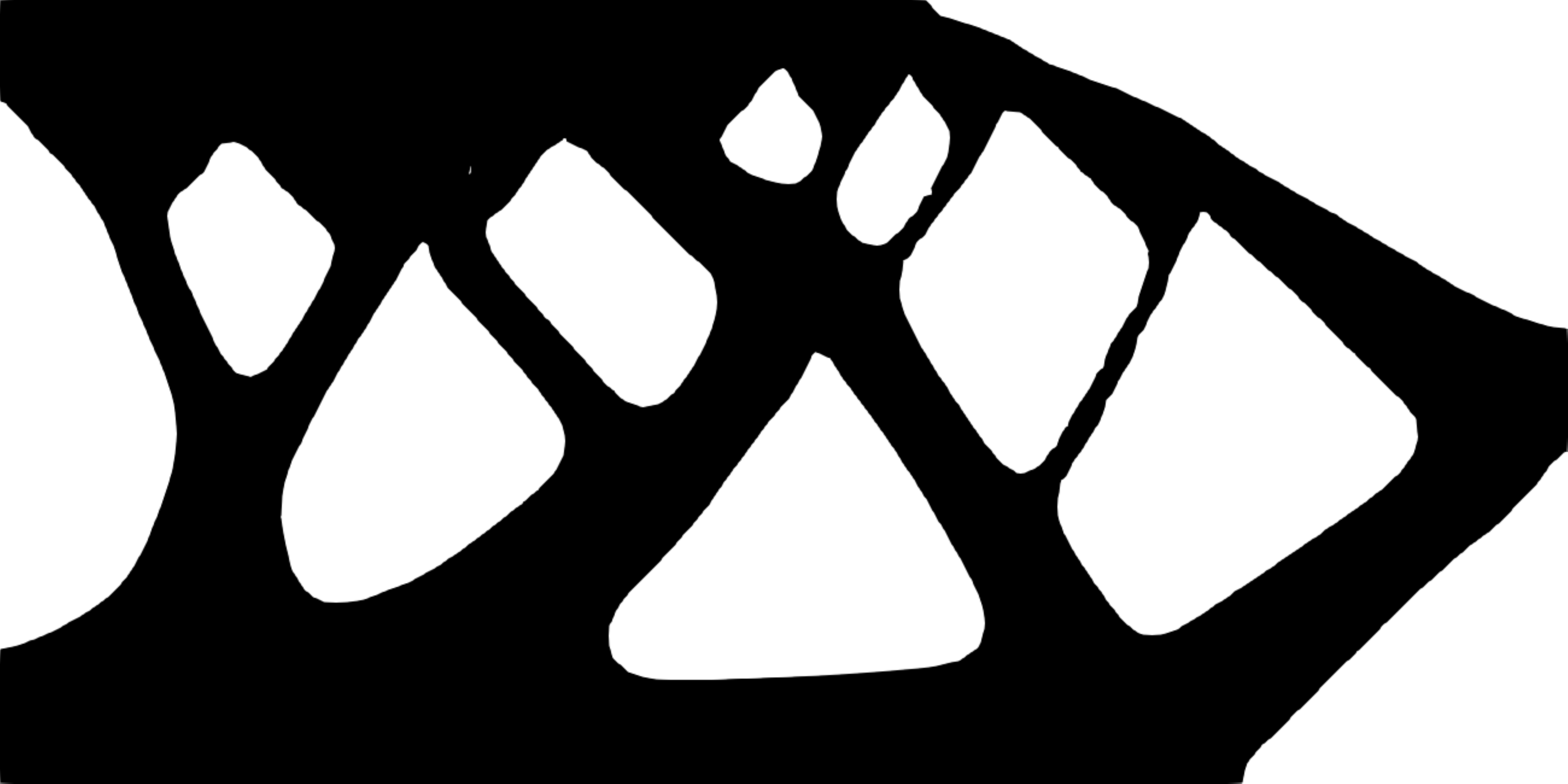}}
		\hspace{0.5cm}
		\subfigure[$\theta_{0}=60^{\circ}$; $J_{v}/J_{v~ref}=114\%$]{\includegraphics[width=6cm]{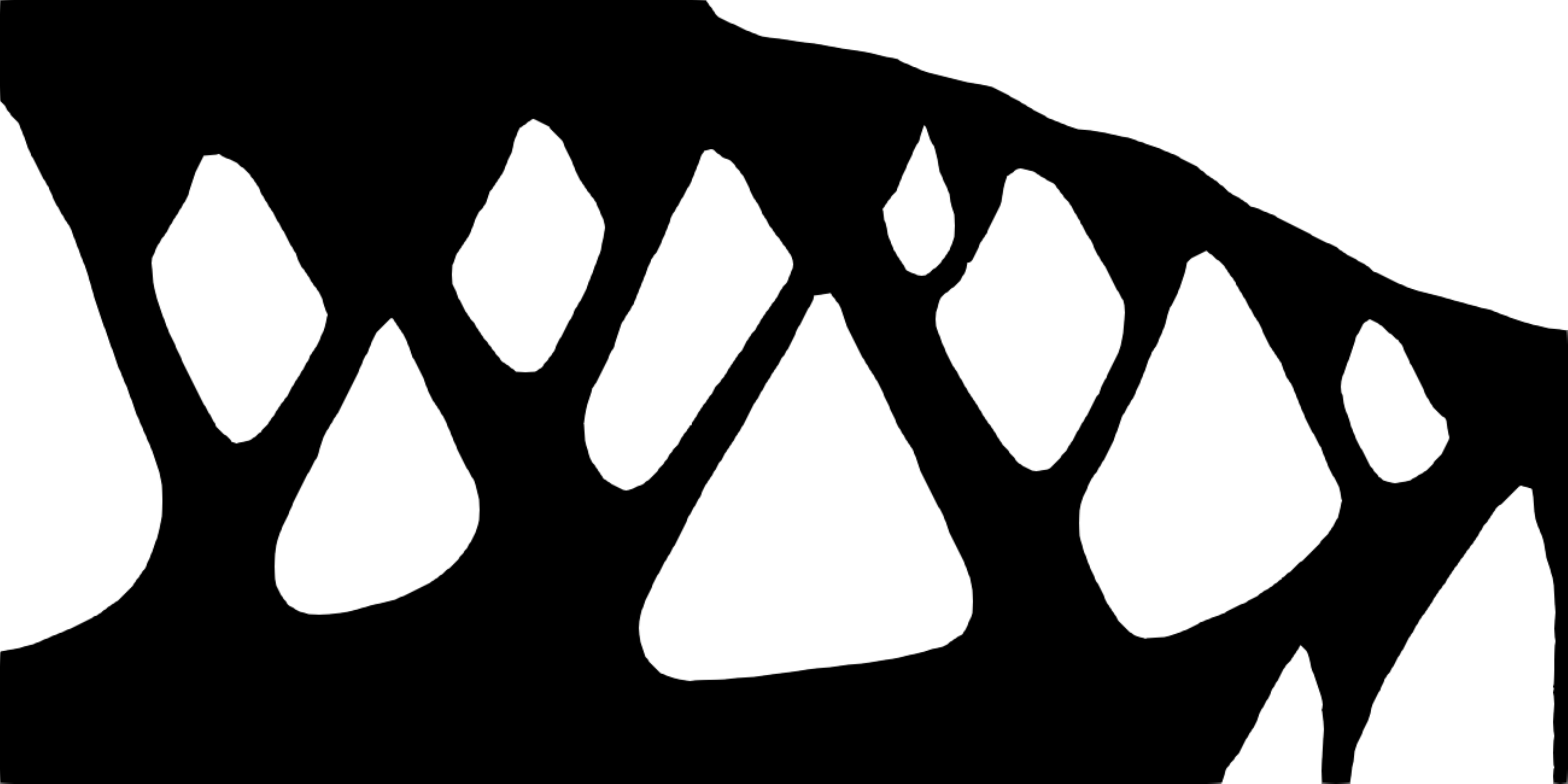}}
		\caption{Optimized cantilever beams under different threshold angles.}
		\label{fig:ne7}
	\end{center}
\end{figure}
Figs. \ref{fig:ne6} and \ref{fig:ne7} show the optimization results obtained under different threshold angles.
In the MBB example, shapes are obtained in which no member below the specified threshold angle is created.
However, in the cantilever example, some members violate the constraint in Fig. \ref{fig:ne7} (d).
This is because the constraint function is treated as a penalty term, which implies that increasing the threshold angle with the same penalty parameters may not fully satisfy the constraint.
Therefore, the penalty parameters should be modified.
Furthermore, no significant deterioration in compliance is observed even when the threshold angle is $\theta_{0}=60^{\circ}$.
These results demonstrate the effectiveness of the proposed methodology for self-support constraints.
\subsection{Combination with the distortion constraint}\label{sec:7.2}
This subsection presents 3D optimization examples that combine a self-support constraint with a distortion constraint.
\subsubsection{3D cantilever beam}\label{sec:7.2.1}
First, we consider the minimum mean compliance problem for a 3D cantilever beam.
\begin{figure}[htbp]
	\begin{center}
		\includegraphics[width=11cm]{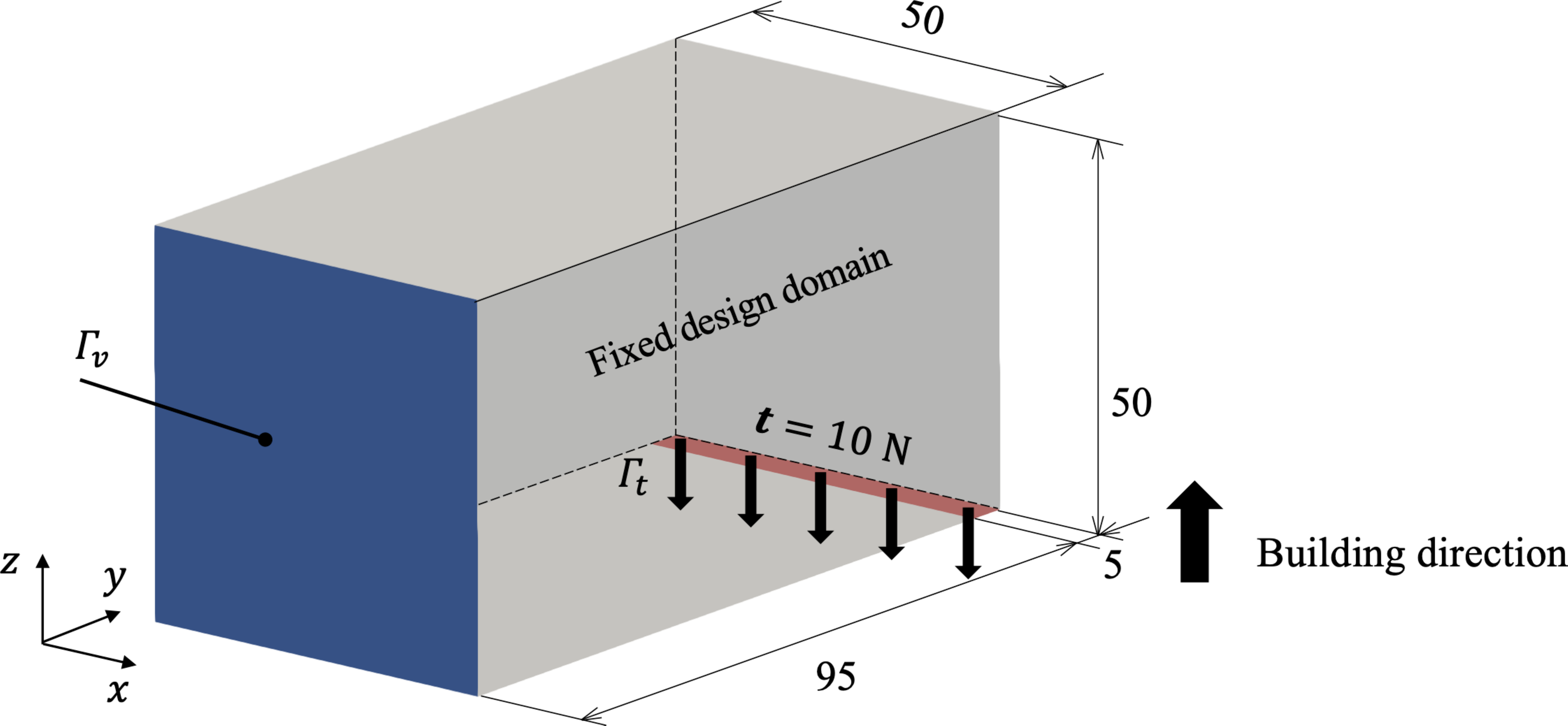}
		\caption{Problem setting for the 3D cantilever beam, with dimensions in mm. The blue surface indicates the fully cramped condition, the red surface indicates the applied traction, and the building direction is the positive z-axis.}
		\label{fig:3DCanti}
	\end{center}
\end{figure}
The fixed design domain and boundary conditions are shown in Fig. \ref{fig:3DCanti}
The upper limit of the allowable volume is set to 20\% of the fixed design domain.
The representative length in Eq. \ref{eq:ovhg1} is set to $L=$ 50 mm.
The threshold overhang angle is set to $\theta_{0}=45^{\circ}$.
The fixed design domain is divided in the building direction into $m = 25$ layers for the thermal constraint and $n = 50$ layers for the distortion constraint.
The weighting parameter, which is related to the distortion constraint in Eq. \ref{eq:op2}, is set to $\alpha=0.05$.
The other parameters are those set in previous optimization examples.
\begin{figure}[htbp]
	\begin{center}
		\centering
		\subfigure[Bird's-eye view]{\includegraphics[width=6cm]{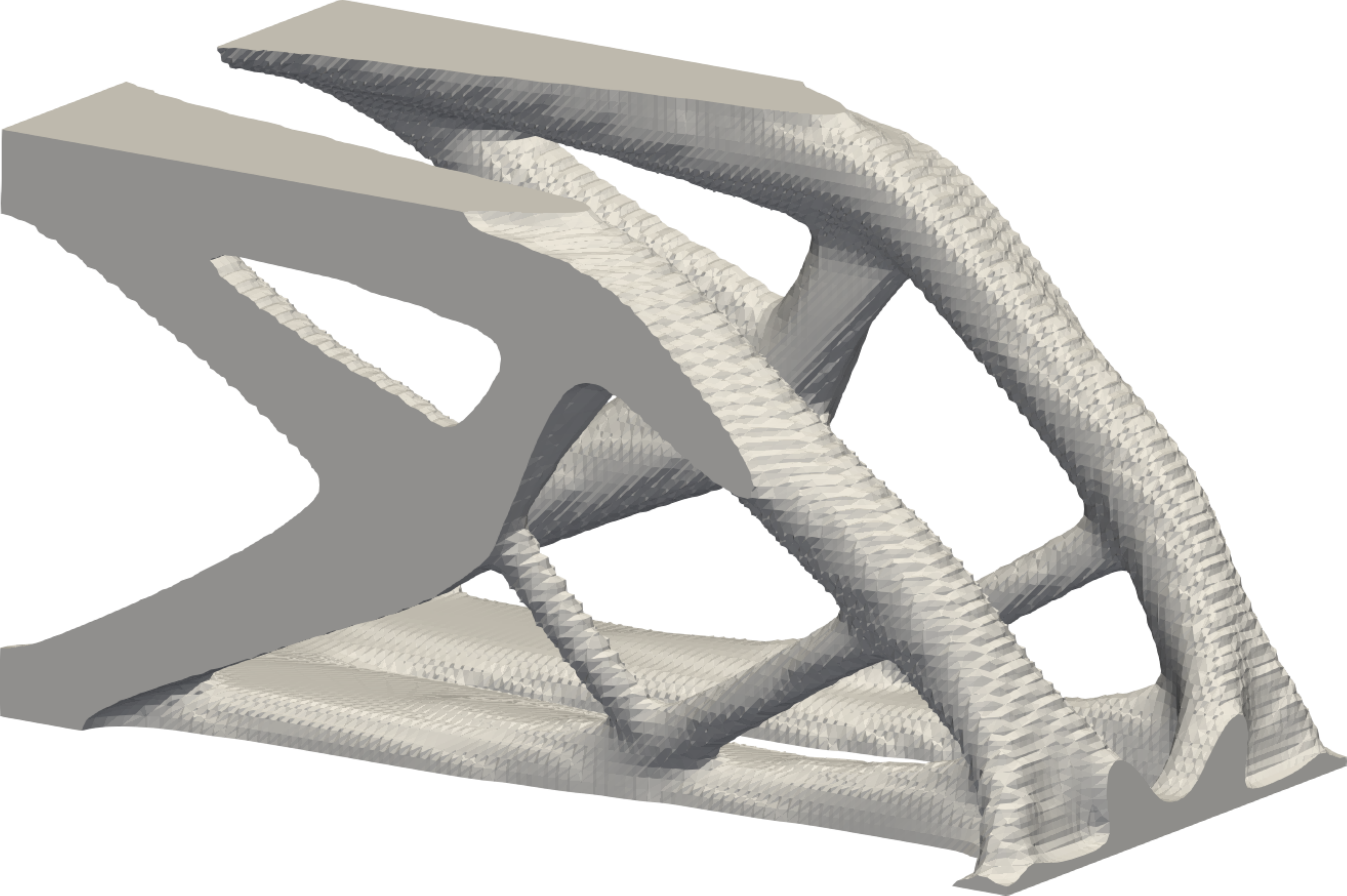}}
		\hspace{0.1cm}
		\subfigure[Rear view]{\includegraphics[width=3.5cm]{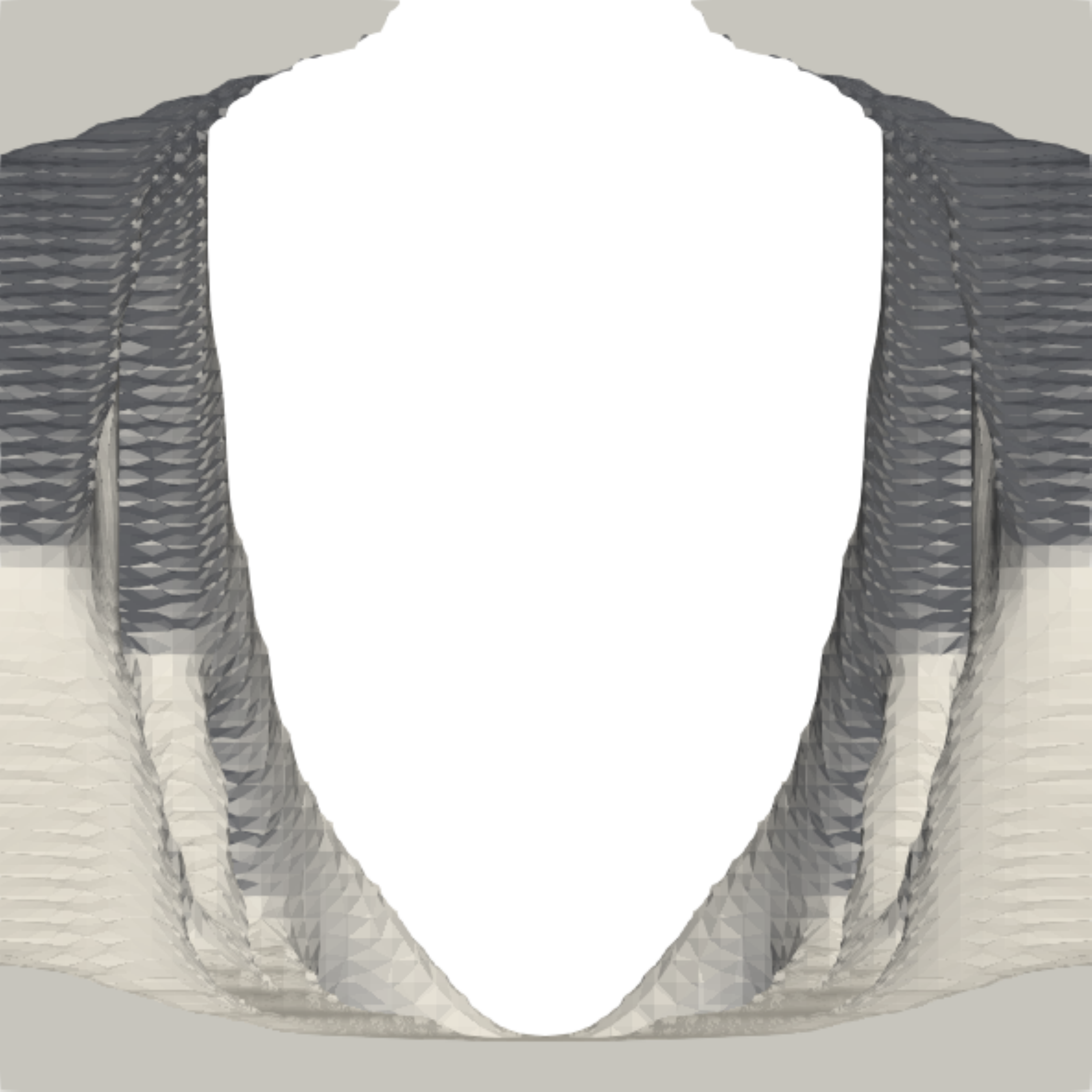}}
		\hspace{0.1cm}
		\subfigure[Front view]{\includegraphics[width=3.5cm]{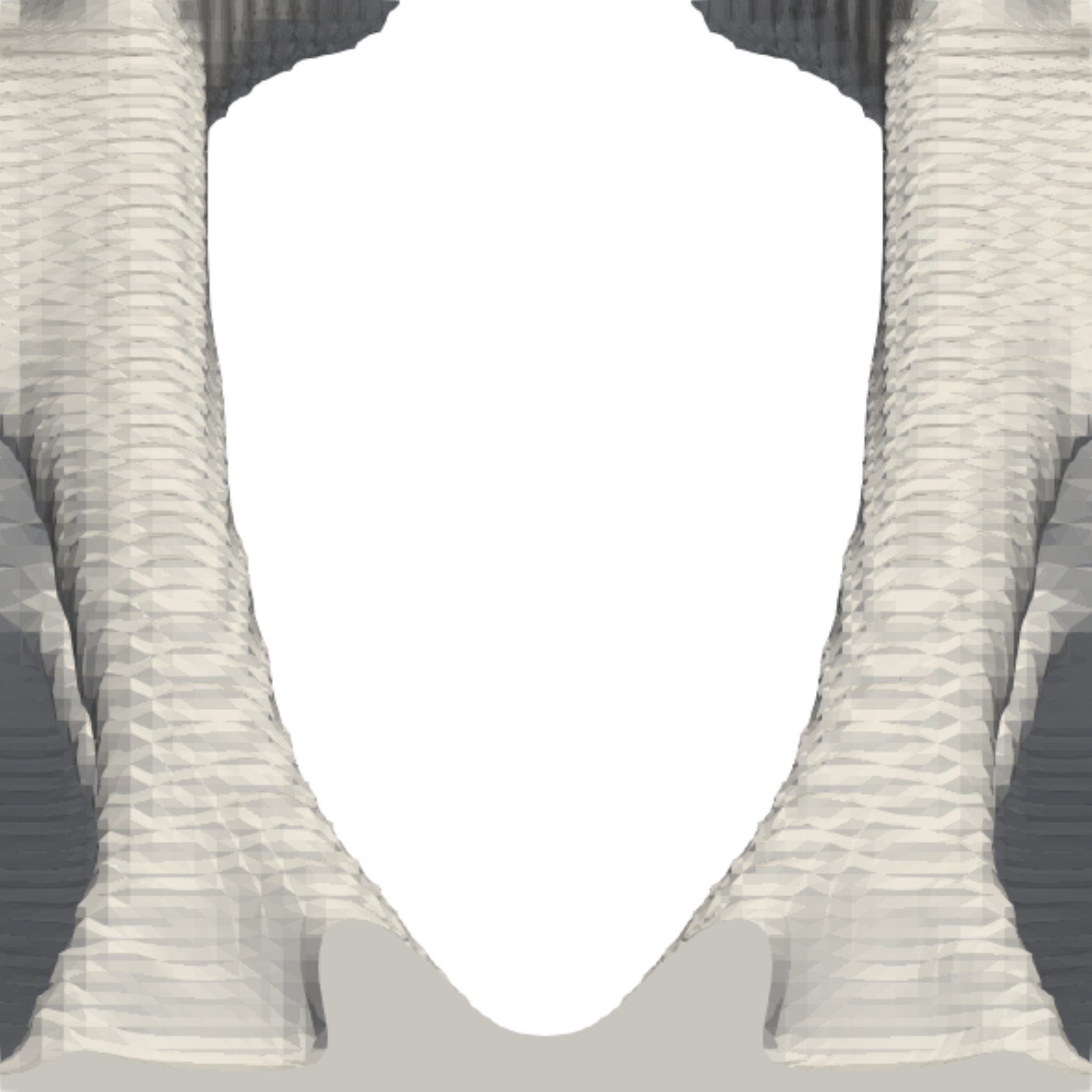}}
		\caption{Optimized 3D cantilever beam without constraints; $J_{v~ref}=0.031$.}
		\label{fig:ne8}
	\end{center}
\end{figure}
\begin{figure}[htbp]
	\begin{center}
		\centering
		\subfigure[Bird's-eye view]{\includegraphics[width=6cm]{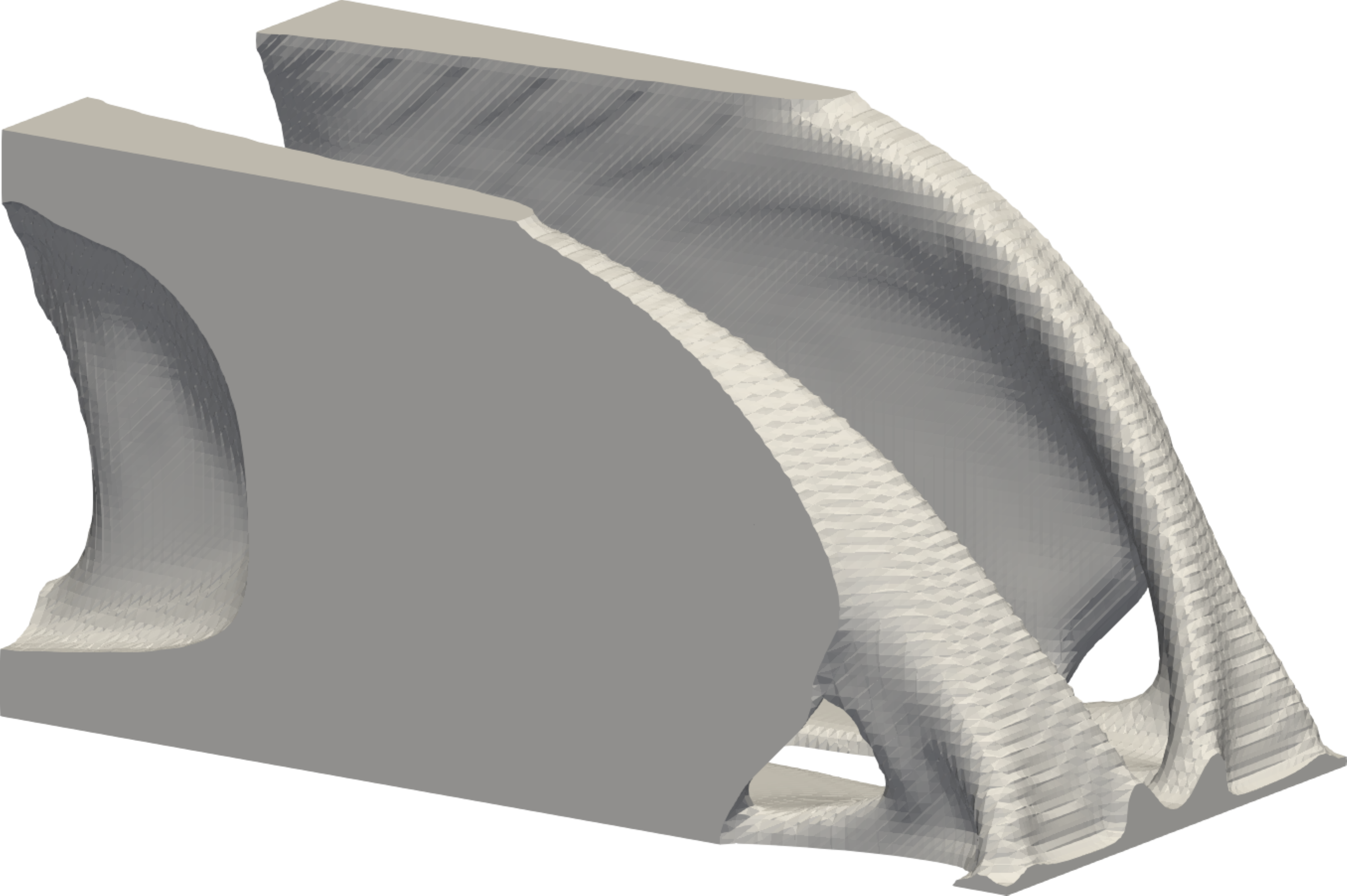}}
		\hspace{0.1cm}
		\subfigure[Rear view]{\includegraphics[width=3.5cm]{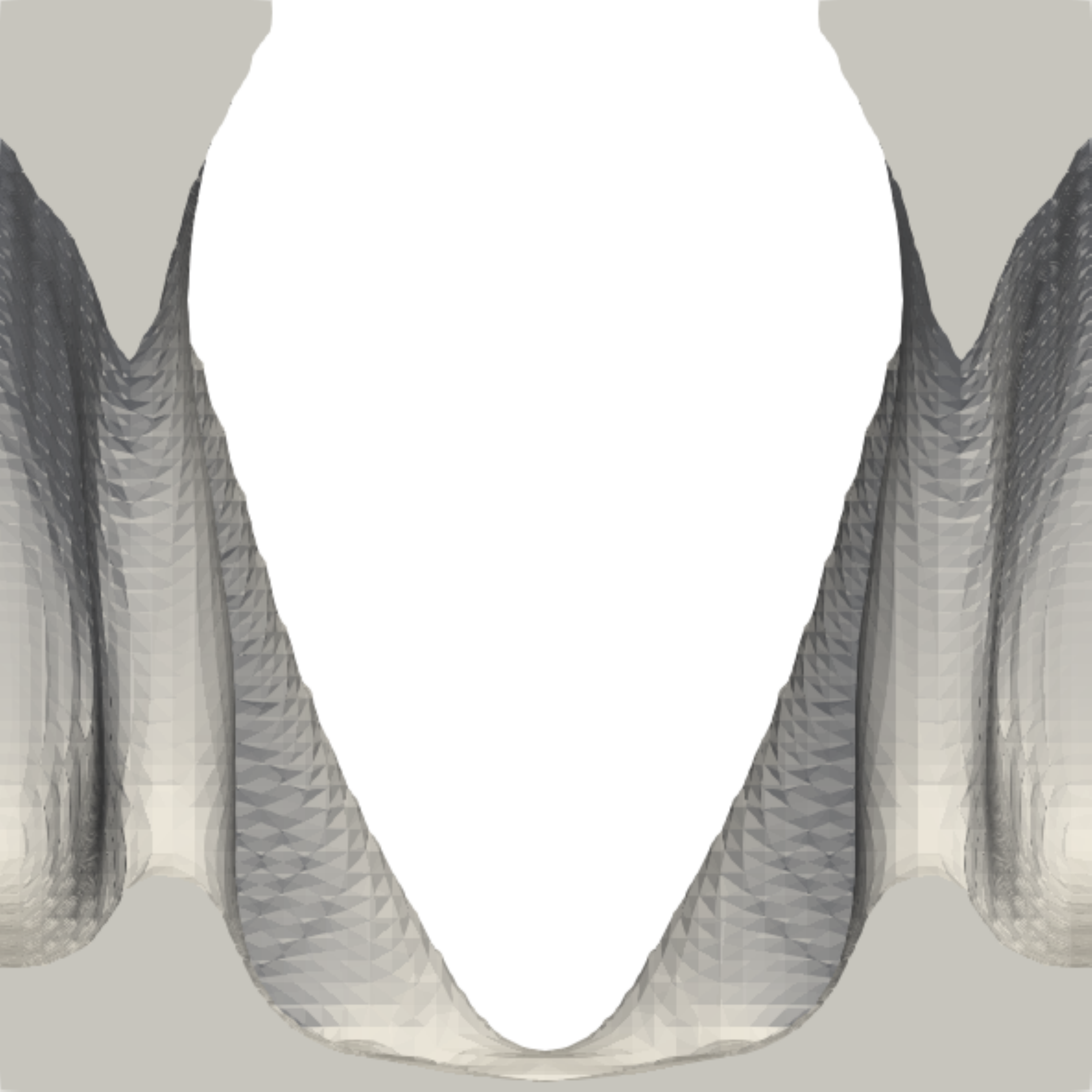}}
		\hspace{0.1cm}
		\subfigure[Front view]{\includegraphics[width=3.5cm]{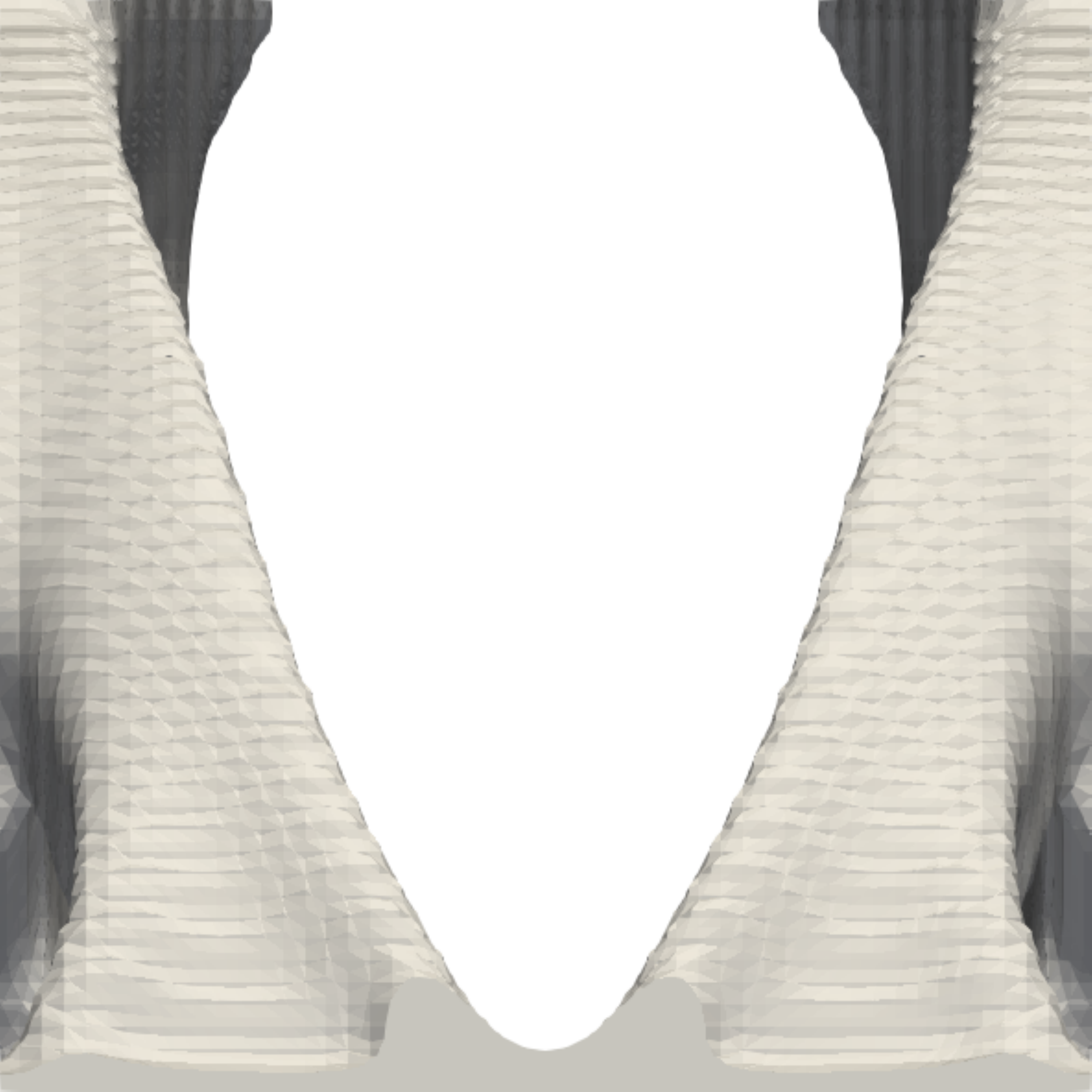}}
		\caption{Optimized 3D cantilever beam with the self-support constraint; $J_{v}/J_{v~ref}=110\%$.}
		\label{fig:ne9}
	\end{center}
\end{figure}
\begin{figure}[htbp]
	\begin{center}
		\centering
		\subfigure[Bird's-eye view]{\includegraphics[width=6.0cm]{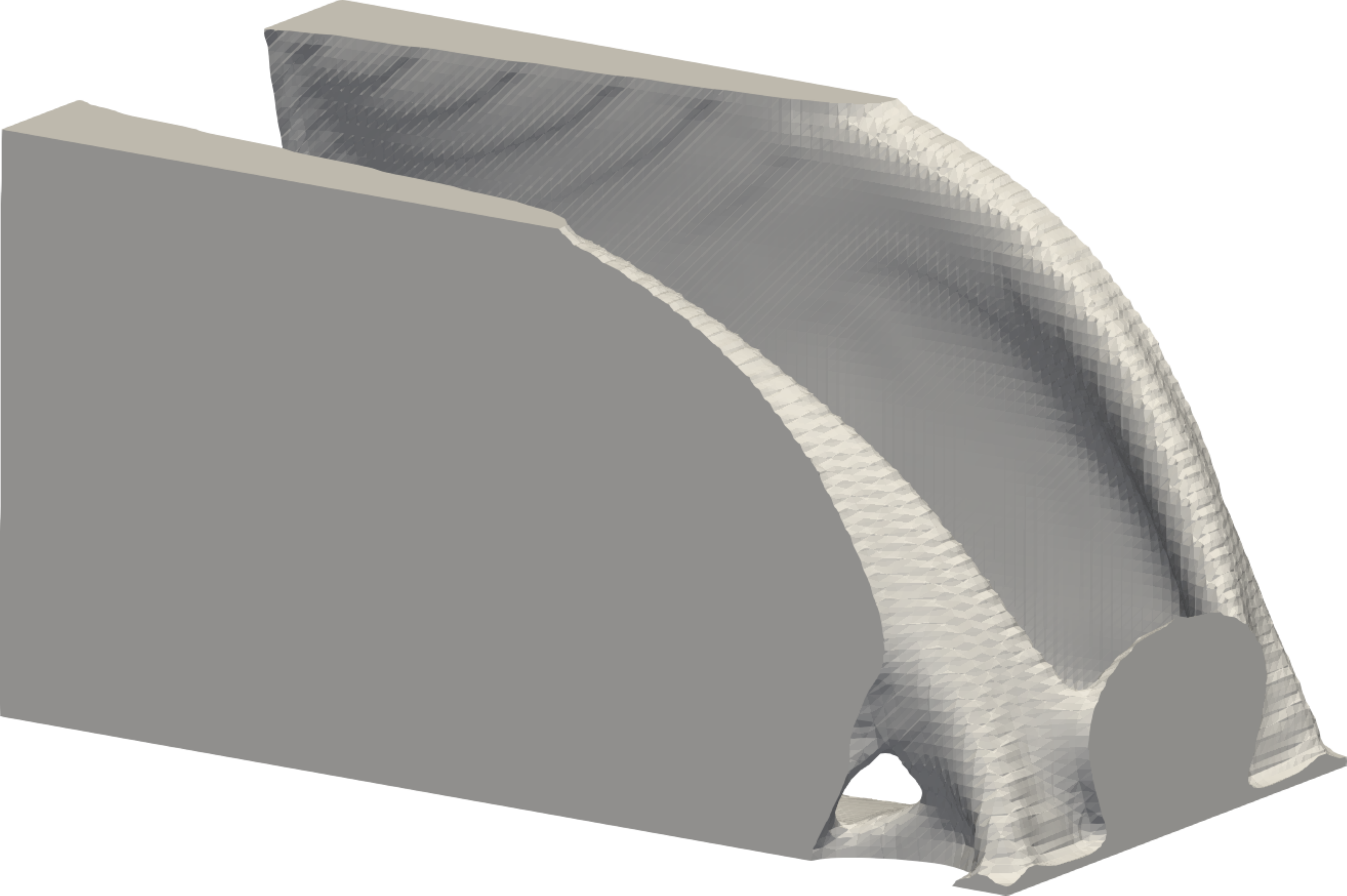}}
		\hspace{0.1cm}
		\subfigure[Rear view]{\includegraphics[width=3.5cm]{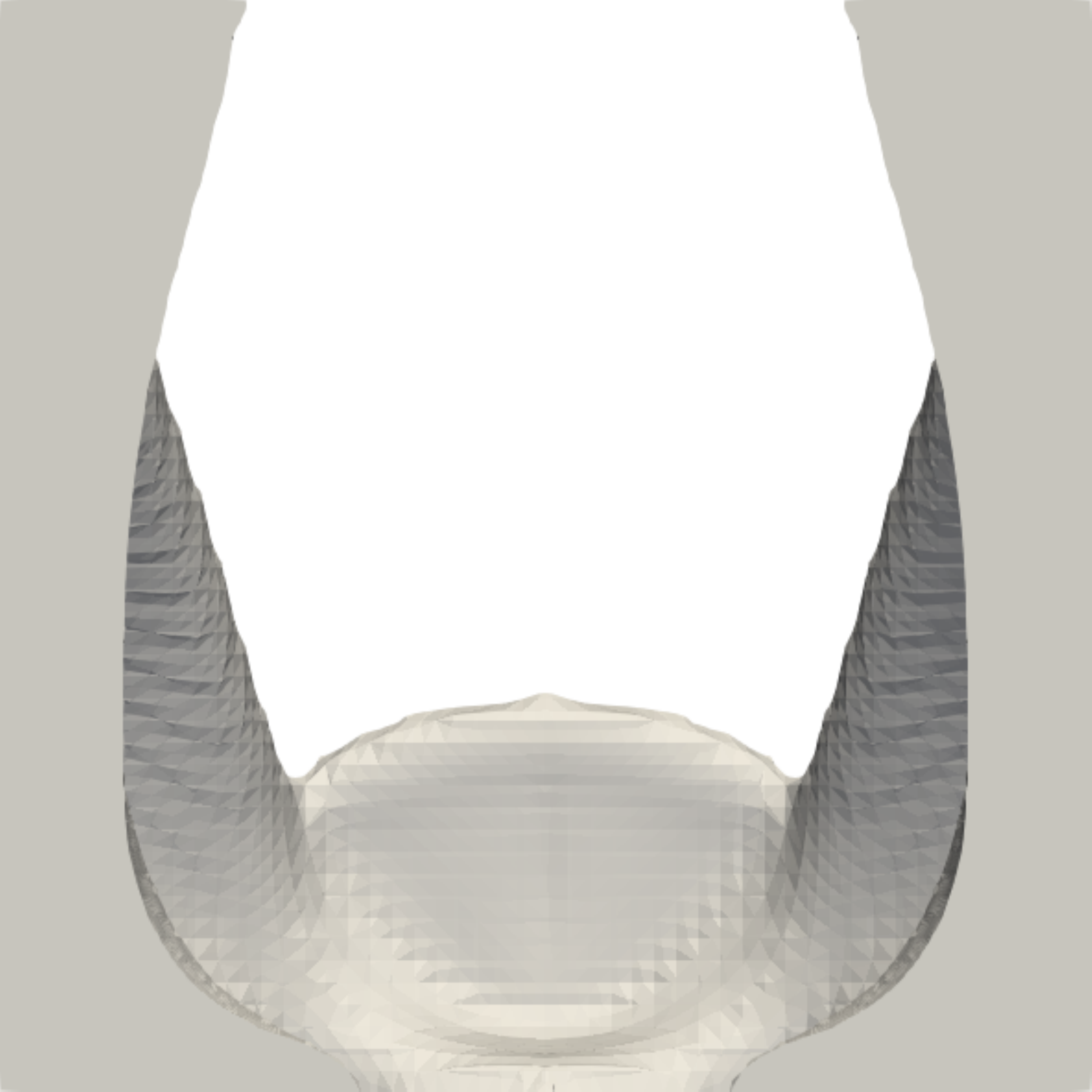}}
		\hspace{0.1cm}
		\subfigure[Front view]{\includegraphics[width=3.5cm]{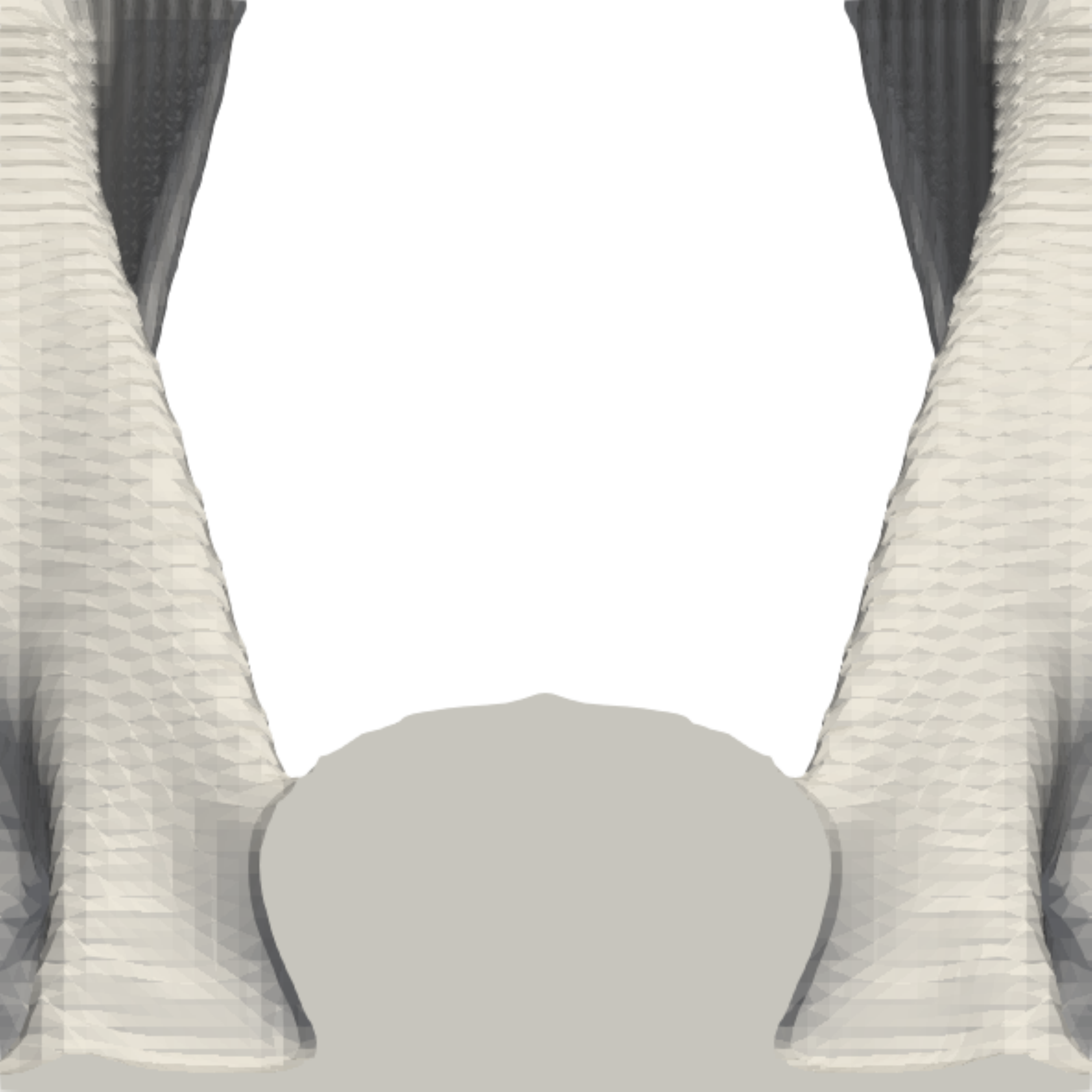}}
		\caption{Optimized 3D cantilever beam with the self-support and distortion constraints; $J_{v}/J_{v~ref}=114\%$.}
		\label{fig:ne10}
	\end{center}
\end{figure}
Figs. \ref{fig:ne8}, \ref{fig:ne9} and \ref{fig:ne10} present the obtained optimization results without a constraint, with the self-support constraint, and with the self-support and distortion constraint, respectively.
Both constraint-imposed shapes suppress the downward convex shapes and satisfy the overhang angle constraint.
Furthermore, comparing the compliance of each shape, it can be observed that the effect of the distortion constraint on the structural performance is minimal.
\begin{figure}[htbp]
	\begin{center}
		\includegraphics[width=14cm]{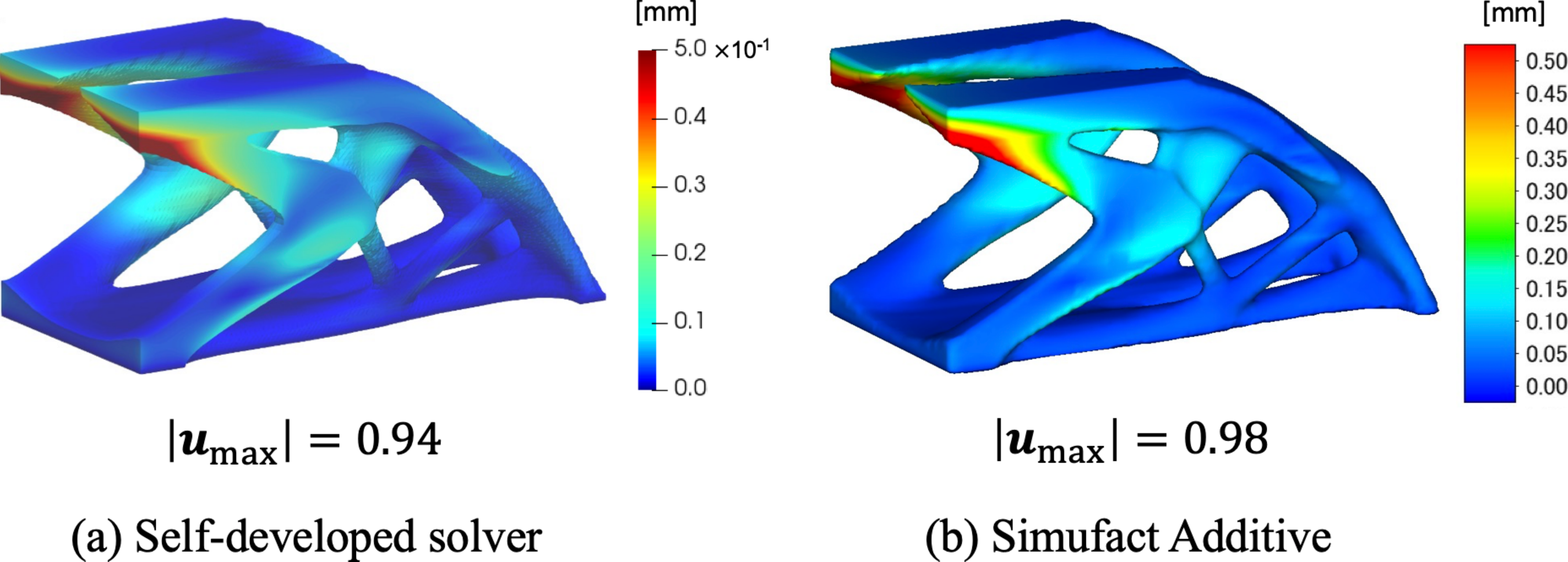}
		\caption{Comparison of distortion obtained from the self-developed solver and Simufact Additive.}
		\label{fig:CantiDistSimu}
	\end{center}
\end{figure}
\begin{figure}[htbp]
	\begin{center}
		\includegraphics[width=14cm]{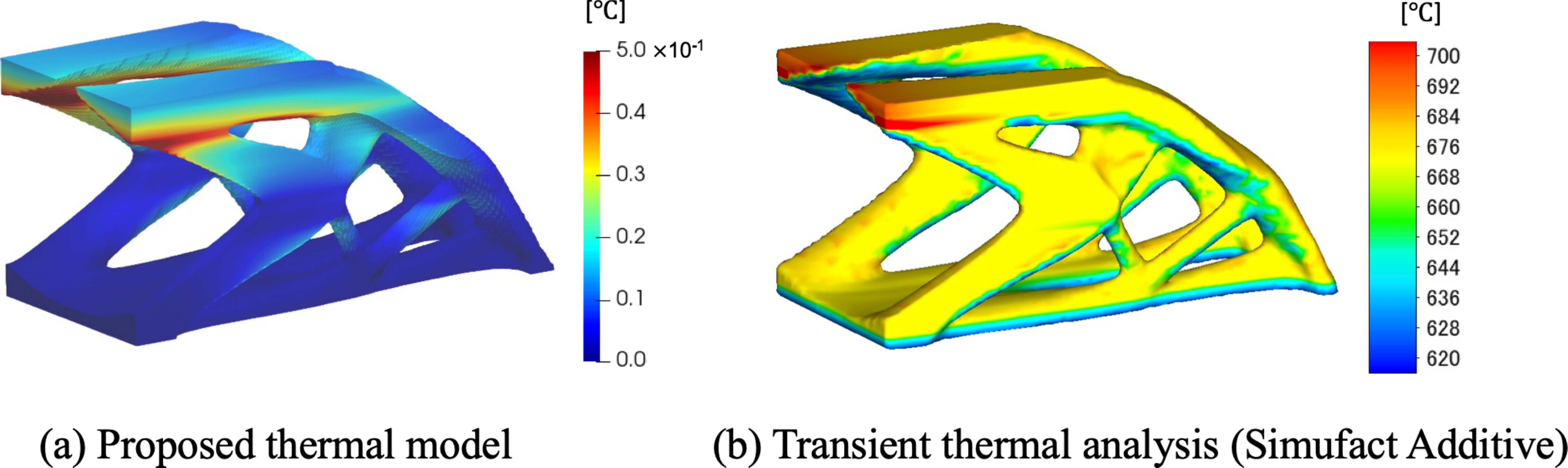}
		\caption{Comparison of peak temperature from the proposed thermal model and transient thermal analysis (Simufact Additive).}
		\label{fig:CantiHSSimu}
	\end{center}
\end{figure}
\begin{figure}[htbp]
	\begin{center}
		\includegraphics[width=14cm]{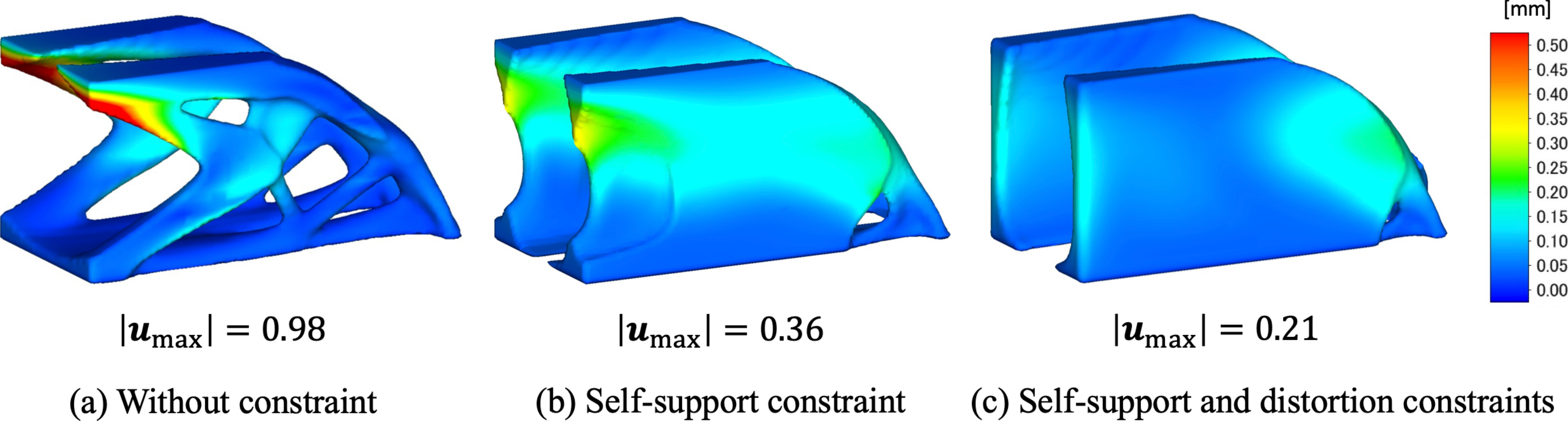}
		\caption{Comparison of distortion induced by the building process for the optimized 3D cantilever beam.}
		\label{fig:CantiDist}
	\end{center}
\end{figure}
\begin{figure}[htbp]
	\begin{center}
		\includegraphics[width=14cm]{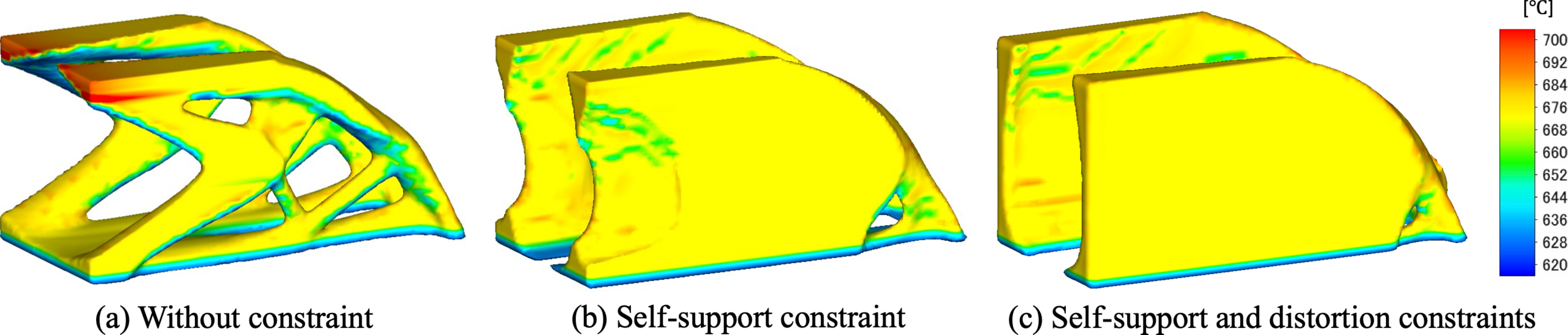}
		\caption{Comparison of the peak temperature for the optimized 3D cantilever beam.}
		\label{fig:CantiHS}
	\end{center}
\end{figure}
Fig. \ref{fig:CantiDistSimu} shows a comparison of distortion obtained with the self-developed inherent strain method and Simufact Additive (Simufact Engineering Gmbh, Hamburg, Germany).
Evidently, the distortion distribution results exhibit an acceptable level of agreement.
Fig. \ref{fig:CantiHSSimu} shows a comparison of the peak temperature for each layer obtained with the proposed thermal model and transient thermal analysis (Simufact Additive).
Note that the temperature field in Fig. \ref{fig:CantiHSSimu} (a) is obtained from the nondimensionalized version of Eq. \ref{eq:drp1}.
The appearance of overheating in the same region indicates that the proposed thermal model detects overheating.
The proposed thermal model is computationally less expensive than a transient thermal analysis; however, if more layers are accumulated, the temperature gradient in the building direction reduces, and overheating may not be evaluated accurately.
Therefore, validation through comparison with optimization results obtained from transient thermal analysis remains a topic for future research.
The following discussion is based on the numerical results obtained from Simufact Additive.
Figs. \ref{fig:CantiDist} and \ref{fig:CantiHS} show a comparison of distortion and peak temperature for each optimal shape.
The optimal shape without a constraint has a large distortion and overheating in the overhanging region.
In contrast, the proposed self-support constraint not only suppresses the creation of members below the threshold angle, but also reduces the distortion and overheating.
The reduction of distortion and overheating are associated with reduced overhanging regions and improved heat dissipation.
In addition, by adding the distortion constraint, the maximum value of distortion $|\bm{u}_{\rm{max}}|$ decreases, and the distortion distribution becomes uniform.
However, the constraints including the building process increase the computational cost because they must be calculated for each layers.
Therefore, in actual design, the process of adding distortion constraints as needed is more realistic than considering all constraints.
\subsubsection{3D heat conduction model}\label{sec:7.2.2}
Next, we consider the thermal diffusivity problem for the 3D heat conduction model.
\begin{figure}[htbp]
	\begin{center}
		\includegraphics[width=11cm]{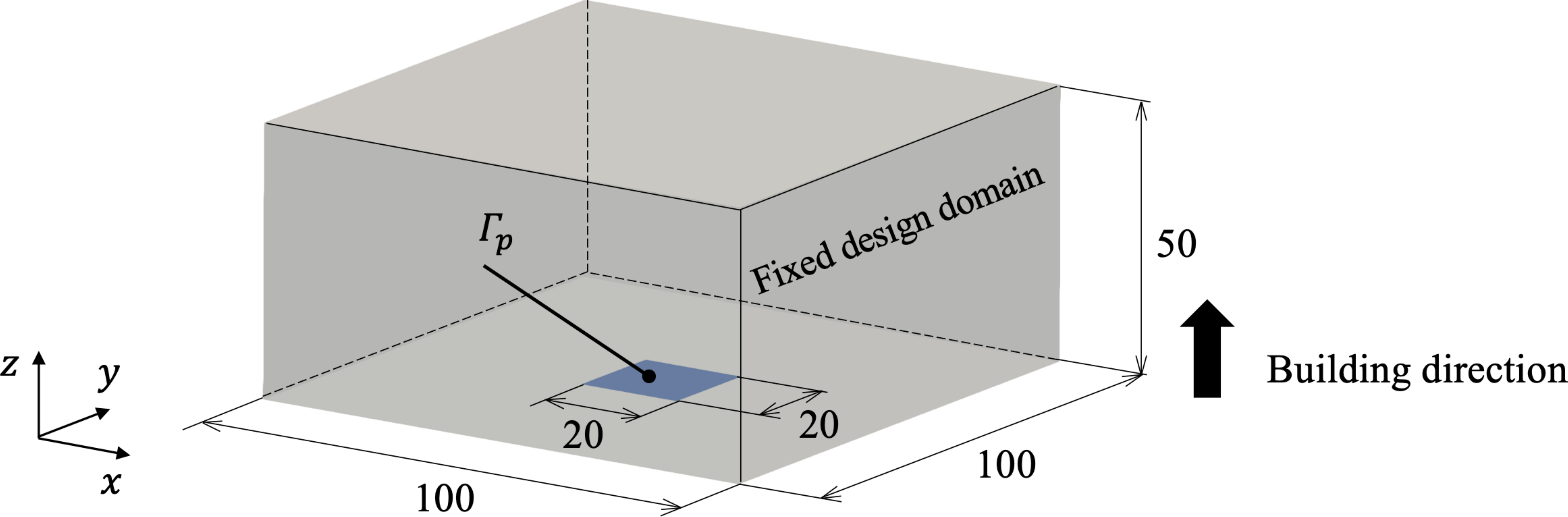}
		\caption{Problem setting for the 3D heat conduction model with the dimensions in mm. The blue surface is the heat sink, and the building direction is the positive z-axis.}
		\label{fig:3DHC}
	\end{center}
\end{figure}
The fixed design domain and boundary conditions are shown in Fig. \ref{fig:3DHC}
The applied heat source $Q$ is set to 10 W, and the temperature is set to $p_{amb}=0^{\circ}C$ in Eqs. \ref{eq:op10} and \ref{eq:op10-1}.
The upper limit of the allowable volume is set to 15\% of the fixed design domain.
The representative length in Eq. \ref{eq:ovhg1} is set to $L=$50 mm.
The threshold overhang angle is set to $\theta_{0}=45^{\circ}$.
The fixed design domain is divided in the building direction into $m = 25$ layers for the thermal constraint and $n = 50$ layers for the distortion constraint.
\begin{figure}[htbp]
	\begin{center}
		\centering
		\subfigure[Bird's-eye view]{\includegraphics[width=7cm]{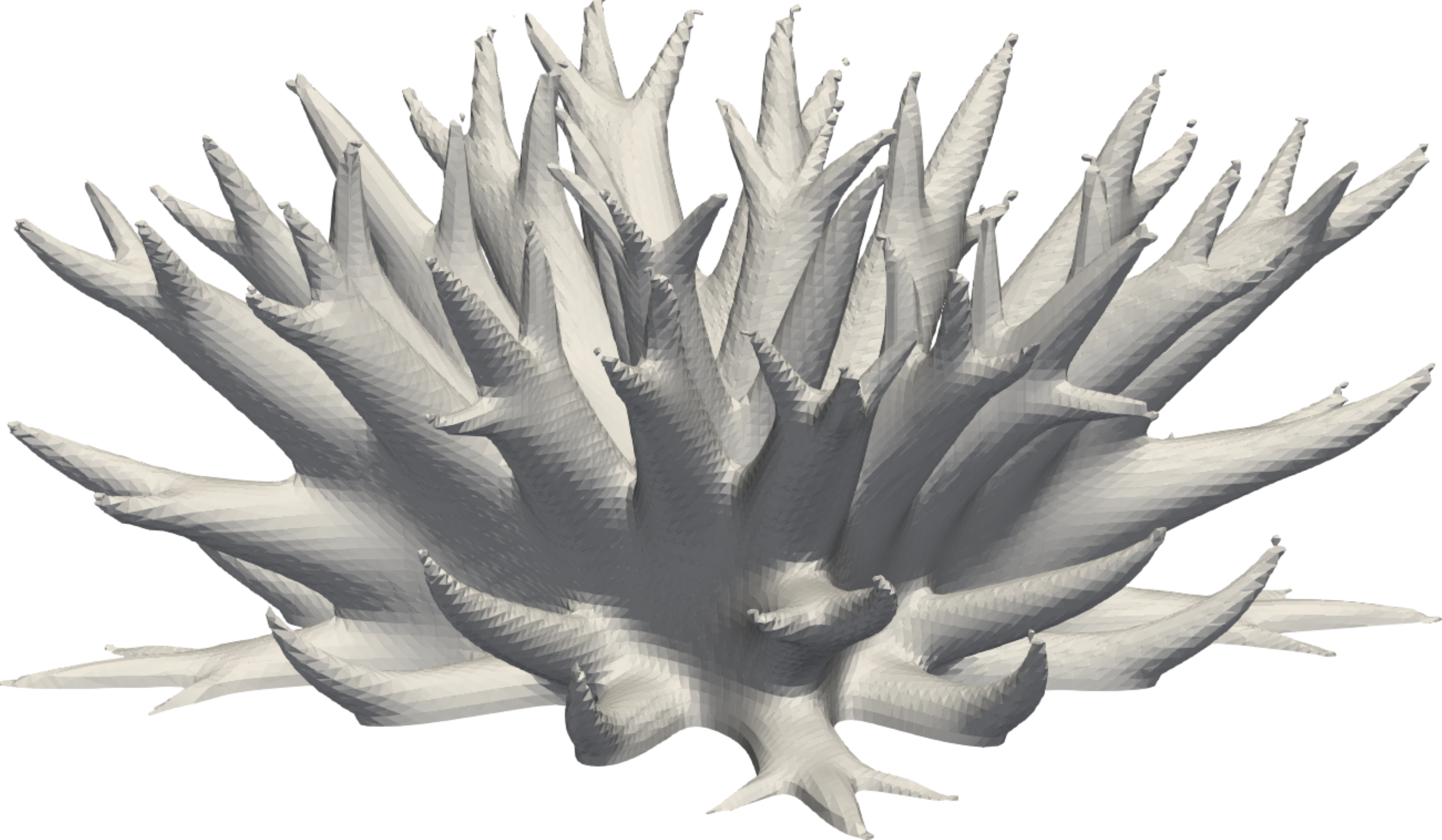}}
		\subfigure[Cross-sectional view]{\includegraphics[width=6cm]{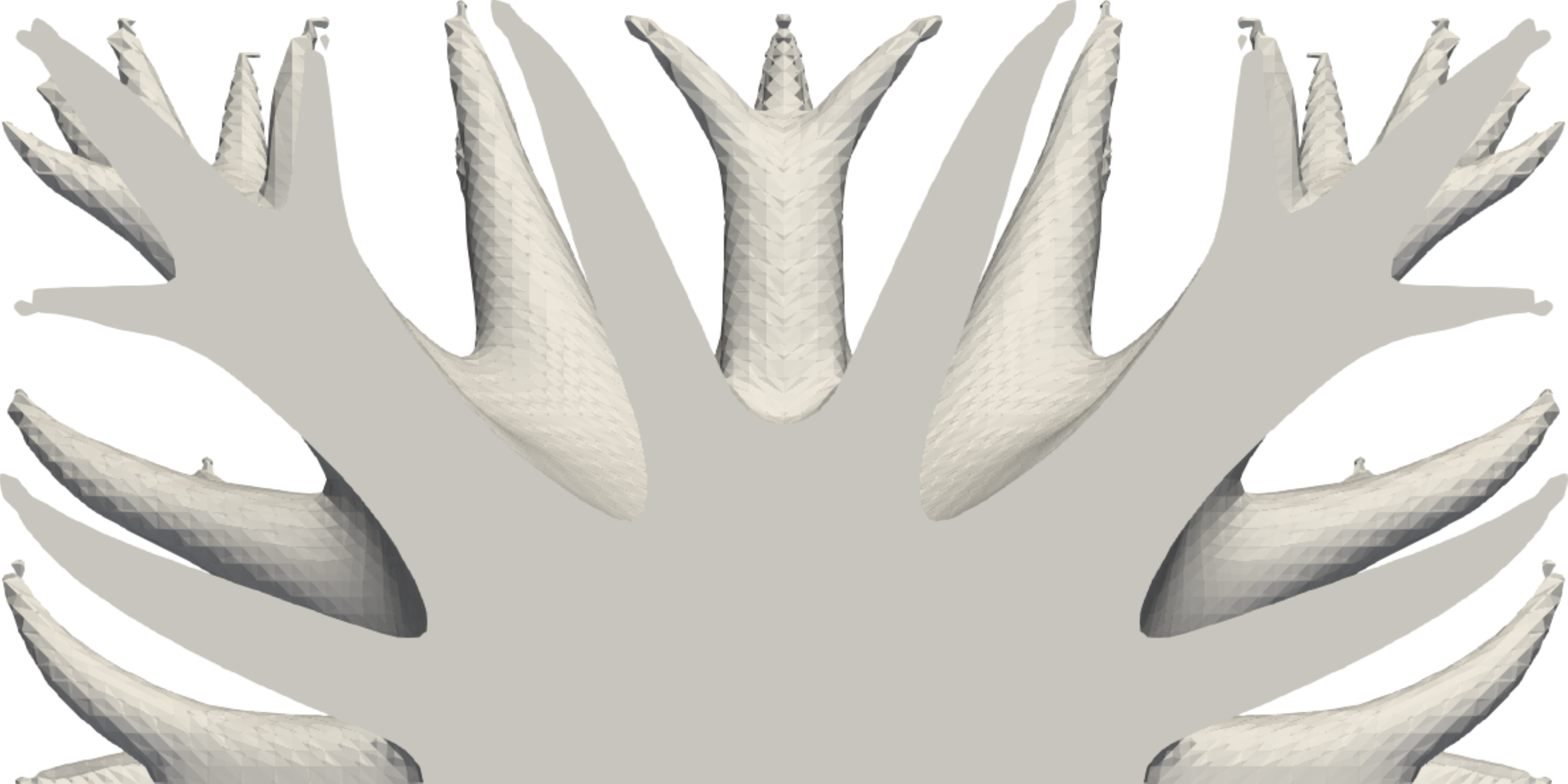}}
		\caption{Optimized 3D heat conduction without constraints; $J_{p~ref}=170495$}
		\label{fig:ne12}
	\end{center}
\end{figure}
\begin{figure}[htbp]
	\begin{center}
		\centering
		\subfigure[Bird's-eye view]{\includegraphics[width=7cm]{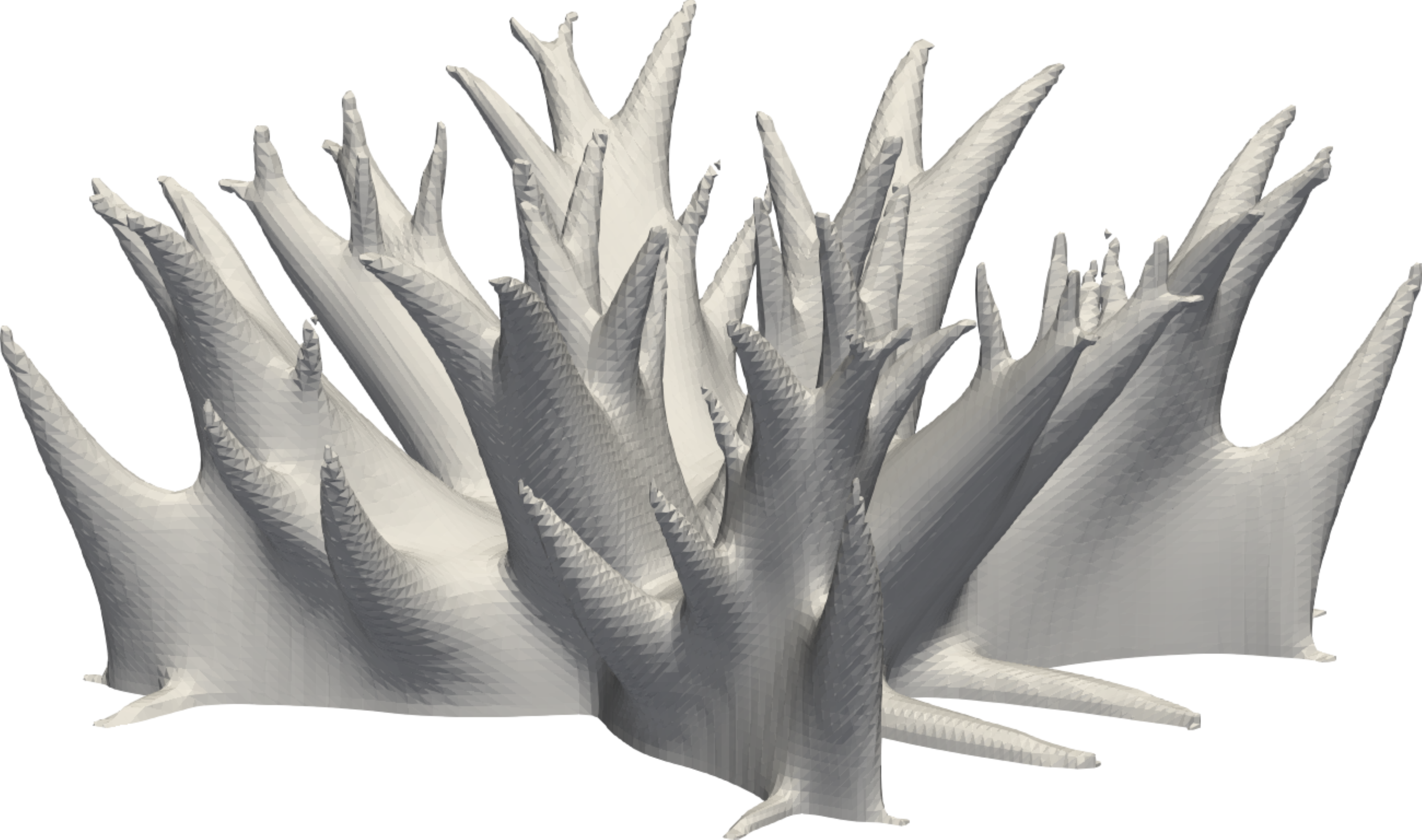}}
		\subfigure[Cross-sectional view]{\includegraphics[width=6cm]{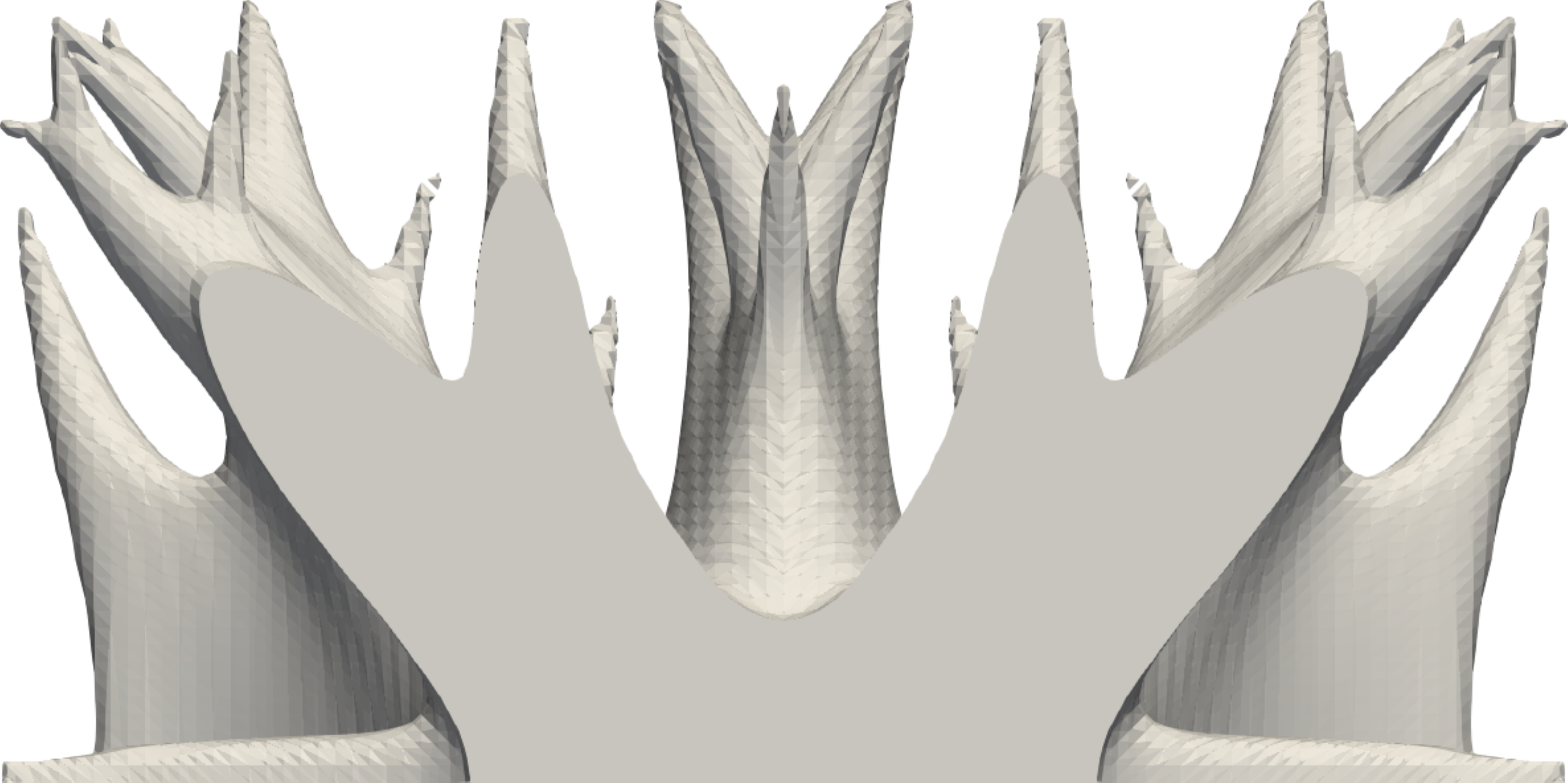}}
		\caption{Optimized 3D heat conduction with the self-support constraint; $J_{p}/J_{p~ref}=109\%$.}
		\label{fig:ne13}
	\end{center}
\end{figure}
\begin{figure}[htbp]
	\begin{center}
		\centering
		\subfigure[Bird's-eye view]{\includegraphics[width=7cm]{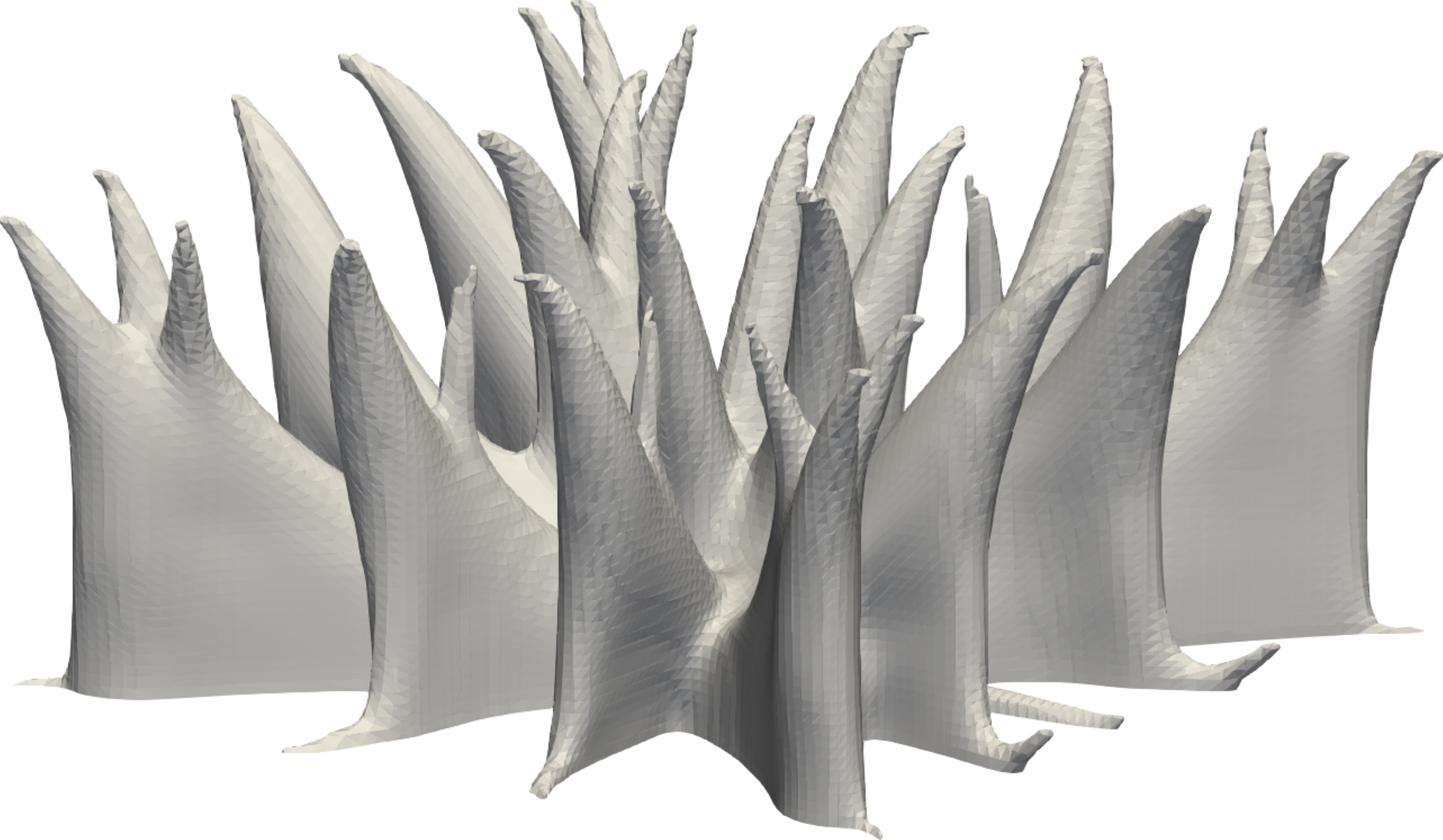}}
		\subfigure[Cross-sectional view]{\includegraphics[width=6cm]{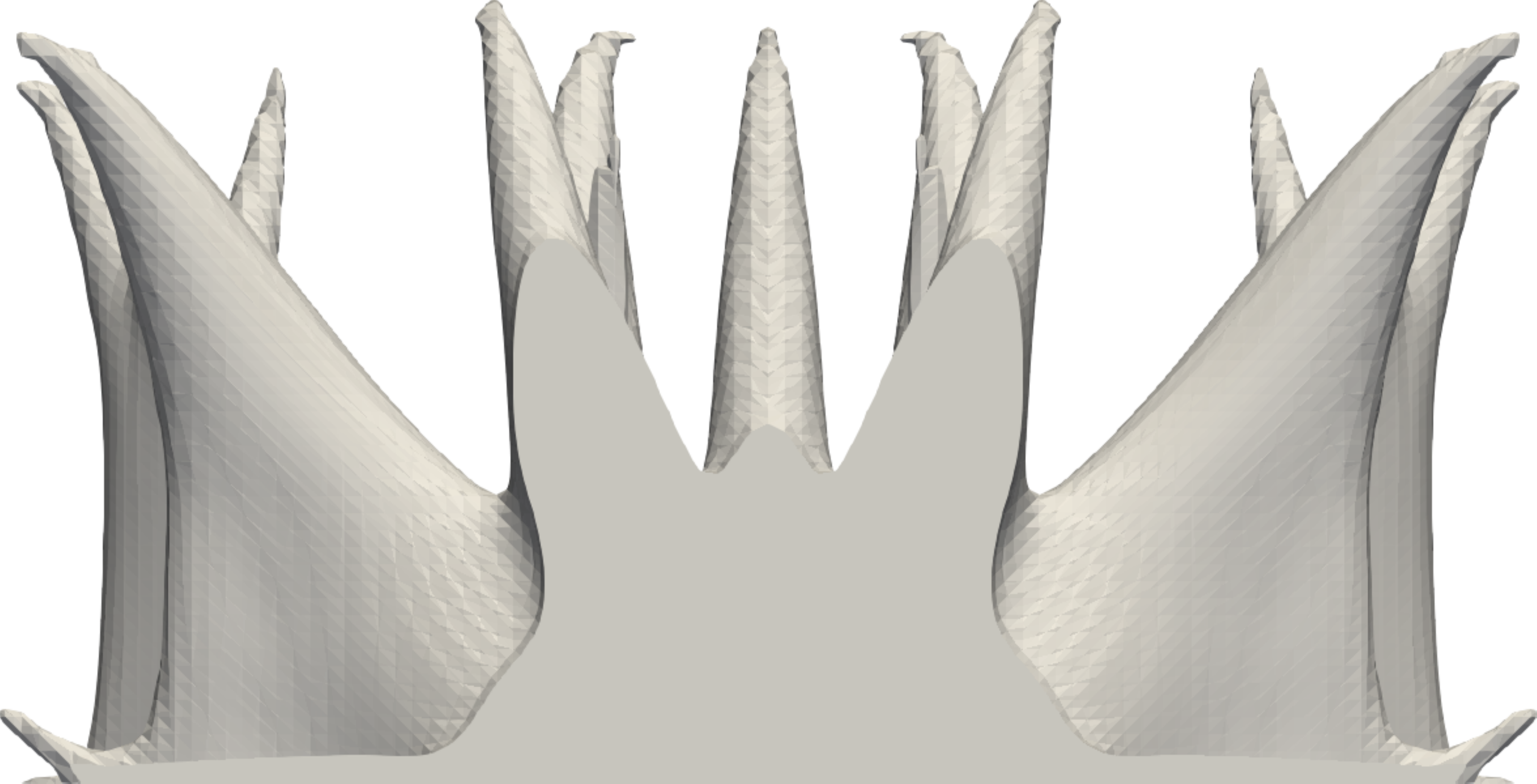}}
		\caption{Optimized 3D heat conduction with the self-support and distortion constraints $J_{p}/J_{p~ref}=112\%$.}
		\label{fig:ne14}
	\end{center}
\end{figure}
\begin{figure}[htbp]
	\begin{center}
		\includegraphics[width=14cm]{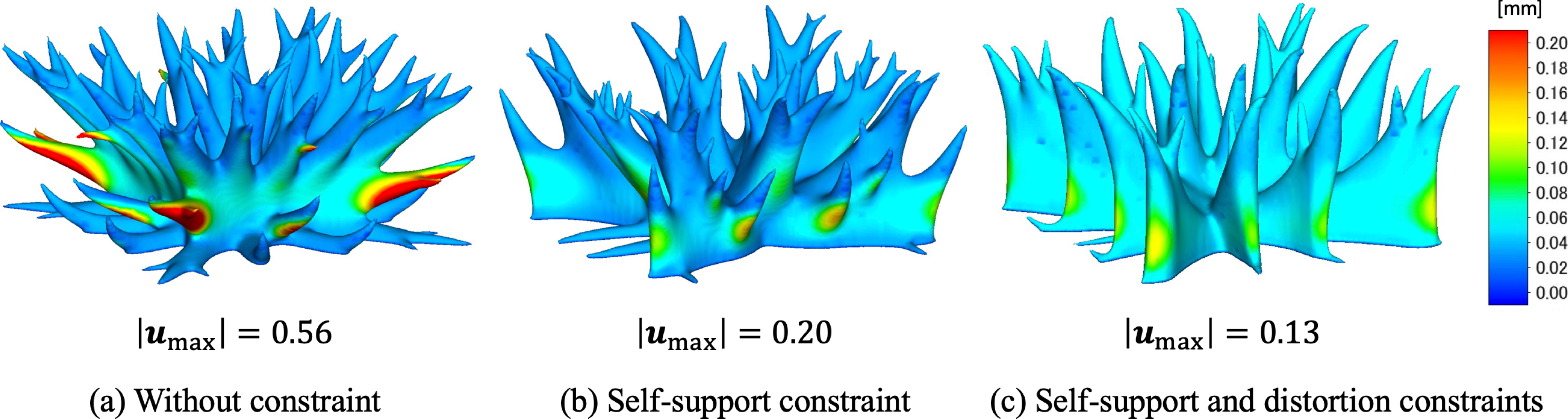}
		\caption{Comparison of the distortion induced by the building process for the optimized 3D heat conduction model.}
		\label{fig:HCDist}
	\end{center}
\end{figure}
\begin{figure}[htbp]
	\begin{center}
		\includegraphics[width=14cm]{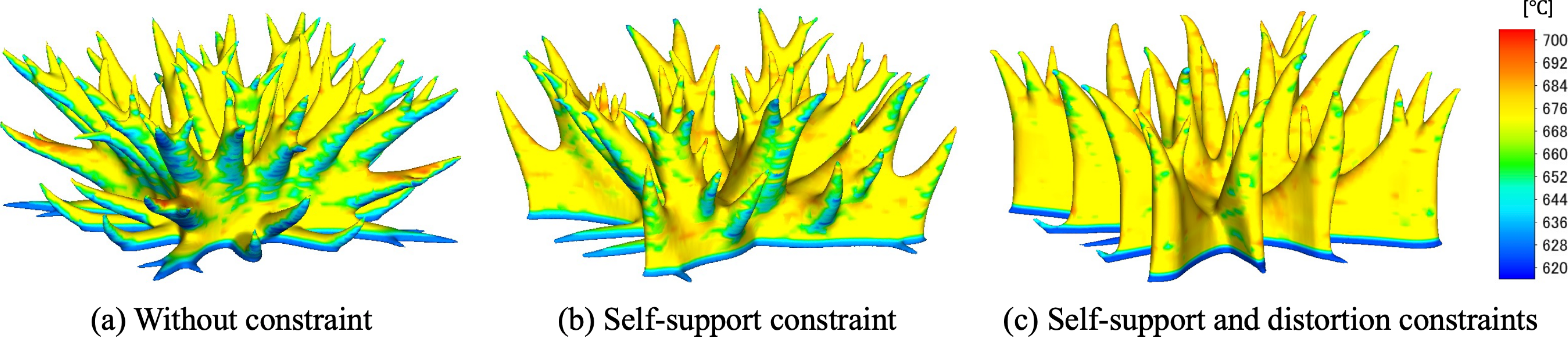}
		\caption{Comparison of peak temperature for the optimized 3D heat conduction model.}
		\label{fig:HCHS}
	\end{center}
\end{figure}
Figs. \ref{fig:ne12}, \ref{fig:ne13}, and \ref{fig:ne14} present the obtained optimization results without a constraint, with the self-support constraint, and with the self-support and distortion constraint, respectively.
The effect of each constraint on the thermal compliance is similar to that of the cantilever beam results.
Figs. \ref{fig:HCDist} and \ref{fig:HCHS} show the numerical results of the distortion and overheating for the optimal shape.
Similar to the cantilever beam, the self-support constraint-imposed shape has no members that violate the overhang angle constraint, reducing distortion and overheating.
This is also true for the distortion constraint.
In the 3D numerical examples, each parameter of the self-support constraint determined through 2D verification is set.
Therefore, each parameter is independent of the problem setting.
These examples demonstrate that the proposed method produces self-supporting shapes that can be manufactured with high precision using AM.
This not only reduces the manufacturing time and costs but also prevents manufacturing failures due to the distortion.
This means that no design changes are required for manufacturability, potentially reducing the product development time and costs.
\section{Conclusion}\label{sec:8}
In this paper, we propose a self-support topology optimization method that considers distortion in the LPBF process.
The main contributions of this study are summarized as follows.
\begin{enumerate}
	\item A Helmholtz-type PDE with adjustable degree of downward convex shapes are proposed, and an overhang angle constraint is formulated using angle vectors and validated through numerical examples.
	Two-dimensional optimization examples show that solely adjusting the penalty parameter $\beta$ and the diffusion coefficient $a$ cannot yield the self-support shape.
	\item A thermal model of the building process is proposed and a thermal constraint that maximizes heat dissipation in each layer is formulated.
	Two-dimensional optimization examples show that the proposed thermal constraint suppresses the downward convex shapes.
	The effect of the constraint parameters on the downward convex shapes and structural performance is also investigated.
	The ability of the proposed thermal model to evaluate overheating is demonstrated through a three-dimensional numerical example.
	\item A mechanical model based on the inherent strain method in the building process is presented, and a constraint to suppress the distortion is formulated.
	\item An unconstrained optimization problem is formulated by including the constraint function as a penalty term in the objective function, and an optimization algorithm is constructed using FEM.
	The method of adjusting each penalty parameter is demonstrated through optimization examples.
	\item Two-dimensional optimization examples indicate the effectiveness of the proposed self-support constraint, and the effect of design changes on structural performance is minimal.
	\item Three-dimensional optimization examples demonstrate that the proposed self-support constraint yields a self-support shape with suppressed distortion and overheating.
	Furthermore, the addition of distortion constraint results in a more uniform distortion distribution.
	The proposed method enables supportless and high-precision manufacturing by AM.
\end{enumerate}
Future research will focus on printing the optimized parts and comparing their effectiveness with and without constraints.
\section{Acknowledgments}
This study was supported by a JSPS grant for Scientific Research (C) JP21K03826.
\bibliography{mybibfile}
\end{document}